\documentclass[a4paper,11pt]{article}
\pdfoutput=1
\usepackage{jcappub}
\usepackage{bbold}
\usepackage{ulem}
 \usepackage{subfigure}

\def\laq{~\raise 0.4ex\hbox{$<$}\kern -0.8em\lower 0.62ex\hbox{$\sim$}~}
\def\gaq{~\raise 0.4ex\hbox{$>$}\kern -0.7em\lower 0.62ex\hbox{$\sim$}~}

\def\beq{\begin{equation}}
\def\eeq{\end{equation}}
\def\bea{\begin{eqnarray}}
\def\eea{\end{eqnarray}}
\def\bean{\begin{eqnarray*}}
\def\eean{\end{eqnarray*}}

\def \dd {\partial}
\def \ra {\rightarrow}

\def \ka {\kappa}

\def \de {\delta}
\def \De {\Delta}

\def \Om {\Omega}
\def \Si {\Sigma}

    \def\be{\begin{equation}}
    \def\ee{\end{equation}}
    \def\ba{\begin{eqnarray}}
    \def\ea{\end{eqnarray}}

\newcommand{\eq}{\begin{equation}}
\newcommand{\eqx}{\end{equation}}
\newcommand{\eqn}{\begin{eqnarray}}
\newcommand{\eqnx}{\end{eqnarray}}
\newcommand{\bx}{{\bf x}}

\newcommand{\nv}{{\bf n}}

\newcommand{\HH}{\mathcal{H} }

\newcommand{\vn}{{\bf n}}
\newcommand{\vk}{{\bf k}}

\newcommand{\bn}{{ \bf n }}
\newcommand{\bk}{{ \bf k }}
\newcommand{\bnabla}{{ \boldsymbol\nabla }}

\newcommand{\Gaunt}[6]{{\cal G}^{#1#2#3}_{#4#5#6}}

\usepackage{mathtools}

\newcommand{\bnx}{\hat{ \bf x }}
\newcommand{\bnk}{\hat{ \bf k }}

\newcommand{\tj}[6]{ \begin{pmatrix}
  #1 & #2 & #3 \\
  #4 & #5 & #6
\end{pmatrix}}

\newcommand{\deta}{r}

\newcommand\spart{\;\raise1.0pt\hbox{$/$}\hskip-6pt\partial}
\newcommand{\sY}[3]{  \ _{#1}\!Y_{#2 #3 }}

\newcommand{\class}{{\sc class}}
\newcommand{\classgal}{{\sc class}gal}

\title{The bispectrum of relativistic galaxy number counts}

\author[a,b,c]{Enea Di Dio }
\author[c]{, Ruth Durrer }
\author[c]{, Giovanni Marozzi }
\author[c,d]{, Francesco Montanari }

\affiliation[a]{
  Osservatorio Astronomico di Trieste, Universit\`a degli Studi di Trieste, Via Tiepolo 11, 34143 Trieste, Italy}
\affiliation[b]{
  INFN, Sezione di Trieste, Via Valerio 2, I-34127 Trieste, Italy
}
\affiliation[c]{
Universit\'e de Gen\`eve, D\'epartement de Physique Th\'eorique and CAP,
24 quai Ernest-Ansermet, CH-1211 Gen\`eve 4, Switzerland
}
\affiliation[d]{
Physics Department, University of Helsinki and Helsinki Institute of Physics\\
P.O. Box 64, FIN-00014, University of Helsinki, Finland
}

\emailAdd{Enea.DiDio@oats.inaf.it}
\emailAdd{Ruth.Durrer@unige.ch}
\emailAdd{Giovanni.Marozzi@unige.ch}
\emailAdd{Francesco.Montanari@helsinki.fi}

\abstract{
We discuss the dominant  terms of the relativistic galaxy number counts to second order in cosmological perturbation theory
on sub-Hubble scales and on intermediate to large redshifts. In particular, we determine their contribution to the bispectrum.
In addition to the terms already known from Newtonian second order perturbation theory, we find that there are a series of additional `lensing-like' terms which contribute to the bispectrum.  We derive analytical expressions for the full leading order bispectrum and we evaluate it numerically for different configurations, indicating how they can be measured with upcoming surveys.
In particular, the  new `lensing-like' terms are not negligible for wide redshift bins
and even dominate the bispectrum at well separated redshifts. This offers us the possibility to  measure them in future surveys.}

\begin{document}
\maketitle

\section{Introduction}
\label{s:intro}
\setcounter{equation}{0}
A present challenge in cosmology is to repeat the big success of CMB observations, see e.g.~\cite{Adam:2015rua,Planck:2015xua}, with large galaxy surveys which are planned and underway~\cite{Samushia:2013yga,Delubac:2014aqe,DES,euclid,SKA}.  The advantage is that galaxies fill  3-dimensional space as compared to the  2-dimensional surface of last scattering, which significantly increases the number of independent modes that can be observed. This is especially important if we want to measure not only the power spectrum (the harmonic transform of the 2-point function) but also the bispectrum (the harmonic transform of the 3-point function).

Primordial non-Gaussianities generated during inflation~\cite{Maldacena:2002vr,Maldacena:2011nz} contain information about physical interactions at inflationary energy scales. However, gravity is itself non-linear and therefore also evolution under gravity after inflation introduces non-Gaussianities even if  primordial fluctuations are Gaussian. It is important to distinguish these gravitational non-Gaussianities from the primordial ones. In order to do so, one has to calculate the bi\-spec\-trum induced at second order in perturbation theory. This has been carried out within Newtonian gravity in great detail, see~\cite{Bernardeau:2001qr} for a review.

In a fully relativistic analysis, we take into account that observations are made on the past light-cone. We cannot observe a constant time hypersurface, but we observe directions and redshifts of galaxies. We then can observe the fluctuation of their number within a small opening angle around a given direction $\bn$ and within a redshift bin around an observed redshift $z$. But also the observed direction and redshift are perturbed. This relativistic number counts have   been determined to first order in perturbation theory in~\cite{Yoo:2009au,Yoo:2010ni,Bonvin:2011bg,Challinor:2011bk}. Here we follow especially the latter two references where the perturbation of the number counts, $\De(\bn,z)$ is consistently considered as a function of observed direction $\bn$ and redshift $z$. These are good coordinates on the past light-cone since they are directly observable.
In a previous paper, Ref.~\cite{DiDio:2014lka}, we have derived the number counts to second order following a purely  geometric approach based on the so-called Geodesic Light-Cone (GLC) coordinates~\cite{Gasperini:2011us} and using the result for the luminosity distance obtained in~\cite{BenDayan:2012wi,Fanizza:2013doa,Marozzi:2014kua}. Results for the second order number counts are also published in~\cite{Bertacca:2014dra,Bertacca:2014wga,Yoo:2014sfa}, and in~\cite{Kehagias:2015tda} the relativistic bispectrum is computed in the squeezed limit (see also~\cite{Yoo:2015uxa}). In~\cite{DiDio:2014lka} we have identified the terms which dominate at sub-Hubble scales and at intermediate to large redshifts. These are quadratic terms which contain in total four transverse derivatives of the Bardeen potentials. Parametrically they are of the same amplitude as the square of density perturbations which means that they can become large.

In~\cite{DiDio:2014lka} we have computed
 the bispectrum from only a few of these dominant terms.
In this paper we  complete the previous analysis. We calculate the bispectrum from all the dominant terms. We also compare our dominant terms with the ones obtained in the Newtonian analysis and find that, apart from the lensing-like terms which are new, they fully agree.
The main difference is that we express the bispectrum, including the `Newtonian terms',  in the directly observable angular and redshift space. Finally, we compare the new lensing terms with the standard Newtonian terms and device a strategy to measure them.

The paper is organized as follows.
In the next section we present the dominant terms of the second order number counts, interpret the different terms and compare our expression with previous Newtonian results. In Section~\ref{s:bipec} we  derive expressions for their contribution to the bispectrum. In Section~\ref{s:res} we then show numerical examples of all the leading contributions to the bispectrum,
and discuss the situations in which the lensing-like terms can be non negligible and even dominate the signal.
In Section~\ref{s:con} we conclude.  Some lengthy derivations are deferred to Appendix~\ref{a:deri}.  We also derive second order expressions for the magnification bias in Appendix~\ref{app:B} and for intensity mapping in Appendix~\ref{app:C}.

\section{Galaxy Number Counts}
\label{Sec2}
\setcounter{equation}{0}
\subsection{Basic ideas}
In galaxy surveys we  observe the number $N$ of galaxies in a redshift bin $dz$ and a solid angle $d\Om$ in terms of the observed redshift $z$ and of the observation direction defined by the unit vector $-\nv$ (in this work $\nv$ denotes the direction of propagation of the  photon).
The fluctuation of galaxy number counts~\cite{Bonvin:2011bg,Challinor:2011bk} is defined as a function of these observable coordinates
\be \label{numcount}
\Delta \left(\nv ,z \right) \equiv \frac{N  \left(\nv ,z \right) - \langle N \rangle \left( z \right)  }{ \langle N \rangle \left( z \right) } \, ,
\ee
where $\langle \cdots \rangle$ denotes the ensemble average over many realisations. In observations this is replaced by an average over observed directions, at a fixed observed redshift.

Traditionally, galaxy density fluctuations have been represented in Fourier space~\cite{Peebles-book}. This has the advantage that in the case of Gaussian fluctuations, the full information is contained in the power spectrum, $P(k)$ and the density fluctuations at different $k$'s,  $\de(\bk_1)$, $\de(\bk_2)$ are, at least theoretically, uncorrelated as a consequence of statistical homogeneity. Correspondingly, the 
bispectrum is also represented in Fourier space, $B(\bk_1,\bk_2)$. It is a function only of the three variables determining the triangle spanned by $\bk_1$ and $\bk_2$. The disadvantage of these variables is the fact that the {\it observed} fluctuations live on the backward light-cone and not on an equal time hypersurface on which we might perform a Fourier transform. To move them onto a fixed time hypersurface one must make assumptions about the time evolution of fluctuations. Furthermore, it is not simple to include line of sight deflection effects (lensing) in such a formalism, as lensing mixes different scales. Furthermore, as mentioned above, the truly measured quantities are redshifts and angles in the sky. To convert them to distances we have to assume a cosmological model, i.e. fix the cosmological parameters. It is then somewhat inconsistent to use the matter power spectrum to determine cosmological parameters, e.g., via the baryon acoustic oscillations (BAO), see~\cite{Anderson:2013oza}.

Because of these arguments, we prefer to describe clustering in the directly observed angle and redshift space. Doing this, already the power spectrum, $C_{\ell}(z_1,z_2)$ depends on three variables, the multipole $\ell$ and two redshifts. As we shall see, the (reduced) 
bispectrum even depends on six variables, $b_{\ell_1\ell_2\ell_3}(z_1,z_2,z_3)$. This of course makes the data more noisy and more difficult to analyze. 
But these spectra also contain much more information.  As for the power spectrum in~\cite{Montanari:2015rga}, we shall  find
for the bispectrum in this work, that we can use number counts not only to determine density fluctuations and redshift space distortions, but when correlating different redshifts, they are also very sensitive to the gravitational potential, especially to the lensing potential. Therefore, this more complex treatment  has in principle the capacity to test General Relativity by  investigating the consistency between the inferred density fluctuations and the lensing potential.

In Ref.~\cite{DiDio:2014lka} we have derived an expression for the number counts for scalar perturbations to second order. The derivation has been performed in the geodesic lightcone  (GLC) gauge~\cite{Gasperini:2011us}, and then expressed in the flat Poisson gauge (or longitudinal gauge) defined by metric perturbations of the form
\bea
ds^2 &=& a^2 \left[ -\left(1 + 2 \Psi \right)  d\eta^2+ \left( 1 - 2 \Phi\right) \delta_{ij} dx^i dx^j \right] \, ,\\
\Phi &=&  \Phi^{(1)} + \Phi^{(2)} \,, \\
\Psi &=& \Psi^{(1)} + \Psi^{(2)} \,.
\eea
Here $\Phi$ and $\Psi$ are the gauge invariant Bardeen potentials which can also be defined in any other gauge.
Note that several publications expand $\Phi =  \Phi^{(1)} + \frac{1}{2}\Phi^{(2)}$, etc. which leads to different factors of $2$ in some of the results. Here, following~\cite{Bernardeau:2001qr}, we do not introduce any factor in front of the second order term. 

Because the perturbed Einstein equations are not known in GLC gauge, the change to longitudinal gauge is necessary to compare the derived results with current and future surveys. Starting from a relatively simple result in  GLC gauge, after  the coordinate transformation, the full expression for $\De^{(2)}(\bn,z)$ obtained in~\cite{DiDio:2014lka} becomes long and cumbersome, filling several pages and we do not want to repeat it here.

However, most of the terms are very small and appreciable only at scales close to the Hubble scale. These terms are most probably not measurable due to cosmic variance.
Our aim is to determine the amplitude of all the terms which appear  at the same parametrical order in $k/\HH$ as the leading Newtonian terms.
At second order these are the terms $\propto (k/\HH)^4\Phi^2 \propto \de^2$. Here $\HH$ denotes the comoving Hubble parameter, $k$ is the wavenumber of the perturbation and $\de$ is the total matter density perturbation.

\subsection{Second order number counts: the dominant terms}
\label{ss:2dom}

We  concentrate here on the `dominant terms'  on intermediate to large redshift and sub-Hubble scales, where the
pure Doppler and potential terms can be neglected. These are the terms with the maximal number of spatial derivatives given in Eq.~(4.45) of Ref.~\cite{DiDio:2014lka}.  All other terms are suppressed by at least one factor $\HH/k$ with respect to these terms.  

We introduce the Weyl potential and the following integrals
\bea
\Phi_W &=& \frac{1}{2}\left(\Phi+\Psi\right)\,,\\
\label{e:lenpot}
\psi(\bn,z) &=& -2\int_0^{r(z)} dr \frac{r(z)-r}{r(z)r}\Phi_W^{(1)}(-r\bn,\eta_0-r)\,, \qquad  -2\ka^{(1)} = \De_2\psi(\bn,z) \,,\\
\label{e:timpot}
\Psi_1(\bn,z)  &=& \frac{2}{r(z) }\int_0^{r(z)} dr \Phi_W^{(1)}(-r\bn,\eta_0-r) \,.
\eea
Here $\eta$ denotes conformal time and $r(z)=\eta_0-\eta(z)$ is the conformal distance to a point on the (background)
 light-cone at redshift $z$. The integral
$\psi(\bn,z)$ is  the (first order) lensing potential and $\ka^{(1)}$ is the first order magnification or convergence, while $\Psi_1(\bn,z)$ is the time delay integral,
$r(z)\Psi_1(\bn,z)$ is the Shapiro time delay. $\De_2$ is the (dimensionless) Laplacian on ${\mathbb S}^2$ with respect to the direction  $\bn$.

The first order number count fluctuations are given by
\bea
\De^{(1)}(\bn,z,) &=& \de^{(1)} +\frac{1}{\HH}\dd^2_rv^{(1)}-2\ka^{(1)} +2\Psi_1
+\frac{1}{\HH}\dot\Phi^{(1)}  - 2\Phi^{(1)} + \Psi^{(1)}  \nonumber \\  &&
+ \left(\frac{{\dot\HH}}{\HH^2}+\frac{2}{r\HH}\right)\left(\Psi^{(1)}+\dd_rv^{(1)}+
 \int_0^{r}\hspace{-0.3mm}d\tilde r(\dot\Phi^{(1)}+\dot\Psi^{(1)})\right)  
   \nonumber \\  &=&  \Si^{(1)} + R^{(1)}  \,,
  \label{e:De1tot}
\eea
where $\Si^{(n)}$ and $R^{(n)}$ identify the leading and subleading terms of nth order, with the leading terms of order  $(k/\HH)^{2n}\Phi^n$ like the density term,
 $\de^{(n)}$.
In these expressions we suppress the arguments which are $\bx(\bn,z) =-r(z)\bn$ and $\eta(z)=\eta_0-r(z)$.
 Here $\de^{(1)}$ is the first order density fluctuation, $v^{(1)}$ is the first order velocity potential so that $v_r=-\dd_rv$ is the radial component of the velocity.

 In this paper we do not include biasing. Neither galaxy nor magnification or evolution bias are considered.  This is not a good approximation, but assuming deterministic bias it is not difficult to insert the corresponding factors $b(z)$, 
 $s(z)$ or $f_{\rm evo}(z)$ for a given survey,
 see e.g.~\cite{Montanari:2015rga}. In this work we set $b(z)=1$ and $s(z)=f_{\rm evo}(z)=0$ in all the plots shown. We do, however discuss the modification of the magnification bias $s(z)$ when second order perturbations are considered
 both in the main text and in Appendix~\ref{app:B}.

The dominant terms of the first order number counts are the first three terms of Eq.~(\ref{e:Si1}). 
They are given by
\be \label{e:Si1}
\Si^{(1)}(\bn,z) = \de^{(1)}(-r(z)\bn,z) + \HH^{-1}(z)\dd_r^2v^{(1)}(-r(z)\bn,z) -2 \ka^{(1)}(-r(z)\bn,z)\,.
\ee
 All other terms are suppressed by at least one factor $\HH/k$. This is correct for $z\gtrsim 0.2$ where $1/r(z) \lesssim k$ and we can neglect also the term proportional to  $(\HH r(z))^{-1}(\dd_rv^{(1)} +\Psi^{(1)})$ which blows up at very small redshift.  The first two terms in (\ref{e:Si1}) are the well known density and redshift space distortion (RSD), while the third term is the lensing term which becomes relevant at high redshift. RSD and lensing represent respectively a radial and a transverse volume distortion, see~\cite{Bonvin:2011bg}.

The second order expression allows an analogous split into leading and sub-leading terms,
\bea
\De^{(2)}(\bn,z,) &=&  \Si^{(2)} + R^{(2)}  - \left\langle \Si^{(2)} + R^{(2)} \right\rangle\,.
  \label{e:De2tot}
\eea
The dominant terms of the number counts at second order can be written in the form $\Si^{(2)}-\langle \Si^{(2)}\rangle$.

The definitions~(\ref{e:lenpot}) and~(\ref{e:timpot}) imply
\bea
\frac{d\psi}{dr} &=& -\frac{1}{r}\Psi_1 \,,
\label{FirstUseEq}
 \\
\frac{d(r\Psi_1)}{dr} &=& 2\Phi^{(1)}_W \,.
\eea

With the help of these identities one can rewrite the expression given in Eq.~(4.45) of~\cite{DiDio:2014lka} in the form
\bea
\Sigma^{(2)}(\bn,z) &=& \delta^{(2)} + \HH^{-1} \partial^2_r v^{(2)}
+ \HH^{-2} \left[ \left( \partial^2_r v\right)^2+ \dd_rv \partial_r^3v \right]
  + \HH^{-1} \left(\dd_r v \partial_r \delta+\partial^2_r v \, \delta\right)
\nonumber \\
&& -2\delta \ka
+ \nabla_a \delta \nabla^a \psi
  +\HH^{-1}\left[-2(\partial^2_r v)  \ka+\nabla_a (\partial^2_r v) \nabla^a \psi \right]
- 2 \ka^{(2)} +2  \ka^2
-2\nabla_b \ka \nabla^b \psi
\nonumber \\
&&
 - \frac{1}{2r(z)} \int_0^{r(z)} \!\!dr\frac{r(z)-r}{r} \Delta_2 \left( \nabla^b \Psi_1 \nabla_b \Psi_1 \right)
- 2 \int_0^{r(z)} \frac{dr}{ r}  \nabla^a \Psi_1\nabla_a \ka \,.
\label{e:Si2}
\eea
Here the
derivatives $\nabla_a$ are covariant derivatives with respect to the direction $-\bn$ and
we have  introduced the second order convergence as
\be
\label{eq:kappa2}
\ka^{(2)} = -\frac{1}{2} \De_2\psi^{(2)} \quad \mbox{ where } \quad
\psi^{(2)}(\bn,z) = -\int_0^{r(z)} dr \frac{r(z)-r}{r(z)r}(\Psi+\Phi)^{(2)}(-r\bn,\eta_0-r) \,.
\ee
Note that our definitions of $\Phi^{(2)}$ and  $\Psi^{(2)}$ here differ by a factor $2$ from those in Ref.~\cite{DiDio:2014lka}. In order to alleviate the notation we suppress the superscript $~^{(1)}$ in all quadratic expressions where it is understood since all our expressions are only correct to second order.

Equation (\ref{e:Si2}) is the most important equation of this section.  We now give the physical interpretation of all its terms. 

First we split it into the Newtonian part given by the first line of (\ref{e:Si2}), which contains only density and velocity terms,
and the rest which contains lensing given by the second and third lines.
The first two terms, $\delta^{(2)} + \HH^{-1} \partial^2_r v^{(2)}$,  are simply the second order Newtonian expression, density and redshift space distortion. The fourth and fifth terms can be written as
$\HH^{-1} \partial_r v\dd_r(\delta + \HH^{-1} \partial^2_r v)$. They take into account, by a first order Taylor expansion, that due to the first order RSD we have to evaluate the fluctuation at a slightly different redshift. The third and sixth terms just account for the radial volume distortion by RSD, which now has to be multiplied also by the first order perturbation, $(\HH^{-1} \partial^2_rv)(\delta + \HH^{-1} \partial^2_r v)$.

The terms which contain a product of Newtonian and lensing contribution can be interpreted like the Newtonian terms, except that now we consider the transverse displacement in the second and forth terms on the second line, $(\nabla^a \psi)\nabla_a \left(\delta +  \HH^{-1} \partial^2_r v\right)$, and the transverse volume distortion in the first and the third terms on the second line, $-2  \ka\left(\delta +  \HH^{-1} \partial^2_r v\right)$.

We now show that all the remaining terms, i.e. the 'pure lensing' contributions to $\Si^{(2)}$, are simply the second order contribution to the determinant of the Jacobian of the lens map,
\bea
{A_a}^b &=& \nabla_a\theta^b \,,\\
{\rm det}\left({A_a}^b\right) &=& (1 +\nabla_1\de\theta^1) (1 +\nabla_2\de\theta^2) - \nabla_1\de\theta^2\nabla_2\de\theta^1  \label{e:detA}
\eea
Here $\theta^a(z,\bn) = -\bn^a +\de\theta^a$ is the angular direction of the source
as a function of the observed direction (to lowest order these are just the polar angles of $-\bn = (\theta^1,\theta^2)^{(0)}$).

To calculate this determinant we remember that our definition of $\ka^{(2)}$ is not  the divergence of the second order deflection angle.  We define this divergence as $1-\ka_{\rm mag}= \nabla_a\theta^a/2$. To first order we obtain $\ka^{(1)}_{\rm mag}=\ka^{(1)}$ while to second order the leading term (i.e.~the term with four derivatives)  for $\ka_{\rm mag}$ is given by
\bea
\ka^{(2)}_{\rm  mag}  &=& \ka^{(2)} +\nabla_b \ka \nabla^b \psi
 + \frac{1}{4}\int_0^{r(z)} \!\!dr\frac{r(z)-r}{r(z)r} \Delta_2 \left( \nabla^b \Psi_1 \nabla_b \Psi_1 \right)  \nonumber \\  &&
+  \int_0^{r(z)} \frac{dr}{ r}  \nabla^a \Psi_1\nabla_a \ka  -\frac{1}{4} (\nabla_b \nabla_a\psi)( \nabla^b \nabla^a\psi)\,.
\label{ka2mag}
\eea
This can be obtained from the corresponding leading deflection contribution computed in~\cite{Bonvin:2015uha,Fanizza:2015swa}, or from the full result which has been calculated in~\cite{BenDayan:2012wi,Fanizza:2015swa}, with the help of the following useful identities
\bea
\int_0^{r(z)} dr \frac{r(z)-r}{r(z)r}\nabla_a\left(\nabla^a\nabla^d\Phi_W\,\nabla_d\psi\right) &=&
\frac{1}{2}\int_0^{r(z)} \frac{dr}{r}\nabla_a\De_2\Psi_1\nabla^a\psi -\frac{1}{4} \nabla_a\nabla_d \psi \nabla^a\nabla^d \psi
\nonumber \\
\label{e:id1}
 &&+\frac{1}{2}\int_0^{r(z)} dr \frac{r(z)-r}{r(z)r}\De_2\left(\nabla_a\Psi_1\nabla^a\Psi_1\right) \,,
 \\
\label{e:id2}
\int_0^{r(z)} \frac{dr}{r}\nabla_a\De_2\Psi_1\nabla^a\psi &=&
-\nabla_a\De_2\psi\nabla^a\psi -\int_0^{r(z)} \frac{dr}{r}\nabla_a\Psi_1\nabla^a\De_2\psi\,.
\hspace{1.2cm}~
\eea

With Eq.~(\ref{ka2mag}) the second order terms of (\ref{e:detA}) are simply
\bea
{\rm det}\left({A_a}^b\right)^{(2)} &=& -2\ka^{(2)}_{(\rm mag)} +\nabla_1\de\theta^1\nabla_2\de\theta^2 - \nabla_1\de\theta^2\nabla_2\de\theta^1 \nonumber \\
&=& - 2\ka^{(2)}_{\rm  mag} +2\ka^2 - \frac{1}{2} (\nabla_b \nabla_a\psi)( \nabla^b \nabla^a\psi)\,  \label{e:Si2lens}\\
&=& \Si_{\rm lens}^{(2)}\,,
\eea
where $\Si_{\rm lens}^{(2)}$ denotes the last three terms of the second line and the full third line of our expression (\ref{e:Si2}), i.e.~the pure lensing terms.
As one naively expects, these pure lensing contributions also to second order are given by the determinant of the Jacobian of the lens map.
Here we neglect vector and tensor perturbations which also appear at second order but are subdominant in our derivative counting.
This pure lensing contribution does not quite agree with the corresponding expression in Ref.~\cite{Bertacca:2014dra}\footnote{While the leading terms of Eq.~(2.10) in \cite{Bertacca:2014dra} do have the same functional form as our Eq.~(\ref{e:Si2lens}), their result for
$\kappa^{(2)}_{\rm mag}$ does not agree with Eq.~(\ref{ka2mag}). The only leading term present in Eq.~(2.12) of \cite{Bertacca:2014dra} for $\kappa^{(2)}_{\rm mag}$ seems to be our $\kappa^{(2)}$.
 The other non-trivial leading terms contributing to $\kappa^{(2)}_{\rm mag}$ are missing.}.

In the remainder of this paper we use Eq.~(\ref{e:Si2}) to determine the bispectrum of the number counts. For this we first have to express the second order perturbations $\de^{(2)},~v^{(2)}$ and $\ka^{(2)}$ in terms of first order quantites.

 The second order density fluctuation, $\de^{(2)}$, velocity potential $v^{(2)}$ and Weyl potential $\Phi_W^{(2)}$ are obtained by solving the perturbed Einstein equations to second order, see~\cite{Acquaviva:2002ud}. The difference of the relativistic solutions from the Newtonian ones are subdominant with respect to our counting, so that we shall use the Newtonian expressions for $\de^{(2)}$, $v^{(2)}$ and  $\Phi_W^{(2)}$ which can be found, e.g.,  in~\cite{Bernardeau:2001qr}. In Fourier space\footnote{We use the following Fourier convention
$$f\left( \bx \right) = \frac{1}{\left( 2 \pi \right)^3 }\int d k^3 \hat f \left( \bk \right) e^{-i \bk \cdot \bx}\, .$$ Note that different Fourier notations can lead to a different sign of the velocity terms.
} one has
\bea\label{e:de2}
\de^{(2)}(\bk,\eta)&=&  \frac{1}{\left( 2 \pi \right)^3} \int d^3k_1 d^3k_2 \delta_D \left( \bk - \bk_1 - \bk_2 \right) F_2 \left( \bk_1, \bk_2 \right) \delta \left( \bk_1 ,\eta\right) \delta\left( \bk_2,\eta \right)\, ,\\
v^{(2)}(\bk,\eta) &=& -\frac{\HH}{k^2}   \frac{f^2(z) }{\left( 2 \pi \right)^3} \int d^3k_1 d^3k_2 \delta_D \left( \bk - \bk_1 - \bk_2 \right) G_2 \left( \bk_1, \bk_2 \right) \delta\left( \bk_1 ,\eta\right) \delta \left( \bk_2,\eta \right)\, ,\label{e:v2} \quad \\
\Phi_W^{(2)}(\bk,\eta) &=& -\frac{3\HH^2\Om_m(\eta)}{2k^2}  \de^{(2)}(\bk,\eta) \,. \label{e:ka2}
\eea
Here $\de_D$ are Dirac-delta functions, $f(z)=d\log D_1/d\log a$ is the growth factor and $D_1$ is the linear growth rate of density perturbations. The kernels $F_2$ and $G_2$ are given by~\cite{Goroff:1986,Bernardeau:2001qr}
\bea
F_2(\bk_1,\bk_2) &=&\frac{5}{7} + \frac{1}{2} \frac{\bk_1 \cdot \bk_2}{k_1 k_2} \left( \frac{k_1}{k_2} + \frac{k_2}{k_1} \right) + \frac{2}{7} \left(\frac{\bk_1 \cdot \bk_2}{k_1 k_2} \right)^2 \nonumber \\   \label{e:F2}
 &=&\frac{17}{21} + \frac{1}{2} \left( \frac{k_1}{k_2} +  \frac{k_2}{k_1} \right) P_1 \left( \hat \bk_1 \cdot \hat \bk_2\right) + \frac{4}{21} P_2 \left( \hat \bk_1 \cdot \hat \bk_2 \right) \,,\\
 G_2(\bk_1,\bk_2) &=&\frac{3}{7} + \frac{1}{2} \frac{\bk_1 \cdot \bk_2}{k_1 k_2} \left( \frac{k_1}{k_2} + \frac{k_2}{k_1} \right) + \frac{4}{7} \left(\frac{\bk_1 \cdot \bk_2}{k_1 k_2} \right)^2  \nonumber \\
 &=&\frac{13}{21} + \frac{1}{2} \left( \frac{k_1}{k_2} +  \frac{k_2}{k_1} \right) P_1 \left( \hat \bk_1 \cdot \hat \bk_2\right) + \frac{8}{21} P_2 \left( \hat \bk_1 \cdot \hat \bk_2 \right) \,,\label{e:G2}
 \eea
where $P_1$ and $P_2$ denote the first and second Legendre polynomials. For more details, see~\cite{Bernardeau:2001qr}.

Note that even though $v^{(2)}$ and $\Phi^{(2)}_W$ are suppressed by factors of $\HH/k^2$ and $\HH^2/k^2$ with respect to $\de^{(2)}$, our terms are $\HH^{-1}\dd_r^2v^{(2)} \propto (k^2/\HH)(\hat\bk\cdot n)^2v^{(2)}$ and $\kappa^{(2)} \sim  \ell^2\Phi_W^{(2)} \sim (k/\HH)^2\Phi_W^{(2)}$.  Hence $\HH^{-1}\dd_r^2v^{(2)}$ and $\ka^{(2)}$ are expected to be of the same order as $\de^{(2)}$.
 Furthermore,  within the Newtonian approximation the potentials  $\Phi$, $\Psi$ and $\Phi_W$, all agree and are equal to the Newtonian potential. In a fully relativistic calculation these potentials differ, but the difference is given by the anisotropic stress which is small, especially for cold dark matter  (of the order of the kinetic energy) and it does not contribute in our determination of the dominant terms.

\subsection{Comparison with the Newtonian calculation}
\label{sec:newtonian}
It is well-known that the evolution of matter perturbations is correctly described by Newtonian gravity on small scales. Nevertheless, we observe correlations by detecting photons, which  travel in a clumpy universe. A proper description of galaxy number counts therefore includes relativistic terms on all scales. On small scales relativistic corrections reduce to lensing-like terms only, as shown in Eq.~(\ref{e:Si2}). Neglecting the lensing-like terms in Eq.~(\ref{e:Si2}), the leading expression simplifies to
\be \label{sigma}
\Sigma_{\text{Newton}} =  \delta^{(2)} + \HH^{-1} \partial_r^2 v^{(2)}+\left( \HH^{-1} \partial^2_r v \right)^2 +\HH^{-2} \partial_r v \partial_r^3 v
 +  \HH^{-1} \partial_r v \partial_r \delta  +  \HH^{-1} \delta \partial_r^2 v \, .
\ee
 We now show,  that this expression is equivalent to the Newtonian result given in
Ref.~\cite{Bernardeau:2001qr} (and references therein). In these references the second order perturbations are given in Fourier space.
We therefore express the real space perturbations in terms of their Fourier modes as follows
\bea
&&\delta^{(2)} \left( \bn , z \right) = \frac{D_1^2}{\left( 2 \pi \right)^6} \int d^3k d^3k_1 d^3k_2\delta_D \left( \bk - \bk_1 - \bk_2 \right) F_2 \left( \bk_1 , \bk_2 \right) \delta\left( \bk_1 \right) \delta \left( \bk_2 \right) e^{ir\left( z \right) \bk \cdot \bn}\, ,
\nonumber
\\
&& \HH^{-1} \partial_r^2 v^{(2)} \left( \bn , z \right) =
  \frac{D_1^2f^2}{\left( 2 \pi \right)^6} \!\int\! d^3k d^3k_1 d^3k_2 \delta_D \left( \bk - \bk_1 - \bk_2 \right) G_2 \left( \bk_1, \bk_2 \right) \delta \left( \bk_1 \right) \delta \left( \bk_2 \right)
  \mu^2 e^{ir\left( z \right) \bk \cdot \bn} \,,
 \nonumber
 \\
&& \left( \HH^{-1} \partial^2_r v\left( \bn , z \right) \right)^2  = \frac{D_1^2f^2}{\left( 2 \pi \right)^6} \int d^3k_1 d^3k_2 \mu_1^2 \mu_2^2   \delta\left( \bk_1 \right) \delta\left( \bk_2 \right) e^{ir\left( z \right) \bk_1 \cdot \bn}e^{ir\left( z \right) \bk_2 \cdot \bn}
  \nonumber
 \\
 &&\hspace{1cm} =\frac{D_1^2f^2}{\left( 2 \pi \right)^6} \int d^3k d^3k_1 d^3k_2\delta_D \left( \bk - \bk_1 - \bk_2 \right) \mu_1^2 \mu_2^2   \delta \left( \bk_1 \right) \delta\left( \bk_2 \right) e^{ir\left( z \right) \bk \cdot \bn} \,,
  \nonumber
  \eea   
  \bea
 && \HH^{-2} \partial_rv \left( \bn , z \right) \partial_r^3 v \left( \bn , z \right)
   \nonumber \\
   &&\hspace{1cm}
    =
  \frac{D_1^2f^2}{2\left( 2 \pi \right)^2} \int d^3k_1 d^3k_2  \frac{\mu_1 \mu_2}{k_1 k_2} \left( \mu_1^2k_1^2+ \mu_2^2k_2^2 \right)  \delta \left( \bk_1 \right) \delta \left( \bk_2 \right) e^{ir\left( z \right) \bk_1 \cdot \bn}e^{ir\left( z \right) \bk_2 \cdot \bn}
  \nonumber \\
   &&\hspace{1cm} =\frac{D_1^2f^2}{2\left( 2 \pi \right)^6} \int d^3k d^3k_1 d^3k_2\delta_D \left( \bk - \bk_1 - \bk_2 \right) \frac{\mu_1 \mu_2}{k_1 k_2} \left( \mu_1^2 k_1^2+ \mu_2^2 k_2^2 \right) \delta \left( \bk_1 \right) \delta \left( \bk_2 \right)   e^{ir\left( z \right) \bk \cdot \bn} \,,
 \nonumber
 \\
&& \HH^{-1} \partial_r v\partial_r \delta=   \frac{D_1^2f}{2\left( 2 \pi \right)^2} \int d^3k_1 d^3k_2 \mu_1 \mu_2 \left(\frac{k_1}{k_2} + \frac{k_2}{k_1} \right)  \delta\left( \bk_1 \right) \delta\left( \bk_2 \right) e^{ir\left( z \right) \bk_1 \cdot \bn}e^{ir\left( z \right) \bk_2 \cdot \bn}
  \nonumber \\
 &&\hspace{1cm} =\frac{D_1^2f}{2\left( 2 \pi \right)^6} \int d^3k d^3k_1 d^3k_2\delta_D \left( \bk - \bk_1 - \bk_2 \right) \mu_1 \mu_2 \left(\frac{k_1}{k_2} + \frac{k_2}{k_1} \right)  \delta\left( \bk_1 \right) \delta \left( \bk_2 \right) e^{ir\left( z \right) \bk \cdot \bn}\,,
   \nonumber
  \eea
 \bea
\hspace*{-0.5cm}   \HH^{-1} \delta\partial_r^2 v  &=&   \frac{D_1^2f}{2\left( 2 \pi \right)^2} \int d^3k_1 d^3k_2 \left(\mu_1^2 + \mu_2^2 \right) e^{ir\left( z \right) \bk_1 \cdot \bn} \delta \left( \bk_1 \right) \delta \left( \bk_2 \right) e^{ir\left( z \right) \bk_2 \cdot \bn}
  \nonumber \\
&=&\frac{D_1^2f}{2\left( 2 \pi \right)^6} \int d^3k d^3k_1 d^3k_2\delta_D \left( \bk - \bk_1 - \bk_2 \right) \left(\mu_1^2 + \mu_2^2 \right)  \delta \left( \bk_1 \right) \delta \left( \bk_2 \right) e^{ir\left( z \right) \bk \cdot \bn}\,,
\eea
where  we have set $\mu \equiv - \hat \bk \cdot \bn,~ \mu_i \equiv - \hat \bk_i \cdot \bn$ and  $v$ is given by the continuity equation to first order~\cite{Bernardeau:2001qr}, $v =- \HH k^{-2} f D_1 \delta $.
Here $\de(\bk)$ is the initial density fluctuation and $\de(\eta,\bk)=D_1(\eta)\de(\bk)$. We consider the matter dominated regime where the growth factor is independent of $k$.
Combining all the terms together we find
\be\label{e2:Newton}
\Sigma_\text{Newton} \left( \bn , z \right) = \frac{D_1^2}{\left( 2 \pi \right)^6} \int d^3k d^3k_1 d^3k_2\delta_D \left( \bk - \bk_1 - \bk_2 \right)Z_2 \left( \bk_1, \bk_2 \right) \delta\left( \bk_1 \right) \delta\left( \bk_2 \right)  e^{ir\left( z \right) \bk \cdot \bn}\,,
\ee
with
\be
Z_2 \left( \bk_1, \bk_2 \right) = F_2 \left( \bk_1 , \bk_2 \right)  + f^2\mu^2 G_2 \left( \bk_1 , \bk_2 \right)  + \frac{f \mu k}{2} \left[ \frac{\mu_1}{k_1} \left( 1+ f \mu_2^2 \right) +\frac{\mu_2}{k_2} \left( 1 + f \mu_1^2 \right) \right]\,.
\ee
This result agrees with the well known  Newtonian analysis, see Eq.~(612) in~\cite{Bernardeau:2001qr}.
As mentioned above, we  neglect the galaxy bias factor in this work.
A detailed discussion of bias  can be found in~\cite{Bernardeau:2001qr} where also a factor $b_2/2$ is included for second order bias. In the relativistic context, care has to be given to the fact that the density in comoving  synchronous gauge has to be multiplied with a bias factor, see~\cite{Bartolo:2015qva}, but we do not address this issue here. Note also that the difference of the density perturbation in different gauges is subleading in our counting.

In addition to galaxy bias, in a realistic survey we have to consider magnification bias which takes into account that a survey usually has a limited sensitivity. Hence if a faint galaxy is lensed by the foreground mass distribution it can make it into a survey even if its apparent luminosity in the unperturbed Universe would be below the sensitivity limit of the catalog. How to include magnification bias to first order is discussed in detail in~\cite{Challinor:2011bk,DiDio:2013bqa}.
The inclusion of magnification bias to second order was recently discussed in \cite{Yoo:2014sfa,Bertacca:2014hwa}.
Since it is not central to the present work, we address this problem within our framework in Appendix~\ref{app:B}, where  we also point out the following interesting feature:
whereas at first order
the lensing term vanishes if the magnification bias is $s=2/5$,
at second order there are always some lensing terms which do not disappear.
Furthermore, new quantities enter the expressions like the perturbation of the magnification bias,
$$-\frac{5}{2}( \de s)^{(1)}\left(z, \bar{L}\right)= (1+\de^{(1)}) \frac{\partial \ln {\rho}}{\partial \ln L} - \frac{\partial \ln \bar{\rho}}{\partial \ln L}
\,,$$
and its derivative
$$
\frac{\partial}{\partial \ln L}\left(\frac{\partial \ln \bar{\rho}}{\partial \ln L}\right)=-\frac{5}{2} t\left(z, \bar{L}\right)\,.
$$
Here $L$ is the luminosity of the source.
If the luminosity dependence of the galaxy density, $\rho(L)$, scales as a power law, $\rho \propto L^p$, we have $s=-(2/5)p$ while $(\de s)^{(1)}=s\,\de^{(1)}$ and $t=0$.

In Appendix~\ref{app:C} we also derive the expressions for our dominant second order terms for the case of intensity mapping, which can be mathematically obtained simply by setting $\rho\propto L^{-1}$, so that $s=2/5$, $(\de s)^{(1)}=2/5\,\de^{(1)}$ and $t=0$.
As mentioned, this corresponds to the configuration for which the first order lensing term disappears, while at second order the transverse deflection $(\nabla^a \psi) (\nabla_a \delta +\HH^{-1} \nabla_a \partial^2_r v)$ survives (see Eq.~(\ref{Leading-Exp-MB-Order2-almost-no-Lensing})).

In  this paper, we  study individually the dominant terms in the bispectrum which for a given survey have to be multiplied by the appropriate bias and magnification bias.


\section{The dominant terms in the bispectrum}
\label{s:bipec}

\subsection{The bispectrum}
\label{ss:bispec}
The bispectrum of the number counts $\De(\bn,z)$ is defined as the spectrum of the connected part of the expectation value
\be
B(\bn_1,\bn_2,\bn_3,z_1,z_2,z_3) \equiv \langle \De(\bn_1,z_1)\De(\bn_2,z_2)\De(\bn_3,z_3)\rangle_c \,.
\ee
We can expand $\De$ in spherical harmonics,
$$\De(\bn,z)=\sum_{\ell=1}^{\infty}\sum_{m=-\ell}^{\ell}a_{\ell m}(z)Y_{\ell m}(\bn)\,, \qquad a_{\ell m}(z)=\int d\Om_{\bn} \De(\bn,z)Y^*_{\ell m}(\bn) $$
and analogously
\be
B(\bn_1,\bn_2,\bn_3,z_1,z_2,z_3) =\sum_{\substack{\ell_1, \ell_2, \ell_3 \\ m_1, m_2,m_3} }  B^{m_1m_2m_3}_{\ell_1\ell_2\ell_3}(z_1,z_2,z_3)Y_{\ell_1m_1}(\bn_1)Y_{\ell_2m_2}(\bn_2)Y_{\ell_3m_3}(\bn_3)  \,,  \label{e:bispecamp}
\ee
where
\bea
B^{m_1m_2m_3}_{\ell_1\ell_2\ell_3}(z_1,z_2,z_3) &=& \left\langle a_{\ell_1 m_1}(z_1)a_{\ell_2 m_2}(z_2)a_{\ell_3 m_3}(z_3)\right\rangle \nonumber \\
&~\hspace*{-3.5cm}=&~\hspace*{-1.5cm}\!\!\!\!\!\! \!\int \!d\Omega_1 d\Omega_2 d\Omega_3 B \left( \nv_1, \nv_2, \nv_3, z_1, z_2 ,z_3 \right) Y^*_{\ell_1 m_1} \!\left( \vn_1 \right) Y^*_{\ell_2 m_2 } \!\left( \vn_2 \right) Y^*_{\ell_3 m_3 }\! \left( \vn_3 \right) \, .
\eea
Statistical isotropy dictates the $m_i$ dependence of the bispectrum,
\be\label{bi_harm_red}
B_{\ell_1\ell_2\ell_3}^{m_1m_2m_3}(z_1,z_2,z_3) =\Gaunt{m_1,}   {m_2,}   {m_3} {\ell_1,} {\ell_2,} {\ell_3} b_{\ell_1,\ell_2,\ell_3}(z_1,z_2,z_3)\,,
\ee
where $\Gaunt{m_1,}   {m_2,}   {m_3} {\ell_1,} {\ell_2,} {\ell_3} $ is the Gaunt integral given by
\bea
\Gaunt{m_1,}   {m_2,}   {m_3} {\ell_1,} {\ell_2,} {\ell_3} &=& \int d \Omega \ Y_{\ell_1 m_1 } \left( \vn \right)Y_{\ell_2 m_2 } \left( \vn \right) Y_{\ell_3 m_3 } \left( \vn \right)  \\
&=&\left(\begin{array}{ccc} \ell_1 & \ell_2 & \ell_3 \\ 0 & 0 & 0\end{array}\right)\left(\begin{array}{ccc} \ell_1 & \ell_2 & \ell_3 \\ m_1 & m_2 & m_3\end{array}\right)
\sqrt{\frac{(2\ell_1+1)(2\ell_2+1)(2\ell_3+1)}{4\pi}} \,.
\eea
On the second line we have expressed the Gaunt integral in terms of the Wigner $3j$ symbols,  see e.g.~\cite{Abram}. The Gaunt integral is non-vanishing only for $m_1+m_2+m_3=0$ and $|\ell_2-\ell_3|\leq \ell_1\leq \ell_2+\ell_3$. Furthermore, the sum $\ell_1+\ell_2+\ell_3$ has to be even.

We assume Gaussian initial condition so that $\langle \De^{(1)}(\bn_1,z_1)\De^{(1)}(\bn_2,z_2)\De^{(1)}(\bn_3,z_3)\rangle_c =0$. We   want to compute the contributions to $b_{\ell_1,\ell_2,\ell_3}(z_1,z_2,z_3)$ coming from one factor $\De^{(2)}$ and two factors $\De^{(1)}$, taking into account only the leading terms $\Si^{(1,2)}$,
i.e.
\be\label{e3:perms}
\langle \De^{(2)} \left(\bn_1, z_1\right)\De^{(1)} \left(\bn_2, z_2\right)\De^{(1)} \left(\bn_3, z_3\right) \rangle_c + \text{permutations}
\ee
In  terms of standard perturbation theory, (\ref{e3:perms}) is the tree-level bispectrum.
 Since the density fluctuation is typically the largest term in $\De^{(1)}$ we approximate $\De^{(1)} \sim \delta^{(1)}$ and we compute this part of the reduced bispectrum, which is given by
\bea
&& b_{\ell_1 \ell_2 \ell_3}  = b^{\delta0}_{\ell_1 \ell_2 \ell_3} + b^{\delta1}_{\ell_1 \ell_2 \ell_3} +b^{\delta2}_{\ell_1 \ell_2 \ell_3}
+b^{v^{(2)'}}_{\ell_1 \ell_2 \ell_3}
+ b^{{v'}^2}_{\ell_1 \ell_2 \ell_3}
+ b^{vv''}_{\ell_1 \ell_2 \ell_3}
+ b^{v \delta'}_{\ell_1 \ell_2 \ell_3}
+ b^{v' \delta}_{\ell_1 \ell_2 \ell_3}
\nonumber \\
&&
+b^{\ka^{(2)}}_{\ell_1 \ell_2 \ell_3}
+b^{\kappa  \delta}_{\ell_1 \ell_2 \ell_3}
+b^{\nabla \delta \nabla \psi}_{\ell_1 \ell_2 \ell_3}
+b^{v' \kappa}_{\ell_1 \ell_2 \ell_3}
+ b^{\nabla v' \nabla \psi}_{\ell_1 \ell_2 \ell_3}
+b^{\kappa^2}_{\ell_1 \ell_2 \ell_3}
+ b^{\nabla \kappa \nabla \psi}_{\ell_1 \ell_2 \ell_3}
+b^{\int \nabla \kappa \nabla \Psi_1}_{\ell_1 \ell_2 \ell_3}
+ b^{\int \Delta_2 \left( \nabla \Psi_1 \nabla \Psi_1 \right)}_{\ell_1 \ell_2 \ell_3}\,.
\nonumber \\
\label{Allterms}
\eea
We have dropped the redshift dependence $(z_1 , z_2 ,z_3)$ for sake of simplicity and  we  denote by $b^{\delta j}_{\ell_1 \ell_2 \ell_3}$
the monopole ($j=0$), dipole ($j=1$, term containing $P_1(\hat\bk_1\cdot\hat\bk_2)$) and quadrupole ($j=2$,
term containing $P_2(\hat\bk_1\cdot\hat\bk_2)$) of
the bispectrum of the second order density term.
We use  $b^{v^{(2)'}}_{\ell_1 \ell_2 \ell_3}$
for the second order redshift space distortion term. As for the density we shall later decompose this term into
$b^{v' j}_{\ell_1 \ell_2 \ell_3}$ with $j=0$ for the monopole,  $j=1$ for the dipole  and $j=2$ for the quadrupole term.
The same decomposition is applied to the other pure second order term
$b^{\ka^{(2)}}_{\ell_1 \ell_2 \ell_3}$ for the second order lensing from
$\Phi^{(2)}_W$. Again, we denote  the monopole ($j=0$), dipole ($j=1$) and quadrupole ($j=2$) terms by  $b^{\ka j}_{\ell_1 \ell_2 \ell_3}$.
The superscripts of the other terms are according to their 2nd order contribution:
\bean
\mbox{the term containing~}~   \left(\dd_r^2 v\right)^2 &~\mbox{ is denoted by }~&  b^{{v'}^2}_{\ell_1 \ell_2 \ell_3}\,, \\
\mbox{the term containing~}~   \dd_rv\dd_r^3v &~\mbox{ is denoted by }~&  b^{vv''}_{\ell_1 \ell_2 \ell_3}\,,\\
\mbox{the term containing~}~    \dd_rv\dd_r\de &~\mbox{ is denoted by }~&  b^{v\de'}_{\ell_1 \ell_2 \ell_3}\,,\\
\mbox{the term containing~}~    \dd^2_rv\de &~\mbox{ is denoted by }~&  b^{v'\de}_{\ell_1 \ell_2 \ell_3}\,,\\
\mbox{the term containing~}~    \ka\de &~\mbox{ is denoted by }~&  b^{\ka\de}_{\ell_1 \ell_2 \ell_3}\,,\\
\mbox{the term containing~}~  \nabla_a\de\nabla^a\psi  &~\mbox{ is denoted by }~&  b^{\nabla\de\nabla\psi }_{\ell_1 \ell_2 \ell_3}\,,\\
\mbox{the term containing~}~   \dd_r^2 v \ \kappa &~\mbox{ is denoted by }~&  b^{v' \kappa}_{\ell_1 \ell_2 \ell_3}\,,\\
\mbox{the term containing~}~  \nabla_a\dd^2_rv\nabla^a\psi   &~\mbox{ is denoted by }~&  b^{\nabla v'\nabla\psi}_{\ell_1 \ell_2 \ell_3}\,,\\
\mbox{the term containing~}~   \kappa^2   &~\mbox{ is denoted by }~&  b^{\kappa^2}_{\ell_1 \ell_2 \ell_3}\,, \\
\mbox{the term containing~}~  \nabla_a\ka\nabla^a\psi  &~\mbox{ is denoted by }~&  b^{\nabla \ka\nabla\psi }_{\ell_1 \ell_2 \ell_3}\,,
\\
\mbox{the term containing~}~ \int_0^{r(z)}\frac{dr}{r}\nabla_a\ka\nabla^a\Psi_1
 &~\mbox{ is denoted by }~&  b^{\int\nabla\ka\nabla\Psi_1 }_{\ell_1 \ell_2 \ell_3}\,,\\
\mbox{the term containing~}~ \int_0^{r(z)}\hspace{-1.2mm}dr \frac{(r(z)-r)}{r(z)r}\De_2(\nabla_a\Psi_1\nabla^a\Psi_1)
 &~\mbox{ is denoted by }~&  b^{\int\De_2(\nabla\Psi_1\nabla\Psi_1)}_{\ell_1 \ell_2 \ell_3} \,.
 \eean

Eq.~(\ref{Allterms}) is the main result of this paper. In the remainder of this section and in Appendix~\ref{a:deri} we show in detail how each of its terms is calculated. In the Section~\ref{s:res} we present numerical results and discuss the importance of the different terms in different configurations.

The terms
$b^{\delta v'}_{\ell_1 \ell_2 \ell_3}$ and $b^{\kappa\delta}_{\ell_1 \ell_2 \ell_3}$
 have already been computed in Ref.~\cite{DiDio:2014lka}. We shall not repeat their calculation here.

The first line of Eq.~(\ref{Allterms}) refers to Newtonian terms, while the second line consists of lensing-like terms, which are not considered in the standard Newtonian analysis in Fourier space, and whose amplitude we want to quantify in this work.

 \subsection{The Newtonian terms}
 To explain the basics, we give here the full derivation of the monopole part of the density-bispectrum $\langle \de^{(2)}(\bn_1,z_1) \de^{(1)}(\bn_2,z_2)\de^{(1)}(\bn_3,z_3)\rangle_c$. The computation of all the other terms is more involved and  is deferred to Section~\ref{ss:2v} and Appendix~\ref{a:deri}. In Eqs.~(\ref{e:de2}) and (\ref{e:F2}) we have seen that $\de^{(2)}$ can be written as the integral of a monopole, a dipole and a quadrupole term.  Let us first just consider the monopole term which we call $\de^{(2)0}$.
We introduce the initial curvature power spectrum from linear perturbation theory by
\be
\label{eq:Pk_def}
\langle R_{\rm in} \left( \bk \right) R_{\rm in} \left( \bk' \right) \rangle = \left( 2 \pi \right)^3 \delta_D \left( \bk + \bk' \right) P_R \left(k \right)\,.
\ee
For a given variable $A$ we define the transfer function $T_A(\eta,k)$ by
\be
A \left( \eta, \bk \right) =T_A(\eta,k)R_{\rm in}(\bk) \,.
\ee
This determines $A$ within linear perturbation theory.
We will also use the angular power spectra given by
\be\label{e:cAB}
c^{AB}_\ell \left( z_1, z_2 \right) = 4 \pi \int \frac{dk}{k} \mathcal{P}_R (k) \Delta^A_\ell (z_1,k)\Delta^B_\ell (z_2,k) = \frac{2}{\pi} \int dk k^2 P_R (k)  \Delta^A_\ell (z_1,k)\Delta^B_\ell (z_2,k)\,,
\ee
where $\mathcal{P}_R (k) = \frac{k^3}{2 \pi^2} P_R (k)$ is the dimensionless primordial power spectrum. $\Delta^A_\ell \left( z, k \right)$ denotes the transfer function in angular and redshift space for the variable $A$.
For example, for the first order density fluctuation $\de$ the Fourier representation
\be
\de(\bn,z) = \frac{1}{\left( 2 \pi \right)^3} \int d^3k \de(k,\eta(z)) e^{ir(z)\bk\cdot\bn}
\ee
yields
\be
\De^\de_{\ell}(z,k) =  T_\de\left(\eta(z) , k \right)j_\ell\left(kr(z)\right)\,,
\ee
where $j_\ell(x)$ is the spherical Bessel function of order $\ell$.
For the monopole term of $\de^{(2)}$,
$$
 \de^{(2)0}
 (\bk_1,\eta) =\frac{17}{21}\frac{1}{\left( 2 \pi \right)^3}\int d^3k\de^{(1)}(\bk,\eta)\de^{(1)}(\bk_1-\bk,\eta) \,,
$$
we then obtain
\bea
& & \langle \de^{(2)0}(\bn_1,z_1)\de^{(1)}(\bn_2,z_2)\de^{(1)}(\bn_3,z_3)\rangle_c \nonumber \\
& &\,\,\,\quad = \frac{17}{21}\frac{2}{\left( 2 \pi \right)^6} \int  d^3 k_2 d^3 k_3   P\left( k_2 \right) P\left( k_3 \right)
T_\delta\left( k_2, \eta_1 \right)T_\delta\left( k_3, \eta_1 \right)
\nonumber\\
&&\,\,\,\quad \quad \times T_\delta \left( k_2, \eta_2 \right) T_\delta \left( k_3 , \eta_3 \right) e^{-i \left(\bk_2 \cdot \bn_1 r_1+ \bk_3 \cdot \bn_1 r_1 \right)} e^{i \left( \bk_2 \cdot \bn_2 r_2 + \bk_3 \cdot \bn_3 r_3 \right)} \, .
\eea
Expanding the Fourier modes in spherical harmonics and Bessel functions
\be
e^{i \vk \cdot {\bf n} r } = 4 \pi \sum_{ \ell m } i^\ell j_\ell \left( k r \right) Y_{\ell m } \left( \nv \right) Y^*_{\ell m } \left( {\bf \hat  k} \right) \, ,
\ee
and integrating over the angles $\hat\bk_2$ and $\hat\bk_3$ applying the orthogonality of spherical harmonics, we find the three-point function
\bea \label{bi_den}
B^{\de 0} \left( \nv_1, \nv_2, \nv_3, z_1, z_2 ,z_3 \right) &=&\frac{4}{\pi^2}  \sum_{\ell, \ell', m , m'}   Y_{\ell m } \left( \vn_1 \right) Y_{\ell' m'} \left( \vn_1 \right) Y^*_{\ell m } \left( \nv_2 \right) Y^*_{\ell' m'} \left( \nv_3 \right)  F_{\ell \ell'} \left( z_1 , z_2 ,z_3 \right) \nonumber \\ &&+ \text{perm.} \, ,
\eea
with
\bea
 F_{\ell \ell'} \left( z_1, z_2 , z_3 \right) &=&\frac{34}{21}\int dk_2 dk_3 k_2^2 k_3^2 P_R \left( k_2 \right) P_R \left( k_3 \right)  T_\delta\left( \eta_1, k_2 \right) T_\delta\left( \eta_1,k_3 \right)T_\delta\left(\eta_2, k_2 \right)  T_\delta\left( \eta_3, k_3\right) \nonumber \\
 && \qquad  \times j_\ell \left( k_2 \deta_1 \right) j_{\ell'} \left( k_3 \deta_1\right) j_\ell \left( k_2 \deta_2 \right) j_{\ell'} \left( k_3 \deta_3 \right)  \, . \label{e:Fll}
\eea
From~(\ref{bi_den}) we infer the reduced bispectrum defined in Eq.~(\ref{bi_harm_red}),
\be
b^{\delta 0}_{\ell_1 \ell_2 \ell_3}=\frac{4}{\pi^2}  F_{\ell_2 \ell_3} \left( z_1 ,z_2 ,z_3 \right) + \text{perm.} \, .
\ee
The double integral in~(\ref{e:Fll}) is simply a product of two single integrals so that we can simplify the result to
\bea
b^{\delta 0}_{\ell_1 \ell_2 \ell_3}&=& \frac{34}{21} \left[\frac{2}{\pi}\int dk_2  k_2^2  P_R \left( k_2 \right)   T_\delta\left( \eta_1, k_2 \right) T_\delta\left(\eta_2, k_2 \right)
  j_{\ell_2} \left( k_2 \deta_1 \right)  j_{\ell_2} \left( k_2 \deta_2 \right) \right] \nonumber \\
 &&\times  \left[\frac{2}{\pi} \int dk_3  k_3^2  P_R \left( k_3 \right)   T_\delta\left( \eta_1, k_3 \right) T_\delta\left(\eta_3, k_3 \right)
  j_{\ell_3} \left( k_3 \deta_1 \right)  j_{\ell_3} \left( k_3 \deta_3 \right) \right] + \text{perm.} \nonumber \\
 &=& \frac{34}{21}\left[ c_{\ell_2}^{\delta\delta} \left( z_1,z_2 \right) c_{\ell_3}^{\delta\delta} \left( z_1,z_3 \right)\!+\! c_{\ell_1}^{\delta\delta} \left( z_1,z_2 \right) c_{\ell_3}^{\delta\delta} \left( z_2,z_3 \right)\!+\!c_{\ell_1}^{\delta\delta} \left( z_1,z_3 \right) c_{\ell_2}^{\delta\delta} \left( z_2,z_3 \right)\right]\, . \quad \label{e:bde0}
\eea

Here $c_\ell^{\de\de}(z_i,z_j)$ is the contribution  to the number count angular power spectrum from density fluctuations. It can be calculated with the publicly available code \classgal{} described in~\cite{DiDio:2013bqa}. The notation `$+$ perm.'
always indicates that all different permutations have to be considered. Since the expressions are always symmetrical in the second and third index, see Eq.~(\ref{e3:perms}), this means that all three values of $(\ell_i,z_i)$ have to be considered once in the first position.
As mentioned, we set the bias $b$ between matter and galaxies equal to one, but it is easy to add a linear and scale-independent bias as outlined in \class{}.

The derivation of the dipole and the quadrupole terms, as well as all the velocity and lensing contributions, are given in Appendix~\ref{a:deri}. We use in particular the definitions in Eq.~(\ref{eq:c_ll}) for the generalized angular spectra $\prescript{n}{}{c}^{AB}_{\ell\ \ell'}(z_1,z_2)$, and Eqs.~(\ref{eq:gb_red}) and (\ref{eq:Q_def}) for the geometrical factors $g_{\ell_1\ell_2\ell_3}$ and $Q_{\ell \ell' \ell''}^{\ell_1 \ell_2 \ell_3}$ describing the bispectra in terms of Wigner 3$j$ and 6$j$ symbols. The results for these contributions to the bispectrum are as follows:
\bea
b^{\delta 1}_{\ell_1 \ell_2 \ell_3}(z_1,z_2,z_3) &=&
\left(g_{\ell_1\ell_2\ell_3}\right)^{-1} \frac{1}{16\pi^2} \sum_{\ell'\ell''} (2\ell'+1)(2\ell''+1) Q_{1\ \ell' \ell''}^{\ell_1 \ell_2 \ell_3}
\nonumber \\
&& \times \left[
  \prescript{1}{}{c}^{\delta \delta}_{\ell'' \ell_1}(z_3,z_1)\,
  \prescript{-1}{}{c}^{\delta \delta}_{\ell' \ell_2}(z_3,z_2)
  +
  \prescript{-1}{}{c}^{\delta \delta}_{\ell'' \ell_1}(z_3,z_1)
  \prescript{1}{}{c}^{\delta \delta}_{\ell' \ell_2}(z_3,z_2)
  \right]
\nonumber \\
&&+\text{perm.} \;,
\label{e:bde1}
\eea
which is non-vanishing only if $\ell'=\ell_2 \pm 1$ and $\ell''=\ell_1 \pm 1$, and
\bea
b^{\delta 2}_{\ell_1 \ell_2 \ell_3}(z_1,z_2,z_3) &=&
\left(g_{\ell_1\ell_2\ell_3}\right)^{-1}
\frac{1}{42\pi^2}
\sum_{\ell'\ell''} (2\ell'+1)(2\ell''+1) Q_{2\ \ell' \ell''}^{\ell_1 \ell_2 \ell_3}
\;\prescript{0}{}{c}^{\delta \delta}_{\ell'' \ell_1}(z_3,z_1) \;\prescript{0}{}{c}^{\delta \delta}_{\ell' \ell_2}(z_3,z_2)
\nonumber \\
&&+\text{perm.} \;,
 \label{e:bde2}
 \eea
which is non-vanishing only if $\ell'=\ell_2\pm 2, \ell_2$ and $\ell''=\ell_1\pm 2$.

Here and in what follows we denote $\eta(z)=\eta$ and $r(z)=\eta_0-\eta(z)=r$. Correspondingly $\eta'=\eta(z')$, etc. .

Defining the transfer functions
\bea
\label{trans_v}
\De^v_{\ell}(z,k)&=&T_v(k,\eta)j_\ell'(kr) \,,\\
\label{trans_v1}
\De^{v'}_{\ell}(z,k)&=&\frac{k}{\HH(z)}T_v(k,\eta)j_\ell''(kr) \,,\\
\label{trans_v2}
\De^{v''}_{\ell}(z,k)&=&\frac{k^2}{\HH^2(z)}T_v(k,\eta)j_\ell'''(kr) \,,
\eea
we can express the reduced bispectrum for the term $\dd_rv\dd_r^3v$ as (see Appendix~\ref{ssa:v3v})
\be
b^{vv''}_{\ell_1 \ell_2 \ell_3}(z_1,z_2,z_3) =\left[c_{\ell_2}^{v \delta} \left( z_1, z_2 \right)c_{\ell_3}^{v'' \delta} \left( z_1, z_3 \right) +c _{\ell_2}^{v'' \delta} \left( z_1, z_2 \right)c_{\ell_3}^{v \delta} \left( z_1, z_3 \right)\right] +\mbox{ perm.}\,,
\label{Newvvpp}
\ee
and (see Appendix~\ref{ssa:v2v2} and \ref{ssa:d1v1} and \cite{DiDio:2014lka} for details)
\bea
b_{\ell_1 \ell_2 \ell_3}^{v'^2}\left(z_1,z_2,z_3\right)&=&
2 c_{\ell_2}^{v'\delta} \left( z_1, z_2 \right)c_{\ell_3}^{v' \delta} \left( z_1, z_3 \right) +  \mbox{perm.}
 \, , \label{e3:v'v'}\\
b^{v\de'}_{\ell_1 \ell_2 \ell_3}(z_1,z_2,z_3) &=&\left[c_{\ell_2}^{v \delta} \left( z_1, z_2 \right)c_{\ell_3}^{\de'\delta} \left( z_1, z_3 \right) +c _{\ell_2}^{\de' \delta} \left( z_1, z_2 \right)c_{\ell_3}^{v \delta} \left( z_1, z_3 \right)\right] +\mbox{ perm.}\,,
\label{Newvdp}\\
\label{rsdxdelta}
 b_{\ell_1 \ell_2 \ell_3}^{v'\delta} \left(z_1,z_2,z_3\right)&=& \left[ c_{\ell_2}^{v' \delta} \left( z_1, z_2 \right)c_{\ell_3}^{\delta \delta} \left( z_1, z_3 \right)
+ c_{\ell_2}^{\delta \delta} \left( z_1, z_2 \right)c_{\ell_3}^{v' \delta} \left( z_1, z_3 \right)\right]
+  \mbox{perm.}
\,,
\eea
where we have also introduced also the transfer function
\be
\label{trans_delta1}
\De^{\de'}_{\ell}(z,k) =\frac{k}{\HH(z)}T_\de(k,\eta)j_\ell'(kr)\,.
\ee
The contributions to the bispectrum given so far can all also be found in~\cite{Bernardeau:2001qr} where they are given in $k$-space in Eq.~(620), which is not directly related to observables. However, the conversion of these results to $k$-space is straightforward and we have checked  that our results are in perfect agreement with Ref.~\cite{Bernardeau:2001qr}. Apart from the $v^{(2)}$ term which we calculate in Section~\ref{ss:2v}, these are all the Newtonian terms.

\subsection{Terms containing lensing}
The contributions which we describe below are new. They are induced by lensing and, as we shall see, are relevant especially for large redshift differences or wide redshift bins. Their translation to $k$-space is not so straightforward as they are integrals over the backward light-cone.  

We express the lensing-like terms by introducing the following transfer functions,
\bea
\label{trans_L}
\De_\ell^\ka(z,k) &=& \ell(\ell+1)\int_0^{r(z)}dr\frac{r(z)-r}{r(z)r}T_{\Psi+\Phi}(\eta_0-r,k)j_\ell(kr)\,,\\
\De_\ell^{\Psi_1}(z,k) &=& \frac{1}{ \ r\left( z \right)} \int_0^{r(z)}dr T_{\Psi+\Phi}(\eta_0-r,k)j_\ell(kr)\,.
\label{trans_psi1}
\eea
With this we find (see Appendix~\ref{ssa:v2ka} to~\ref{ssa:int} for details)
\bea
\label{lenxden}
b^{\ka\de}_{\ell_1 \ell_2 \ell_3}\left(z_1,z_2,z_3\right)&=& \left[ c_{\ell_2}^{\ka \delta}( z_1, z_2) c_{\ell_3}^{\delta \delta}( z_1 , z_3)  +c_{\ell_3}^{\ka \delta}( z_1, z_3) c_{\ell_2}^{\delta \delta}( z_1, z_2) \right]  + \mbox{perm.} \,, \\
b_{\ell_1 \ell_2 \ell_3}^{ v' \kappa}\left(z_1,z_2,z_3\right)&=& \left[
c_{\ell_2}^{v' \delta} \left( z_1, z_2 \right)c_{\ell_3}^{\ka  \delta} \left( z_1, z_3 \right)
+ c_{\ell_2}^{\ka  \delta} \left( z_1, z_2 \right)c_{\ell_3}^{v' \delta} \left( z_1, z_3 \right)
\right] +  \mbox{perm.}\,, \label{e3:v'k}
\\
b^{\ka^2}_{\ell_1 \ell_2 \ell_3}\left(z_1,z_2,z_3\right) &=&  c_{\ell_2}^{\ka \delta}( z_1, z_2) c_{\ell_3}^{\ka \delta}( z_1, z_3)
+   \mbox{perm.}\,,  \label{e3:bkappa2}
\eea
\bea
b^{\nabla\de\nabla\psi}_{\ell_1 \ell_2 \ell_3}\left(z_1,z_2,z_3\right)&=&A_{\ell_1 \ell_2 \ell_3} \left[\sqrt{\frac{\ell_3 \left( \ell_3 + 1 \right)}{\ell_2 \left( \ell_2 + 1 \right)}} c_{\ell_2}^{\ka \delta} \left( z_1, z_2\right) c_{\ell_3}^{\delta \delta} \left( z_1 , z_3 \right)
 \right. \nonumber \\  && \qquad   \left.
~+~ \sqrt{\frac{\ell_2 \left( \ell_2 + 1 \right)}{\ell_3 \left( \ell_3 + 1 \right)}} c_{\ell_3}^{\ka \delta} \left( z_1, z_3\right) c_{\ell_2}^{\delta \delta} \left( z_1 , z_2 \right) \right]
+ \mbox{perm.}\,, \label{e3:nabla-d-nabla-psi}
\\
b^{\nabla v'\nabla\psi}_{\ell_1 \ell_2 \ell_3}\left(z_1,z_2,z_3\right) &=&
A_{\ell_1 \ell_2 \ell_3} \left[\sqrt{\frac{\ell_3 \left( \ell_3 + 1 \right)}{\ell_2 \left( \ell_2 + 1 \right)}} c_{\ell_2}^{\ka \delta} \left( z_1, z_2\right) c_{\ell_3}^{v' \delta} \left( z_1 , z_3 \right) \right. \nonumber \\  && \qquad   \left.
~ + ~  \sqrt{\frac{\ell_2 \left( \ell_2 + 1 \right)}{\ell_3 \left( \ell_3 + 1 \right)}} c_{\ell_3}^{\ka \delta} \left( z_1, z_3\right) c_{\ell_2}^{v' \delta} \left( z_1 , z_2 \right) \right]
+ \mbox{perm.}\,, \label{e3:nabla-v-nabla-psi}
\eea  \bea
b^{\nabla \ka\nabla\psi}_{\ell_1 \ell_2 \ell_3}\left(z_1,z_2,z_3\right)&=&\!
A_{\ell_1 \ell_2 \ell_3} \frac{\ell_2\left( \ell_2 + 1 \right) + \ell_3 \left( \ell_3 + 1 \right)}{\sqrt{\ell_2 \left( \ell_2 + 1 \right) \ell_3 \left( \ell_3 + 1 \right)}} c_{\ell_2}^{\ka \delta} \left( z_1, z_2\right) \!c_{\ell_3}^{\ka  \delta} \left( z_1 , z_3 \right)
\!  + \! \mbox{perm.}, \label{e3:nabla-k-nabla-psi}
\\
b^{\int\nabla\ka\nabla\Psi_1}_{\ell_1 \ell_2 \ell_3 } \left(z_1,z_2,z_3\right)&=&-
A_{\ell_1 \ell_2 \ell_3 } \sqrt{\ell_2 \left( \ell_2 +1 \right)} \sqrt{\ell_3 \left( \ell_3 +1 \right)}
\nonumber \\
&&\hspace{-1cm}
\times
 \int_0^{r_1} \frac{dr}{{  r}}\bigg[
c_{\ell_2}^{\ka  \delta} \left( z, z_2 \right) c_{\ell_3}^{\Psi_1 \delta}\left( z, z_3\right)
 +
 c_{\ell_3}^{\ka  \delta} \left( z, z_3 \right) c_{\ell_2}^{\Psi_1  \delta}\left( z, z_2\right) \bigg]
   + \mbox{perm.}\,,   \label{e3:int-nabla-k-nabla-psi1}
\eea
 \bea
 b^{\int\De_2(\nabla\Psi_1\nabla\Psi_1)}_{\ell_1 \ell_2 \ell_3} \left(z_1,z_2,z_3 \right) &=& -\bigg\{
 A_{\ell_1 \ell_2 \ell_3 } \sqrt{\ell_2 \left( \ell_2 +1 \right) \ell_3 \left( \ell_3 + 1 \right)}\left[ \ell_2 \left( \ell_2 + 1 \right) + \ell_3 \left( \ell_3 + 1 \right) \right]
\nonumber \\
&&
 \quad +C_{\ell_1 \ell_2 \ell_3} \sqrt{\frac{\left( \ell_2 + 2 \right)!}{\left( \ell_2 - 2 \right)!}}\sqrt{\frac{\left( \ell_3 + 2 \right)!}{\left( \ell_3 - 2 \right)!}}
 +\ell_2 \left( \ell_2+1 \right) \ell_3 \left( \ell_3 + 1\right)
 \bigg\}
\nonumber \\
&& \times
 \int_0^{r_1} dr\frac{r_1-r}{r_1 { r}}c_{\ell_2 }^{\Psi_1 \delta} \left( z, z_2 \right)c_{\ell_3}^{\Psi_1 \delta} \left( z  , z_3 \right) + \mbox{perm.}\,.
     \label{e:last}
\eea
Again $A_{\ell_1\ell_2\ell_3}$ and $C_{\ell_1 \ell_2 \ell_3}$ are combinations of generalised Gaunt factors which are defined in
Appendix~\ref{a:deri}, Eqs.~(\ref{def:Afactor}) and (\ref{def:Cfactor}). In the integrals of the last two equations $z\equiv z(r)$. The derivation of Eqs.~(\ref{rsdxdelta}) and~(\ref{lenxden}) can be found in~\cite{DiDio:2014lka}, all other results are derived in Appendix~\ref{a:deri}.

\subsection{Bispectrum for the  second order  velocity and lensing terms}
\label{ss:2v}

All the contributions to the bispectrum computed so far can be written in terms of products of two point functions. But  this is not the case for the second order velocity, $v^{(2)}$, and the second order lensing,  $\ka^{(2)}$ terms. Because of the pre-factors $k^{-2}=(\bk_1+\bk_2)^{-2}$ in Eqs.~(\ref{e:v2}) and~(\ref{e:ka2}), which can not be written as products of functions of $\bk_1$ and $\bk_2$.

Here we derive analytical expression for these second order terms.
They turn out to be numerically much more involved as they require not two single integrals but a full quadruple integral. For the
$\ka^{(2)}$ term the additional integration over the background light-cone even leads to a quintuple integral.

However, since we are interested mainly in orders of magnitudes, we shall simplify these expressions by using the Limber approximation~\cite{Kaiser:1991qi,LoVerde:2008re}.  This is equivalent to the flat sky approximation and should be reasonable for $\ell_i>30$ or so (of course we have in principle 3 different flat skies as we consider 3 different redshifts).
We shall also repeat the derivation of the bispectrum from $\de^{(2)}$ with the Limber approximation. This gives us an indication of the precision of this approach.  The Limber approximation has  been efficiently used for second order weak lensing calculations~\cite{Bernardeau:2011tc} and our treatment is inspired by Ref.~\cite{Bernardeau:2011tc}.

\subsubsection{Second order redshift space distortion}\label{ss:rsd2}
Let us first compute the bispectrum for the redshift space distortion  at second order. We consider
\be
\left\langle \left( \HH^{-1} \partial_{r_1}^2 v^{(2)} \right) \left( \bn_1, z_1 \right) \delta^{(1)} \left( \bn_2 , z_2 \right)  \delta^{(1)} \left( \bn_3 , z_3 \right)  \right\rangle_c
\equiv \langle \cdots \rangle
\ee
where $v^{(2)}$ is given in Eqs. (\ref{e:v2}) and (\ref{e:G2}).
We find
\bea
&& \hspace{-0.8cm} \langle \cdots \rangle
=
-\frac{f^2(z_1)}{\left( 2 \pi \right)^{12}} \int d^3k d^3 k' d^3k_1 d^3 k_2 d^3 k_3 \delta_D \left( \bk_1 - \bk - \bk' \right) \frac{G_2 \left( \bk, \bk' \right)}{k_1^2} \left( \partial_{r_1}^2 e^{i  \bk_1 \cdot \bn_1r_1 } \right)
\nonumber \\
&&
\times \
 e^{i \left( \bk_2 \cdot \bn_2r_2 +  \bk_3 \cdot \bn_3r_3 \right)}
T_\delta \left( k, \eta_1 \right) T_\delta\left( k', \eta_1 \right) T_\delta \left( k_2, \eta_2 \right) T_\delta \left( k_3 , \eta_3 \right)
\langle R \left( \bk \right) R \left( \bk' \right) R \left( \bk_2 \right) R \left( \bk_3 \right) \rangle
\nonumber \\
&=& -\frac{2f^2(z_1)}{\left( 2 \pi \right)^6} \int d^3k_1 d^3 k_2 d^3 k_3 \delta_D \left( \bk_1 +  \bk_2 + \bk_3 \right) G_2 \left( - \bk_2, - \bk_3 \right)    \left( \partial^2_{( k_1 r_1)} e^{i  \bk_1 \cdot \bn_1r_1}\right)
\nonumber \\
&&\times \ e^{ i\left( \bk_2 \cdot \bn_2r_2 +  \bk_3 \cdot \bn_3r_3 \right)} T_\delta \left( k_2, \eta_1 \right) T_\delta\left( k_3, \eta_1 \right) T_\delta \left( k_2, \eta_2 \right) T_\delta \left( k_3 , \eta_3 \right) P \left( k_2 \right) P \left( k_3 \right)
\nonumber
\\
&=&
-\frac{2f^2(z_1)}{\pi^3} \sum_{\ell_i, m_i} \int  d^3k_1 d^3 k_2 d^3 k_3  \delta_D \left( \bk_1 +  \bk_2 + \bk_3 \right) G_2 \left(  \bk_2,  \bk_3 \right) P \left( k_2 \right) P \left( k_3 \right)
\nonumber \\
&&\times \
i^{\ell_1+ \ell_2 + \ell_3} j''_{\ell_1} \left( k_1 r_1 \right) j_{\ell_2} \left( k_2 r_2 \right) j_{\ell_3} \left( k_3 r_3 \right)  Y^*_{\ell_1 m_1 } \left( \hat \bk_1 \right)Y^*_{\ell_2 m_2 } \left( \hat \bk_2 \right)Y^*_{\ell_3 m_3 } \left( \hat \bk_3 \right)
\nonumber \\
&&\times \
Y_{\ell_1 m_1 } \left( \bn_1 \right)Y_{\ell_2 m_2 } \left( \bn_2 \right)Y_{\ell_3 m_3 } \left( \bn_3 \right)
T_\delta \left( k_2, \eta_1 \right) T_\delta\left( k_3, \eta_1 \right) T_\delta \left( k_2, \eta_2 \right) T_\delta \left( k_3 , \eta_3 \right) \,. \label{e:v222}
\eea
Note that here,  when using the Dirac-delta to eliminate $\bk_1$, we obtain $ j''_{\ell_1} \left( |\bk_2+\bk_3| r_1 \right)$ which cannot be written as a product of functions of $\bk_2$ and $\bk_3$ as we always had it  so far.  To continue we
 rewrite the Dirac-delta distribution as
\bea
 \delta_D \left( \bk_1 +  \bk_2 + \bk_3 \right) &=& \frac{1}{(2\pi)^3}\int d^3x e^{i( \bk_1 +  \bk_2 + \bk_3)\bx} \nonumber\\
 &=&8 \sum_{\ell'_i , m'_i} i^{\ell'_1 + \ell'_2 +\ell'_3} \left( -1 \right)^{\ell'_1 + \ell'_2 + \ell'_3 } \Gaunt{m'_1,}   {m'_2,}   {m'_3} {\ell'_1,} {\ell'_2,} {\ell'_3} Y_{\ell'_1 m'_1} \left( \hat \bk_1 \right) Y_{\ell'_2 m'_2} \left( \hat \bk_2 \right) Y_{\ell'_3 m'_3} \left( \hat \bk_3 \right)
 \nonumber \\
 &&
 \times \int_0^\infty d\chi \chi^2 j_{\ell'_1} \left( k_1 \chi \right)j_{\ell'_2} \left( k_2 \chi \right)j_{\ell'_3} \left( k_3 \chi \right)\,,
\eea
and the kernel $G_2$ as
\be
G_2 \left( k_1 , k_2 ,k_3 \right) \equiv G_2 \left( \bk_2 , \bk_3 \right) = \frac{3}{7} + \frac{1}{4} \frac{k_1^2-k_2^2-k_3^2}{k_2 k_3} \left( \frac{k_2}{k_3} + \frac{k_3}{k_2} \right) + \frac{1}{7} \left(\frac{k_1^2-k_2^2-k_3^2}{k_2 k_3} \right)^2 \, .
\label{G2k1k2k3}
\ee
When inserting the third side of our triangle in $k$-space, $k_1=|\bk_2+\bk_3|$, the angular dependence of $G_2$ disappears (or rather is hidden in $k_1$). With this we find
\bea
\langle \cdots \rangle &=&
-\frac{16f^2(z_1)}{\pi^3} \sum_{\ell_i, m_i}\Gaunt{m_1,}   {m_2,}   {m_3} {\ell_1,} {\ell_2,} {\ell_3} \int  dk_1 dk_2 d k_3 k_1^2 k_2^2 k_3^2
 G_2 \left( k_1, k_2,  k_3 \right) P \left( k_2 \right) P \left( k_3 \right)
\nonumber \\
&& \times \
j''_{\ell_1} \left( k_1 r_1 \right) j_{\ell_2} \left( k_2 r_2 \right) j_{\ell_3} \left( k_3 r_3 \right)  T_\delta \left( k_2, \eta_1 \right) T_\delta\left( k_3, \eta_1 \right) T_\delta \left( k_2, \eta_2 \right) T_\delta \left( k_3 , \eta_3 \right)
\nonumber \\
&&
 \times \int_0^\infty d\chi \chi^2 j_{\ell_1} \left( k_1 \chi \right)j_{\ell_2} \left( k_2 \chi \right)j_{\ell_3} \left( k_3 \chi \right)
Y_{\ell_1 m_1 } \left( \bn_1 \right)Y_{\ell_2 m_2 } \left( \bn_2 \right)Y_{\ell_3 m_3 } \left( \bn_3 \right)\,.
\eea
With this we obtain for the bispectrum defined in (\ref{e:bispecamp})
\bea
B^{m_1 m_2 m_3}_{\ell_1 \ell_2 \ell_3} \left( z_1 , z_2 , z_3 \right) &=&
-\frac{16f^2(z_1)}{\pi^3}  \Gaunt{m_1,}   {m_2,}   {m_3} {\ell_1,} {\ell_2,} {\ell_3} \int  dk_1 dk_2 d k_3 k_1^2 k_2^2 k_3^2
 G_2 \left( k_1, k_2,  k_3 \right) P \left( k_2 \right) P \left( k_3 \right)
\nonumber \\
&&
\hspace{-.5cm}
\times \
j''_{\ell_1} \left( k_1 r_1 \right) j_{\ell_2} \left( k_2 r_2 \right) j_{\ell_3} \left( k_3 r_3 \right)  T_\delta \left( k_2, \eta_1 \right) T_\delta\left( k_3, \eta_1 \right) T_\delta \left( k_2, \eta_2 \right) T_\delta \left( k_3 , \eta_3 \right)
\nonumber \\
&&
\hspace{-.5cm}
  \times\int_0^\infty d\chi \chi^2 j_{\ell_1} \left( k_1 \chi \right)j_{\ell_2} \left( k_2 \chi \right)j_{\ell_3} \left( k_3 \chi \right)+ \text{perm.} \,,
  \eea
and the reduced bispectrum becomes
\bea \label{bisp_RSD2}
b^{v^{(2)'}}_{\ell_1 \ell_2 \ell_3 } \left( z_1 , z_2 , z_3 \right) &=&
-\frac{16f^2(z_1)}{\pi^3}  \int  dk_1 dk_2 d k_3 k_1^2 k_2^2 k_3^2
 G_2 \left( k_1, k_2,  k_3 \right) P \left( k_2 \right) P \left( k_3 \right)
\nonumber \\
&&\times \
j''_{\ell_1} \left( k_1 r_1 \right) j_{\ell_2} \left( k_2 r_2 \right) j_{\ell_3} \left( k_3 r_3 \right)  T_\delta \left( k_2, \eta_1 \right) T_\delta\left( k_3, \eta_1 \right) T_\delta \left( k_2, \eta_2 \right) T_\delta \left( k_3 , \eta_3 \right)
\nonumber \\
&&
 \times\int_0^\infty d\chi \chi^2 j_{\ell_1} \left( k_1 \chi \right)j_{\ell_2} \left( k_2 \chi \right)j_{\ell_3} \left( k_3 \chi \right)
~ +  ~ \text{perm.} \,.
\eea
In this expression it is important  to first evaluate the integral over $\chi$. If we would naively change the order of integration and first integrate over $k_j$, the integral would diverge.
However, when integrating over $\chi$ in the situation where,
for example, $k_1>k_2+k_3$ there exists an analytic solution of the $\chi$-integral  (see e.g.~\cite{GradRyz}, integral no. 1 of 6.578)  which can be written in the form
\be \label{int_chi}
\int_0^\infty d\chi \chi^2 j_{\ell_1} \left( k_1 \chi \right)j_{\ell_2} \left( k_2 \chi \right)j_{\ell_3} \left( k_3 \chi \right) \propto \frac{\alpha\left(\frac{k_2}{k_1},\frac{k_3}{k_1};\ell_1,\ell_2,\ell_3\right)}{\Gamma \left( -\frac{\ell_3 + \ell_2 - \ell_1}{2} \right) }\,,
\ee
where the numerator $\alpha$ is finite.
Furthermore, we need to consider only the combinations of $(\ell_1,\ell_2,\ell_3)$ for which the Gaunt factor $\Gaunt{m_1,}   {m_2,}   {m_3} {\ell_1,} {\ell_2,} {\ell_3}$ does not vanish, namely $\ell_1 + \ell_2+ \ell_3$ has to be a even integer and $ \left| \ell_3 - \ell_2 \right| \leq \ell_1 \leq \ell_3 + \ell_2 $. But this implies that the argument of the Gamma function is an non-positive integer. Because the Gamma function has  poles at non-positive integer arguments, the integral~(\ref{int_chi}) vanishes when the triangle inequality for the $k_i$ is violated.
We therefore can impose the condition $k_1<k_2+k_3$ on the integral over $k_1$.
Similarly, we can also impose the conditions $k_2<k_1+k_3$ and $k_3<k_1+k_2$.
Imposing these bounds for the integrations over $k_i$,  the integral converges unconditionally and we are allowed to change the order of integration at will.
We then define
\bea
\label{eq:I_vel}
\mathcal{I}(k_2, k_3 , \chi, r_1,\ell_1) &=& \int dk_1 k_1^2 \Theta\left(k_2+k_3-k_1\right)
\Theta\left(k_1+k_3-k_2\right)
\Theta\left(k_1+k_2-k_3\right) \nonumber \\
& & \times \
G_2(k_1, k_2, k_3) j''_{\ell_1} (k_1 r_1)j_{\ell_1} (k_1 \chi)
\eea
where the $\Theta$ denotes the Heaviside step function, and re-write the reduced bispectrum~(\ref{bisp_RSD2}) as
\bea
b^{v^{(2)'}}_{\ell_1 \ell_2 \ell_3 } \left( z_1 , z_2 , z_3 \right) &=& -\frac{16f^2(z_1)}{\pi^3} \int d\chi  dk_2 dk_3 \chi^2 k_2^2 k_3^2 P_R(k_2) P_R(k_3)  j_{\ell_2} \left( k_2 r_2 \right) j_{\ell_3} \left( k_3 r_3 \right)
 \nonumber \\
&&  ~~\hspace{-3.1cm}   \times
j_{\ell_2} \left( k_2 \chi \right)j_{\ell_3} \left( k_3 \chi \right)T_\delta \left( k_2, \eta_1 \right) T_\delta\left( k_3, \eta_1 \right) T_\delta \left( k_2, \eta_2 \right) T_\delta \left( k_3 , \eta_3 \right) \mathcal{I}(k_2, k_3 , \chi, r_1,\ell_1)
\nonumber \\
 && + \text{perm.}
\,.
\eea
This is a nested quadruple integral which would have to be evaluated numerically with Monte Carlo techniques.  At the present time this goes beyond the capacities of our numerical code and to obtain a first order of magnitude result
 we use the Limber approximation for  the integrals over $k_2$ and $k_3$.
This means we approximate
\be\label{e:limber}
\frac{2}{\pi}\int dkk^2f(k)j_\ell(k x_1)j_\ell(k x_2) \simeq \frac{\de_D(x_1-x_2)}{x_1^2}f\left(\frac{\ell+1/2}{x_1}\right)\,.
\ee

Applying Eq.~(\ref{e:limber}) twice we obtain
\bea
 b^{v^{(2)'}}_{\ell_1 \ell_2 \ell_3 } \left( z_1 , z_2 , z_3 \right) &\simeq&
 -\frac{4f^2(z_1)}{\pi} \frac{\delta_D (r_2 - r_3)}{r_2^2} {P_R}(\nu_2) { P_R}(\nu_3) T_\delta \left( \nu_2, \eta_1 \right)T_\delta \left( \nu_3, \eta_1 \right)  \nonumber \\
 &&   \times \ T_\delta \left( \nu_2, \eta_2 \right) T_\delta \left( \nu_3, \eta_3 \right)
\mathcal{I}(\nu_2, \nu_3 , r_3, r_1,\ell_1)
 + \text{perm.} \, ,
 \label{RSDorder2FR}
\eea
where $\nu_i \equiv \frac{\ell_i +1/2}{r_i}$.
Hence the 2nd order RSD bispectrum of Eq.~(\ref{RSDorder2FR}) has a `contact term'. It is non-vanishing only if the redshifts $z_2$ and $z_3$ are equal. Also $\mathcal{I}(\nu_2, \nu_3 , r_3, r_1,\ell_1)$ is rapidly oscillating if $r_3\neq r_1$ so this is close to another Dirac-delta of $r_3-r_1$.  This is certainly an approximation, but it reflects the fact that the RSD bispectrum rapidly decreases with increasing redshift difference. Note also that we have to convolve $b^{v^{(2)'}}$ with a window function in redshift space to obtain a finite result in this approximation.

We could try to use the Limber approximation also to evaluate Eq.~(\ref{eq:I_vel}).
We start from the identity
\bea
j_{\ell_1}''(k_1 r_1) &=& \frac{1}{(k_1 r_1)^2}\left\{\left[ \ell_1 \left(\ell_1-1\right) - (k_1 r_1)^2\right]j_{\ell_1}(k_1 r_1)+2\, k_1 r_1 \,
j_{\ell_1+1}(k_1 r_1)\right\}
\eea
and, following \cite{Adamek:2015mna} we approximate $j_{\ell_1+1}(x)\simeq j_{\ell_1}(x)$.  Since the Limber approximation is valid for $\ell_1\gg1$ and it implies $\ell_1\sim k_1 r_1$,  applying the Limber approximation of Eq.~(\ref{e:limber}) to Eq.~(\ref{eq:I_vel}) the leading
terms in $\ell_1$ cancel. This is a strong indication that the contribution (\ref{RSDorder2FR}) of the second order RSD to the total bispectrum  is actually subleading with respect to the second order density contribution,  which is given below in Eq.~(\ref{RedBispDeltaOrd2Alt}) in the Limber approximation.

\subsubsection{Second order lensing}
We now compute the bispectrum for the lensing to second order. We start from
\be
\left\langle -2\kappa^{(2)} \left( \bn_1, z_1 \right) \delta^{(1)} \left( \bn_2 , z_2 \right)  \delta^{(1)} \left( \bn_3 , z_3 \right)  \right\rangle_c
\ee
where $\kappa^{(2)}$ is given in Eq.~(\ref{eq:kappa2}).
Following the same procedure used above for the RSD at second order, we obtain the following expression
\bea
b^{\ka^{(2)}}_{\ell_1 \ell_2 \ell_3 } \left( z_1 , z_2 , z_3 \right) &=&
-\frac{48}{\pi^3}
\int_0^{r_1} dr \frac{r_1-r}{r_1r} \HH^2(\eta) \Omega_m(\eta)
\int d\chi dk_2 dk_3 \chi^2 k_2^2 k_3^2 P_R(k_2) P_R(k_3)
\nonumber \\
&&
 ~~\hspace{-3cm}
 \times
 j_{\ell_2}(k_2\chi) j_{\ell_3}(k_3\chi)  j_{\ell_2}(k_2r_2) j_{\ell_3}(k_3r_3)
T_{\delta}(k_2,\eta') T_{\delta}(k_3,\eta') T_{\delta}(k_2,\eta_2) T_{\delta}(k_3,\eta_3)
\nonumber \\
&&
~~\hspace{-3 cm}
\times
{\mathcal{J}}(k_2, k_3 , \chi, r,\ell_1)
+\ \text{perm.} \;,   \label{e:ka2full}
\eea
where the two additional permutations are, as usual, on the redshift and multipole pairs $\{z_i,\ell_i\}$, and
\bea
{\mathcal{J}}(k_2, k_3 , \chi, r,\ell_1) &=&  \ell_1 (\ell_1+1)\int dk_1
\Theta\left(k_2+k_3-k_1\right)
\Theta\left(k_1+k_3-k_2\right)
\Theta\left(k_1+k_2-k_3\right) \nonumber \\
& & \times \
F_2(k_1, k_2, k_3) j_{\ell_1} (k_1 r)
j_{\ell_1} (k_1 \chi)\,,
\label{JNonsolved}
\eea
with
\be
F_2 \left( k_1 , k_2 ,k_3 \right) \equiv F_2 \left( \bk_2 , \bk_3 \right) = \frac{5}{7} + \frac{1}{4} \frac{k_1^2-k_2^2-k_3^2}{k_2 k_3} \left( \frac{k_2}{k_3} + \frac{k_3}{k_2} \right) + \frac{1}{14} \left(\frac{k_1^2-k_2^2-k_3^2}{k_2 k_3} \right)^2 \, .
\label{F2k1k2k3}
\ee

Finally, the Limber approximation for the $k_1$, $k_2$ and $k_3$ integrations in Eqs.~(\ref{e:ka2full}) and~(\ref{JNonsolved}), leads to
\bea
b^{\ka^{(2)}}_{\ell_1 \ell_2 \ell_3 } \left( z_1 , z_2 , z_3 \right) &=& - 24
\Theta(z_1-z_3)
\frac{r_1-r_3}{r_1 r_3} \mathcal{H}^2(\eta_3) \Omega_m(\eta_3)
 \frac{\delta_D (r_2 - r_3)}{ r_3^2} \nonumber
 \\
& &  \times \
 P_R(\nu_2) P_R(\nu_3)T^2_\delta \left( \nu_2, \eta_3 \right)
  T^2_\delta \left( \nu_3, \eta_3 \right)
 \frac{\ell_1 (\ell_1+1)}{(2 \ell_1+1)^2} F_2\left(\frac{\ell_1+1/2}{r_3}, \nu_2, \nu_3\right)
 \nonumber \\
 & &
 + \text{perm.} \,,
  \label{RedBispLensOrd2}
\eea
The three Heaviside step functions of Eq.~(\ref{JNonsolved}) reduce to $\ell_i+\ell_j-\ell_n>0$ which is trivially satisfied for all terms with non-vanishing Gaunt factor.

\subsubsection{Comparison with the second order density}
We have already evaluated the second order density bispectrum exactly. In
Appendix~\ref{a:deri} we obtain the three terms (\ref{e:bde0}),  (\ref{e:bde1})  and  (\ref{e:bde2}).
Here we re-compute the bispectrum for the  density to second order using the Limber approximation in order to evaluate the difference.

Following the same procedure as above we find
\bea
b^{\delta^{(2)}}_{\ell_1 \ell_2 \ell_3 } \left( z_1 , z_2 , z_3 \right) &=&
\frac{16}{\pi^3}
\int d\chi dk_2 dk_3 \chi^2 k_2^2 k_3^2 P_R(k_2) P_R(k_3)  j_{\ell_2}(k_2\chi) j_{\ell_3}(k_3\chi) \nonumber \\
&&
 ~~\hspace{-3cm}
 \times
  j_{\ell_2}(k_2r_2) j_{\ell_3}(k_3r_3)
T_{\delta}(k_2,\eta_1) T_{\delta}(k_3,\eta_1) T_{\delta}(k_2,\eta_2) T_{\delta}(k_3,\eta_3){\mathcal{K}}(k_2, k_3 , \chi, r_1,\ell_1)
\nonumber \\
&&
~~\hspace{-3 cm}
+\ \text{perm.} \,,
\eea
where
\bea
{\mathcal{K}}(k_2, k_3 , \chi, r_1,\ell_1) &=& \int dk_1 k_1^2
\Theta\left(k_2+k_3-k_1\right)
\Theta\left(k_1+k_3-k_2\right)
\Theta\left(k_1+k_2-k_3\right) \nonumber \\
& & \times \
F_2(k_1, k_2, k_3) j_{\ell_1} (k_1 r_1)
j_{\ell_1} (k_1 \chi)\,.
\label{KNonsolved}
\eea
Applying the Limber approximation, Eq.~(\ref{e:limber}), we then obtain
\bea
b^{\delta^{(2)}}_{\ell_1 \ell_2 \ell_3 } \left( z_1 , z_2 , z_3 \right) &=&
2
 \frac{\delta_D (r_2 - r_3)\delta_D (r_1 - r_3)}{r_3^4}P_R(\nu_2) P_R(\nu_3)T^2_\delta \left( \nu_2, \eta_3 \right) \nonumber   \\
& &\!\!\!\!\!
\times\,  T^2_\delta \left( \nu_3, \eta_3 \right)
 F_2\left(\frac{\ell_1+1/2}{r_3}, \nu_2, \nu_3\right)  \hspace{0.1cm} + \text{perm.} \,.
   \label{RedBispDeltaOrd2Alt}
\eea

The result in Eq.~(\ref{RedBispDeltaOrd2Alt}) is useful for two reasons.
First, comparing it with the exact bispectrum obtained in Eqs.~(\ref{e:bde0}),~(\ref{e:bde1}) and~(\ref{e:bde2}), we can test  the accuracy of the Limber approximation when applied to the calculation of the three-point
function.
Secondly,  comparing Eqs.~(\ref{RedBispLensOrd2}) and~(\ref{RedBispDeltaOrd2Alt}), integrated  over window functions  tells us for which situations
 the lensing contribution to second order is comparable to the density contribution to second order. From the above expressions it is clear that the 2nd order lensing bispectrum is significant when two redshifts are equal and the third is larger, while the 2nd order density
 (or RSD) bispectrum only contributes when all redshifts are (nearly) equal.
 Interestingly
 $b^{\ka^{(2)}}_{\ell_1 \ell_2 \ell_3 } \left( z_1 , z_2 , z_3 \right)$ can be obtained from $b^{\de^{(2)}}_{\ell_1 \ell_2 \ell_3 } \left( z_1 , z_2 , z_3 \right)$  upon replacing
 $$
 \de_D(r_1-r_3) \ra - 12 (r_1-r_3)\frac{r_3}{r_1}\Theta(r_1-r_3)\frac{\ell_1(\ell_1+1)}{(2 \ell_1+1)^2}\HH^2(z_3)\Om_m(z_3) \,.
 $$
For window functions of radial size $\HH^{-1}$, we therefore expect that the two contributions are of the same order.


The results in Eqs.~(\ref{RSDorder2FR}),~(\ref{RedBispLensOrd2}) and~(\ref{RedBispDeltaOrd2Alt}) can be split into their
monopole, dipole and quadrupole contribution (comparing Eqs.~(\ref{F2k1k2k3}) and~(\ref{G2k1k2k3}) with Eqs.~(\ref{e:F2}) and~(\ref{e:G2})) in the following way

\bea
b^{v'j}_{\ell_1 \ell_2 \ell_3 } \left( z_1 , z_2 , z_3 \right) &=&  -\frac{4f^2(z_1)}{\pi}   \frac{\delta_D (r_2 - r_3)}{ r_3^2}P_R(\nu_2) P_R(\nu_3) T_\delta \left( \nu_2, \eta_1 \right)T_\delta \left( \nu_3, \eta_1 \right) T_\delta \left( \nu_2, \eta_2 \right) \nonumber\\
&& \qquad  \times \ T_\delta \left( \nu_3, \eta_3 \right)
\mathcal{I}_j(\nu_2, \nu_3 , r_3, r_1,\ell_1) + \text{perm.} \,,
\label{RSDorder2FRj}
\eea
with
\bea
\mathcal{I}_0(k_2, k_3, r_3, r_1,\ell_1)&=&  \int  dk_1 k_1^2
\Theta\left(k_2+k_3-k_1\right)
\Theta\left(k_1+k_3-k_2\right)
\Theta\left(k_1+k_2-k_3\right) \nonumber \\
& & \times \
\frac{13}{21} j''_{\ell_1} (k_1 r_1)j_{\ell_1} (k_1r_3)\,,
\\
\mathcal{I}_1(k_2, k_3, r_3, r_1,\ell_1)&=& \int dk_1 k_1^2
\Theta\left(k_2+k_3-k_1\right)
\Theta\left(k_1+k_3-k_2\right)
\Theta\left(k_1+k_2-k_3\right) \nonumber \\
& & \times \
\frac{1}{4}\frac{k_1^2-k_2^2-k_3^2}{k_2 k_3}\left(
\frac{k_2}{k_3}+\frac{k_3}{k_2}\right) j''_{\ell_1} (k_1 r_1)j_{\ell_1} (k_1r_3)\,,\\
\mathcal{I}_2(k_2, k_3, r_3, r_1,\ell_1)&=& \int dk_1 k_1^2
\Theta\left(k_2+k_3-k_1\right)
\Theta\left(k_1+k_3-k_2\right)
\Theta\left(k_1+k_2-k_3\right) \nonumber \\
& & \times
\left[\frac{1}{7}\left(\frac{k_1^2-k_2^2-k_3^2}{k_2 k_3}\right)^2
\!\!-\frac{4}{21}\right]j''_{\ell_1} (k_1 r_1)j_{\ell_1} (k_1r_3).
\eea
While for lensing and density we have
\bea
b^{\ka j}_{\ell_1 \ell_2 \ell_3 } \left( z_1 , z_2 , z_3 \right) &=&
- 24
\Theta(z_1-z_3)
\frac{r_1-r_3}{r_1 r_3} \mathcal{H}^2(\eta_3) \Omega_m
 \frac{\delta_D (r_2 - r_3)}{ r_3^2}P_R(\nu_2) P_R(\nu_3)
  \nonumber \\
 & & \times \
 T^2_\delta \left( \nu_2, \eta_3 \right)
 T^2_\delta \left( \nu_3, \eta_3 \right)
 \frac{\ell_1 (\ell_1+1)}{(2 \ell_1+1)^2} F^{(j)}_2\left(\frac{\ell_1+1/2}{r_3}, \nu_2, \nu_3\right)+ \text{perm.}   \,,
 \nonumber \\
 & &
 \label{RedBispLensOrd2j}
\\
b^{\delta j}_{\ell_1 \ell_2 \ell_3 } \left( z_1 , z_2 , z_3 \right) &=&
 2
 \frac{\delta_D (r_2 - r_3)\delta_D (r_1 - r_3)}{ r_3^4}P_R(\nu_2) P_R(\nu_3)
 \nonumber \\
 & & \times \
 T^2_\delta \left( \nu_2, \eta_1 \right)
 T^2_\delta \left( \nu_3, \eta_1 \right)
 F_2^{(j)}\left(\frac{\ell_1+1/2}{r_3}, \nu_2, \nu_3\right) + \text{perm.}   \label{RedBispDeltaOrd2Altj}
 \eea
  with
\bea
F_2^{(0)}(k_1,k_2,k_3)&=&\frac{17}{21}\,,
\\
F_2^{(1)}(k_1,k_2,k_3)&=&\frac{1}{4}\frac{k_1^2-k_2^2-k_3^2}{k_2 k_3}\left(
\frac{k_2}{k_3}+\frac{k_3}{k_2}\right)\,,
\\
F_2^{(2)}(k_1,k_2,k_3)&=&\left[\frac{1}{14}\left(\frac{k_1^2-k_2^2-k_3^2}{k_2 k_3}\right)^2
-\frac{2}{21}\right] \,.
\eea
Due to the two Dirac-$\de$'s, the second order density term (\ref{RedBispDeltaOrd2Altj}) is non-vanishing only if all three redshifts coincide. For the lensing term instead is sufficient if two redshifts coincide while the third one is larger than these two. Clearly in this configuration the two density fluctuations in the foreground lens the background density.  In Appendix~\ref{a:deri} we also give the Limber approximation for the $\ka^2$-term, see Eq.~(\ref{ZA8Limber}). In this case there are two line of sight integrals leading to two Heaviside-$\Theta$ functions and the signal remains substantial also when all three redshifts are different. Of course, the Limber approximation is not exact and the signal for the second order density term does not disappear if the redshifts are different, but as we shall see in the next section, for the Newtonian (non-lensing) contributions the signal decays rapidly with increasing redshift difference. Even though for the velocity term (\ref{RSDorder2FRj}) there is only one Dirac-$\de$, the integrals $\mathcal{I}_j$ in practice act like an additional $\de(r_3-r_1)$.

In the next section we evaluate numerically the reduced bispectrum associated to the different contributions given in Sec.
\ref{ss:bispec} and \ref{ss:2v}.
For this purpose we will divide our terms in three different types: Newtonian terms, Newtonian~$\times$~lensing terms and pure lensing contribution.
The Newtonian terms are the one given by the first line of Eq.~(\ref{Allterms}), the Newtonian~$\times$~lensing ones are combinations of lensing terms with Newtonian ones and are given by the first five terms of the second line of Eq.~(\ref{Allterms}), including the $\kappa^{(2)}$-term, and the pure lensing terms are the last four of the second line of Eq.~(\ref{Allterms}).

Let us, in particular, explain why we consider $\kappa^{(2)}$ as a Newtonian~$\times$~lensing term.
From a naive point of view, one might be tempted to consider $\kappa^{(2)}$ as a pure second order lensing term.
On the other hand, considering its form and  Eq.~(\ref{RedBispLensOrd2}) it is evident that this term resembles more to a
combination of a lensing term with the density.
To clarify this let us apply the Limber approximation to the bispectrum of the $\kappa \delta$ term.
Starting from the result of~\cite{DiDio:2014lka}, where the reduced bispectrum of this term was given in detail. Using the Poisson equation~(\ref{e:ka2}) and applying the
Limber approximation we obtain
\bea
b^{\ka \delta}_{\ell_1 \ell_2 \ell_3 } \left( z_1 , z_2 , z_3 \right) &= & - 12
\Theta(z_1-z_3)
\frac{r_1-r_3}{r_1 r_3} \mathcal{H}^2(\eta_3) \Omega_m(\eta_3)
 \frac{\delta_D (r_1 - r_2)}{ r_2^2} \nonumber
 \\ & &   \qquad \times
 P_R(\nu_2) P_R(\nu_3)T^2_\delta \left( \nu_2, \eta_2 \right)
  T^2_\delta \left( \nu_3, \eta_3 \right)
 \frac{\ell_3 (\ell_3+1)}{(2 \ell_3+1)^2}
  \nonumber \\
 & & -12
\Theta(z_1-z_2)
\frac{r_1-r_2}{r_1 r_2} \mathcal{H}^2(\eta_2) \Omega_m(\eta_2)
 \frac{\delta_D (r_1 - r_3)}{ r_3^2} \nonumber
 \\
& &  \qquad \times
 P_R(\nu_2) P_R(\nu_3)T^2_\delta \left( \nu_2, \eta_2 \right)
  T^2_\delta \left( \nu_3, \eta_3 \right)
 \frac{\ell_2 (\ell_2+1)}{(2 \ell_2+1)^2}
  \nonumber \\
 & &
 + \text{perm.} \,.
  \label{KapaDeltaLimber}
\eea
Comparing Eq.~(\ref{KapaDeltaLimber}) to Eq.~(\ref{RedBispLensOrd2}), clearly the
bispectrum from $\kappa \delta$ and $\kappa^{(2)}$ are very similar (apart from the kernel $F_2$, which is not present in
(\ref{KapaDeltaLimber})).
The difference is simply that the $\delta$ term has fixed redshift for $\kappa \delta$, while it is
inside the integral along the line of sight for $\kappa^{(2)}$.
As we shall see in the next section, Eq.~(\ref{RedBispLensOrd2}) produces similar results as the other
 Newtonian~$\times$~lensing terms
when we consider our bispectrum integrated over window functions (see Sec.~\ref{WinFunction}).
As a consequence, we  consider $\kappa^{(2)}$  as a Newtonian~$\times$~lensing term, and not as a
pure lensing terms.
This also indicates that this second order lensing effect plays a role also when two redshifts are equal, a configuration where the other pure lensing terms are negligible as we shall see.

\section{ Numerical examples and how to measure lensing terms}
\label{s:res}

In this section we show some numerical examples of our results. The figures have been generated with the  cosmological parameters $h=0.67$, $\omega_b= 0.022$, $\omega_{cdm} = 0.12$, and vanishing curvature. The primordial curvature power spectrum has the amplitude $A_s= 2.215 \times 10^{-9}$, the pivot scale $k_{pivot} =0.05 \ \text{Mpc}^{-1}$, the spectral index  $n_s = 0.96$ and no running.
All the transfer functions needed to compute the spectra are evaluated with a modified version of the \class{} code \cite{Blas:2011rf,DiDio:2013bqa}.

\begin{figure}[htbp]
  \begin{center}
    \includegraphics[width=7.5cm]{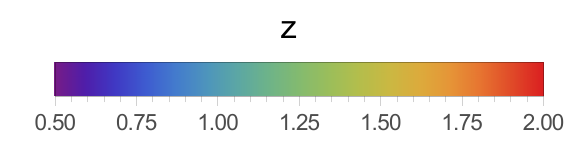} \hspace{7.1cm} \phantom{.} \\
    \includegraphics[width=7.5cm]{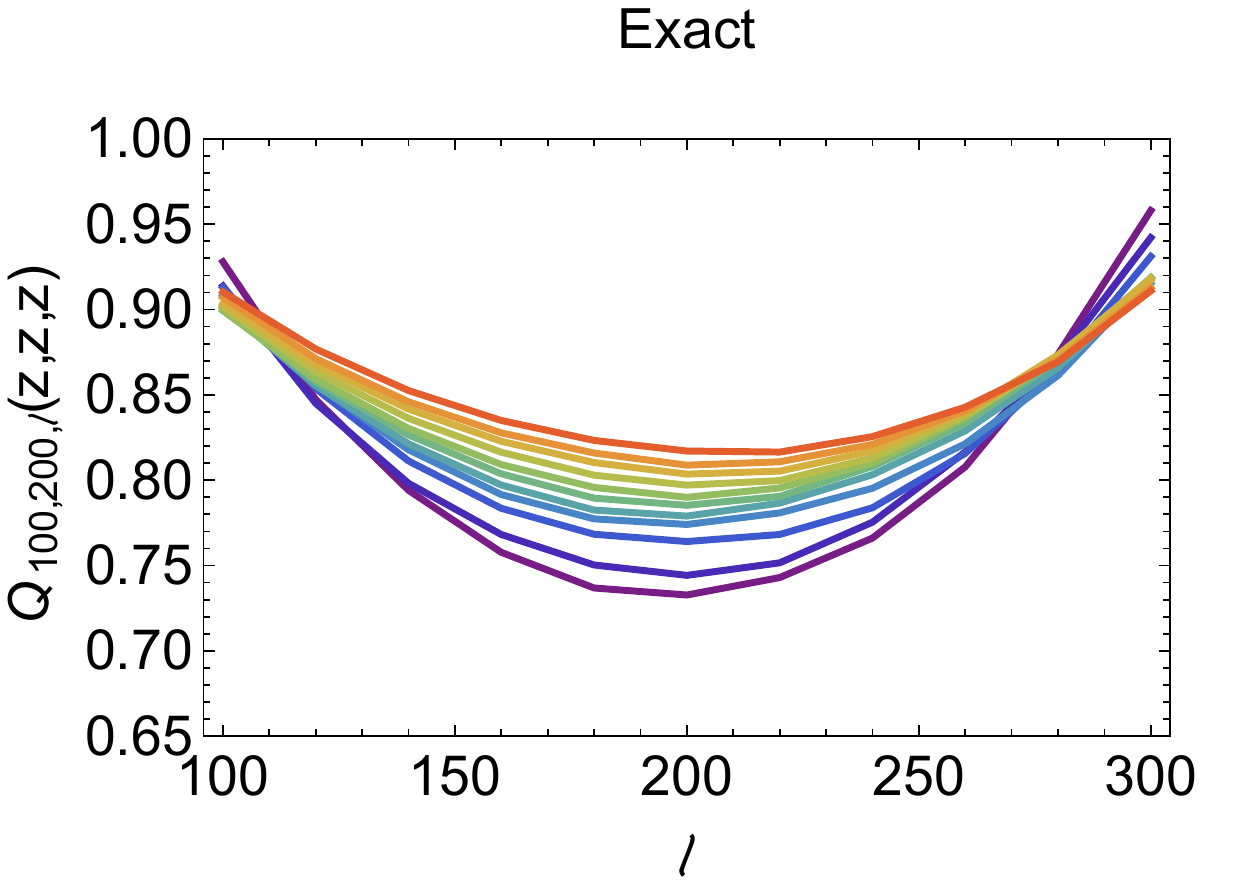}\quad
    \includegraphics[width=7.5cm]{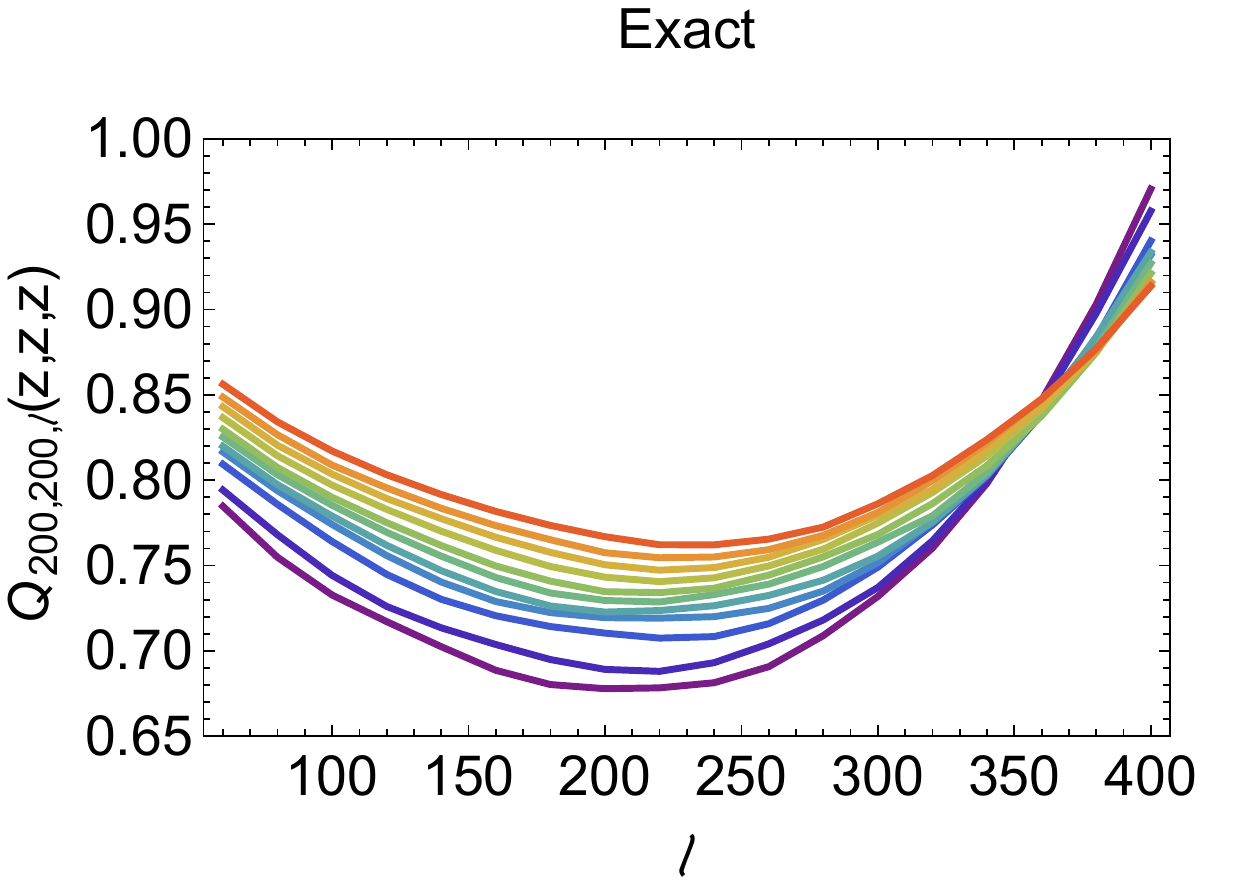}\\
    \vspace{3mm}
    \includegraphics[width=7.5cm]{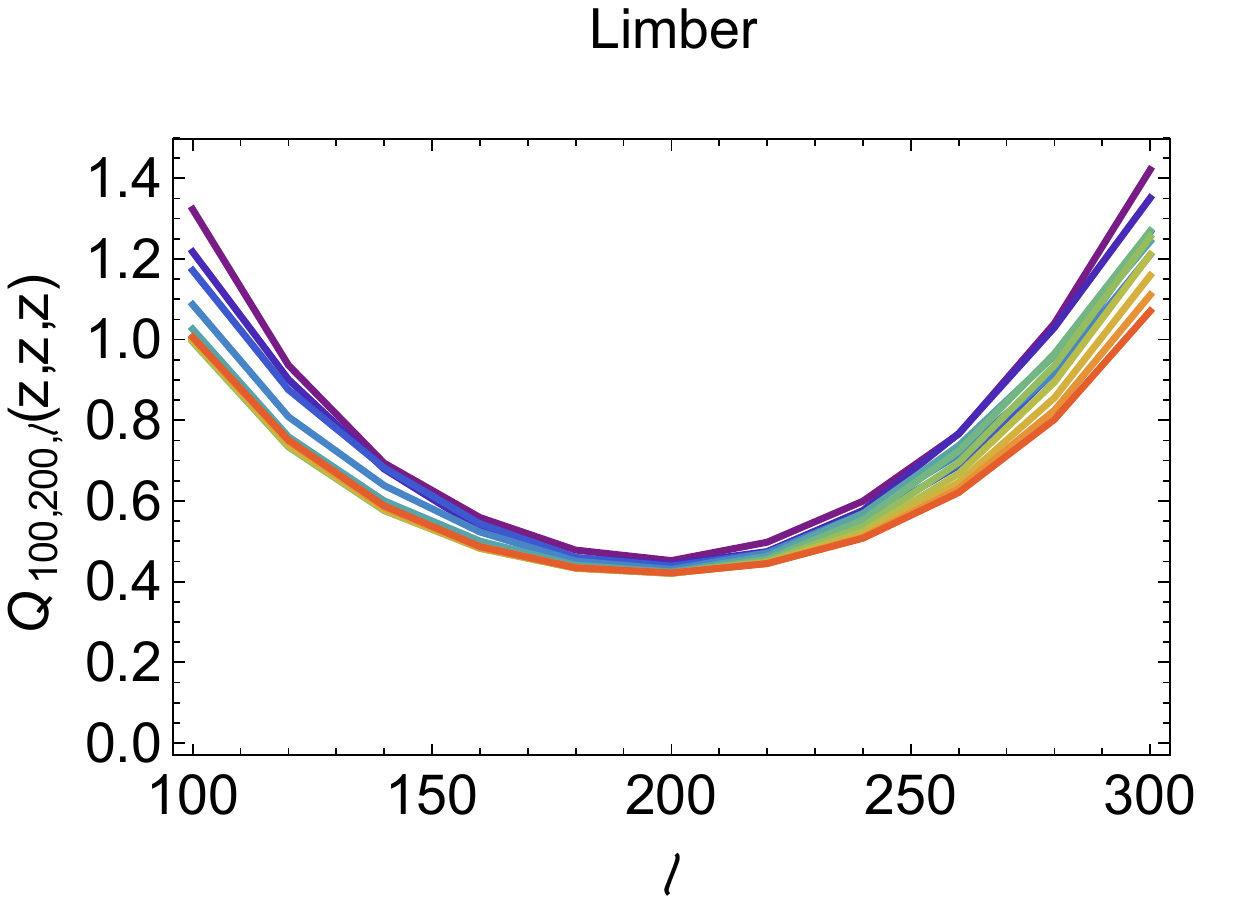}\quad
    \includegraphics[width=7.5cm]{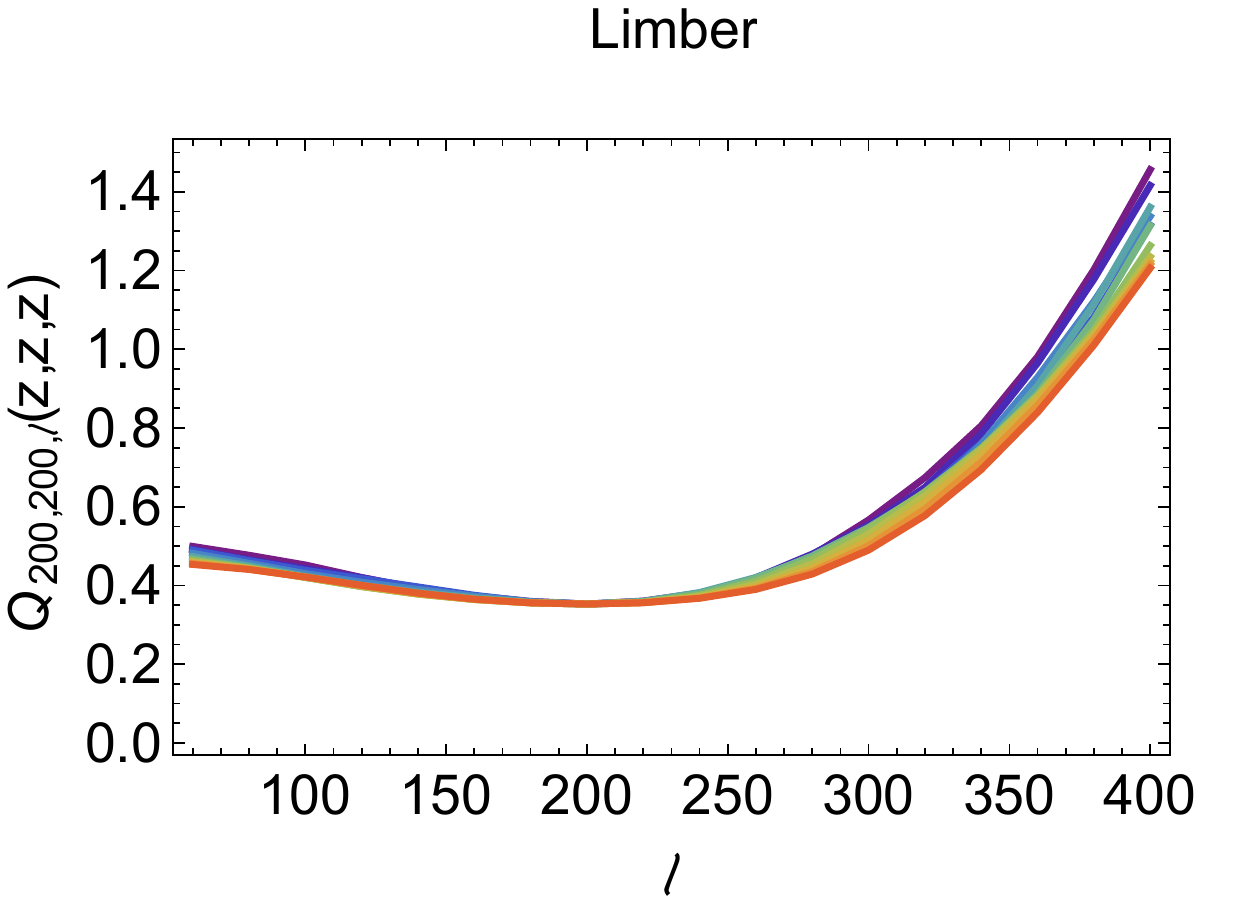}
    \caption{We plot $Q_{\ell_1 \ell_2 \ell_3 }\left( z_1 , z_2 , z_3 \right) \equiv 1+ \left( b^{\delta 1}_{\ell_1 \ell_2 \ell_3}+b^{\delta 2}_{\ell_1 \ell_2 \ell_3}\right) / b^{\delta 0}_{\ell_1 \ell_2 \ell_3}$. The upper panels show the exact computation, and for comparison in the lower panels the Limber approximation has been applied. Different lines denote different redshift from $z=0.5$ (purple) to $z=2$ (red). The deviation from the monopole increases at low redshift, because non-Gaussianities generated by non-linear evolution grow in time. Whereas the analogous quantity plotted in Fourier space is by construction redshift-independent. The Limber approximation (lower panel) gives the right order of magnitude and roughly the correct shape of the ratio $Q$ as a function of $\ell$ but does not agree by about 30\% to 40\% also at high $\ell\gtrsim 100$.
    }
    \label{fig:delta}
  \end{center}
\end{figure}

\begin{figure}[htbp]
  \begin{center}
    \includegraphics[width=7.6cm]{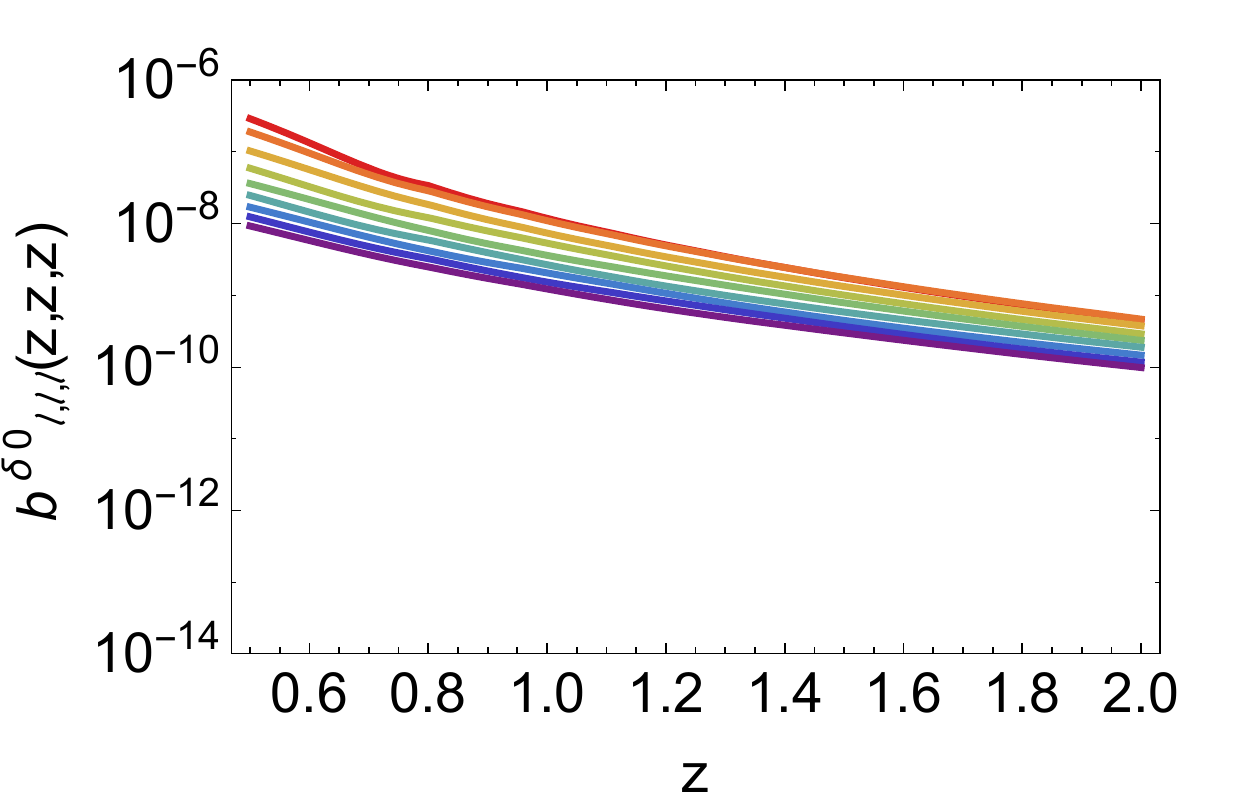}\quad
    \includegraphics[width=1.3cm]{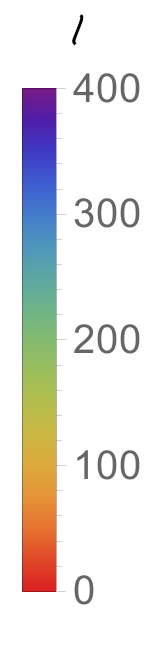}\\
    \includegraphics[width=7.6cm]{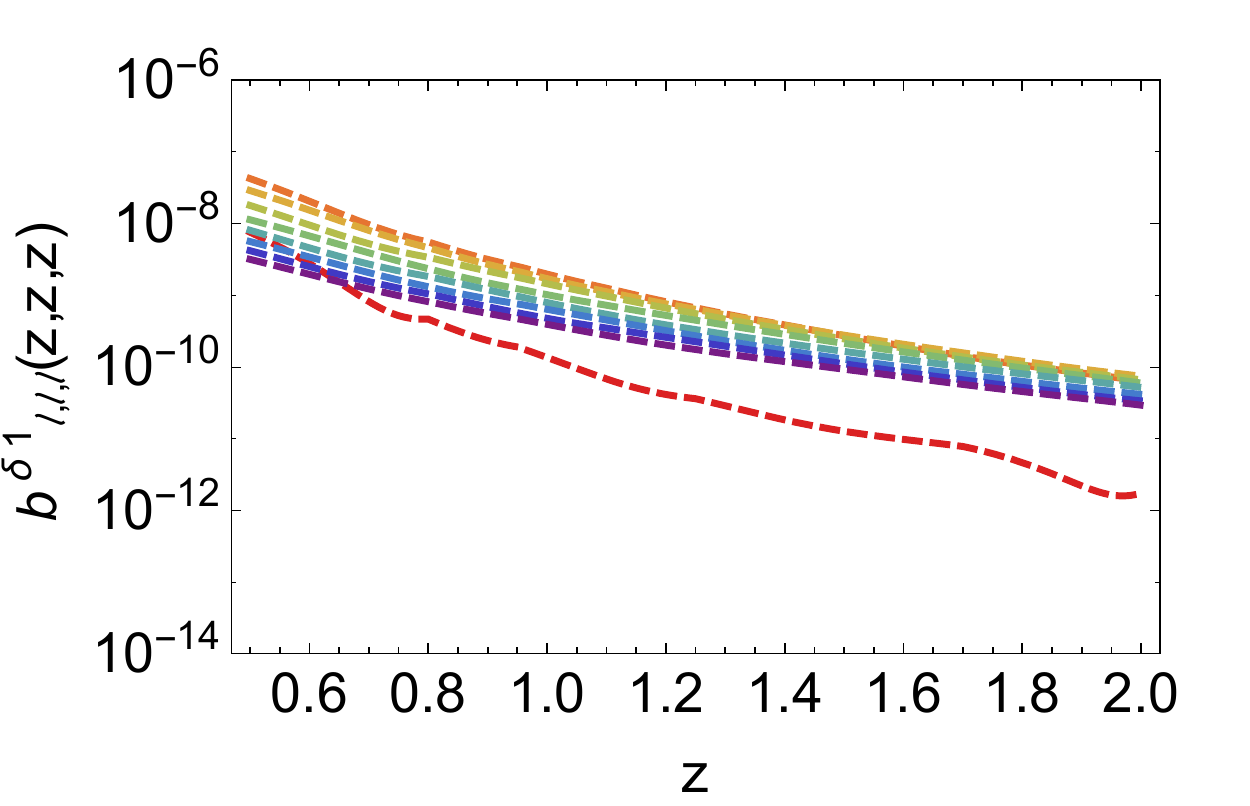}
    \includegraphics[width=7.6cm]{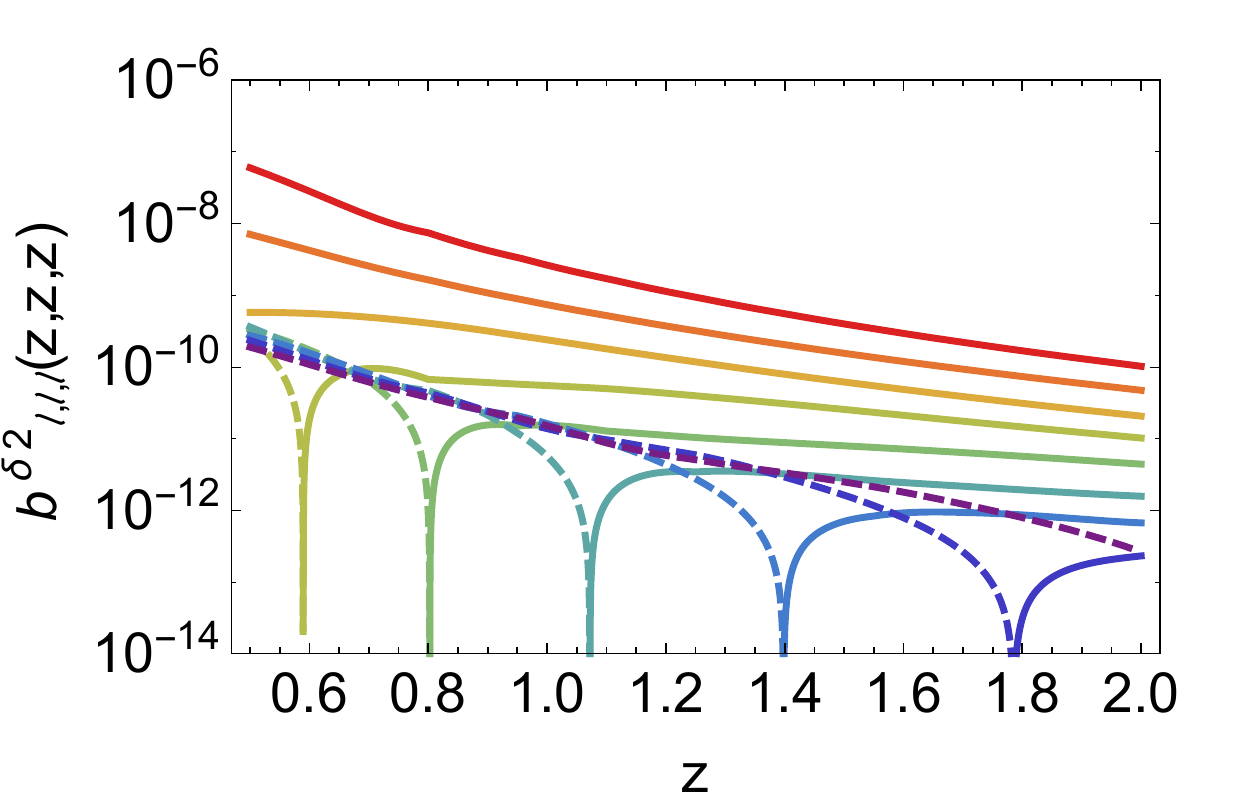}
    \caption{
      The monopole, dipole and quadrupole contributions to the reduced bispectrum for $\ell_1=\ell_2=\ell_3=\ell$ as functions of $z_1=z_2=z_3=z$ are shown.
      Different colors correspond to values from $\ell=4$ (red) to $\ell=404$ (purple) with steps $\Delta\ell=50$.
      Dashed lines correspond to negative values.
      For a fixed redshift, the dipole shows a turnover at the lower values of $\ell$ (red and orange lines).
      The quadrupole becomes negative at large $\ell$'s and low redshifts.
    }
    \label{fig:delta2}
  \end{center}
\end{figure}

In Figs.~\ref{fig:delta} and~\ref{fig:delta2} we show the contributions to the bispectrum from $\de^{(2)}$ which are expected to dominate the result. We compare the exact result with the Limber approximation and find that even though the order of magnitude agrees, the results from the Limber approximation can differ by nearly a factor of 2.  At closer inspection we have found that the Limber approximation for the monopole is excellent, whereas the global difference is due to the relatively slow convergence of the integrands associated to the dipole and quadrupole terms in the UV tail of our region of integration.
Nevertheless, for a first indication of the amplitude we use the Limber approximation for  $v^{(2)}$ and $\ka^{(2)}$ for comparison with the new lensing terms.
For $\delta^{(2)}$  we just consider the monopole in the following figures.

In Fig.~\ref{fig:delta} one sees how the curvature of $Q= 1 +(b^{\de 1}+ b^{\de 2})/b^{\de  0}$ increases with decreasing redshift, i.e. when gravitational structure grows and enhances the asymmetry of the 3-point function. These shapes and their evolution  with redshift are excellent tests of (Newtonian) gravity.  They are due to the fact that gravity tends to `flatten' triangular configurations.
For equal redshifts this means that one opening angle at the observer position should be roughly equal to the sum or the difference of the  other two to maximize the bispectrum. We have also investigated different configurations in $\ell$-space and the accuracy of the Limber approximation remains similar. The order of magnitude is correct but the result can be up to 40\% off also for  large $\ell$.

In Fig.~\ref{fig:delta2} we show the density monopole (Eq.~(\ref{e:bde0}), first term of Eq.~(\ref{Allterms}), first panel), the dipole (Eq.~(\ref{e:bde1}), second term of Eq.~(\ref{Allterms}), second panel) and the quadrupole  (Eq.~(\ref{e:bde2}), third term of Eq.~(\ref{Allterms}), third panel). The negative dipole term $b^{\de 1}$ is mainly responsible for the shape of Fig.~\ref{fig:delta} .

\begin{figure}[htbp]
  \begin{center}
    \includegraphics[width=7.5cm]{legend_z_row.pdf} \hspace{7.1cm} \phantom{.}\\
    \includegraphics[width=7.5cm]{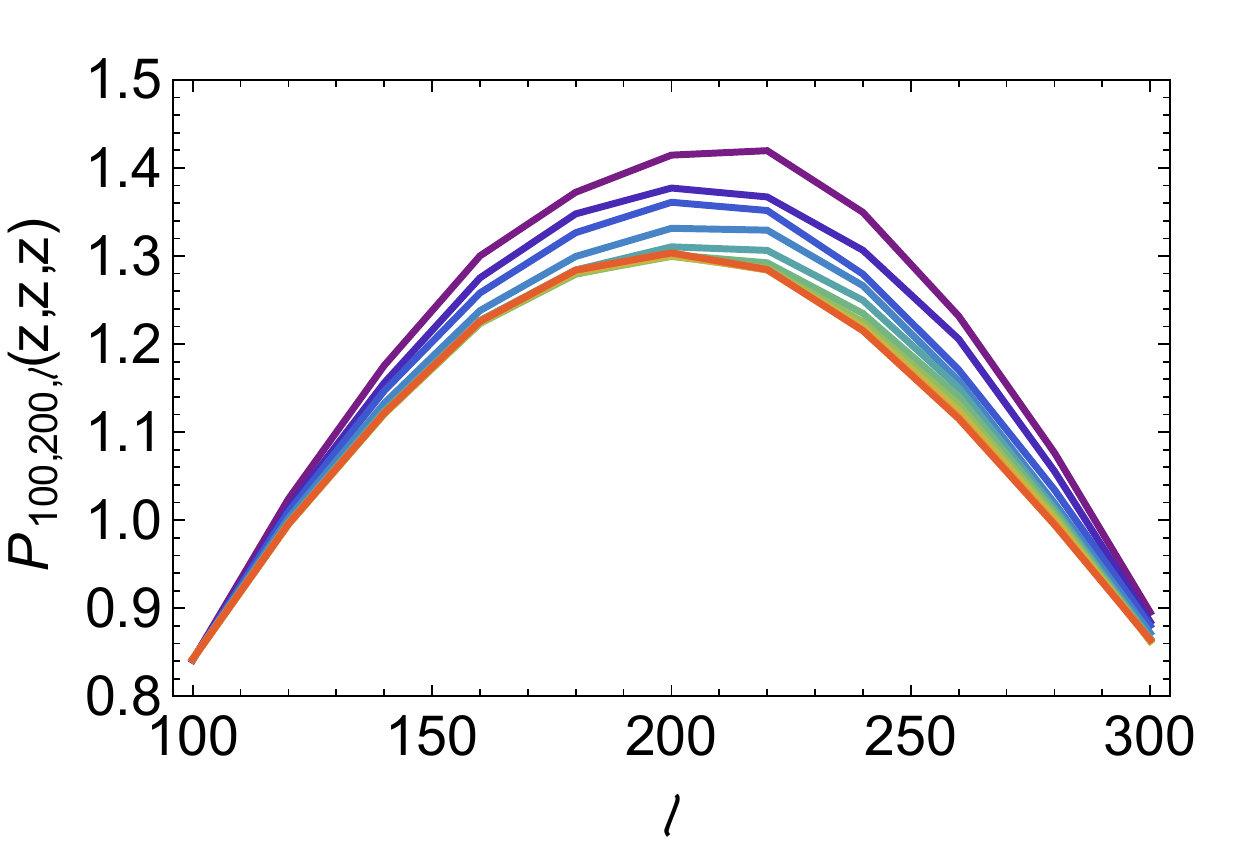}\quad
    \includegraphics[width=7.5cm]{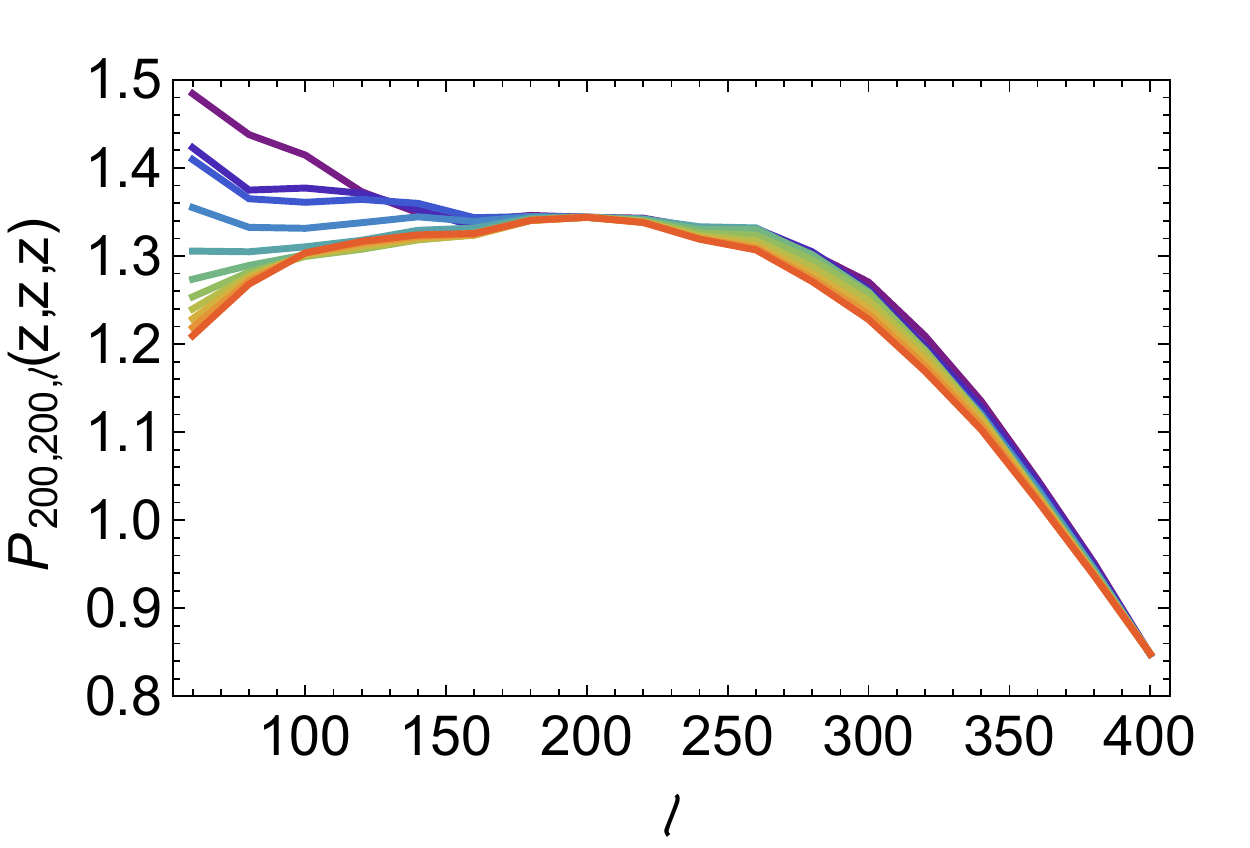}
    \caption{We plot $P_{\ell_1 \ell_2 \ell_3 }\left( z_1 , z_2 , z_3 \right) \equiv 1+ \left( b^{v' 1}_{\ell_1 \ell_2 \ell_3}+b^{v'  2}_{\ell_1 \ell_2 \ell_3}\right) / b^{v' 0}_{\ell_1 \ell_2 \ell_3}$. Different lines denote different redshift from $z=0.5$ (purple) to $z=2$ (red). The deviation from the monopole increases at low redshift, because non-Gaussianities generated by non-linearity evolution grow in time. This calculation uses the Limber approximation which cannot be assumed to be very accurate for this term.}
    \label{fig:rsd2}
  \end{center}
\end{figure}

In Fig.~\ref{fig:rsd2} we show  the curvature of the second order velocity term using Limber approximation, $P= 1 +(b^{v' 1}+ b^{v' 2})/b^{v'  0}$. As explained in section~\ref{ss:rsd2} also the second order RSD (forth term in Eq.~(\ref{Allterms})) can be split into a monopole dipole and quadrupole term denoted by $b^{v'  0}$, $b^{v'  1}$ and $b^{v'  2}$  which are given (within the Limber approximation) in Eq.~(\ref{RSDorder2FRj}). While the Limber approximation should give roughly the right shape of the curvature $P$ (like for the second order density case,
compare top and bottom panels of Fig.~\ref{fig:delta}), the relative amplitude of the dipole and quadrupole terms 
with respect to the monopole term may be quite inaccurate
both in magnitude and in sign (similarly to the case of the second order density bispectrum).
In the bispectrum from the second order velocity term, the dipole has again the opposite sign but the quadrupole is significantly larger and increases the bispectrum especially for $1/\ell_2 +1/\ell_3=1/\ell_1$.

\begin{figure}[htbp]
  \begin{center}
    \includegraphics[width=0.48\textwidth]{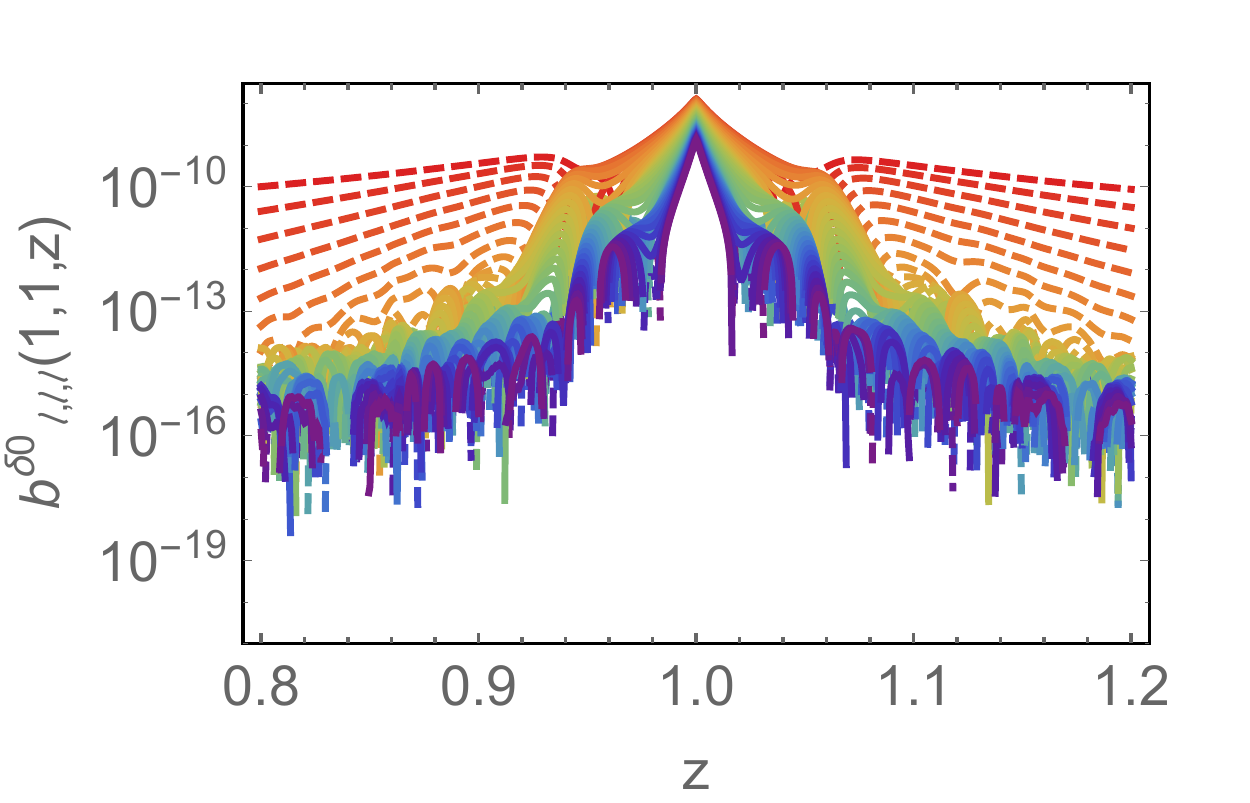}
    \includegraphics[width=1.3cm]{legend_l_col.pdf}\\
    \includegraphics[width=0.48\textwidth]{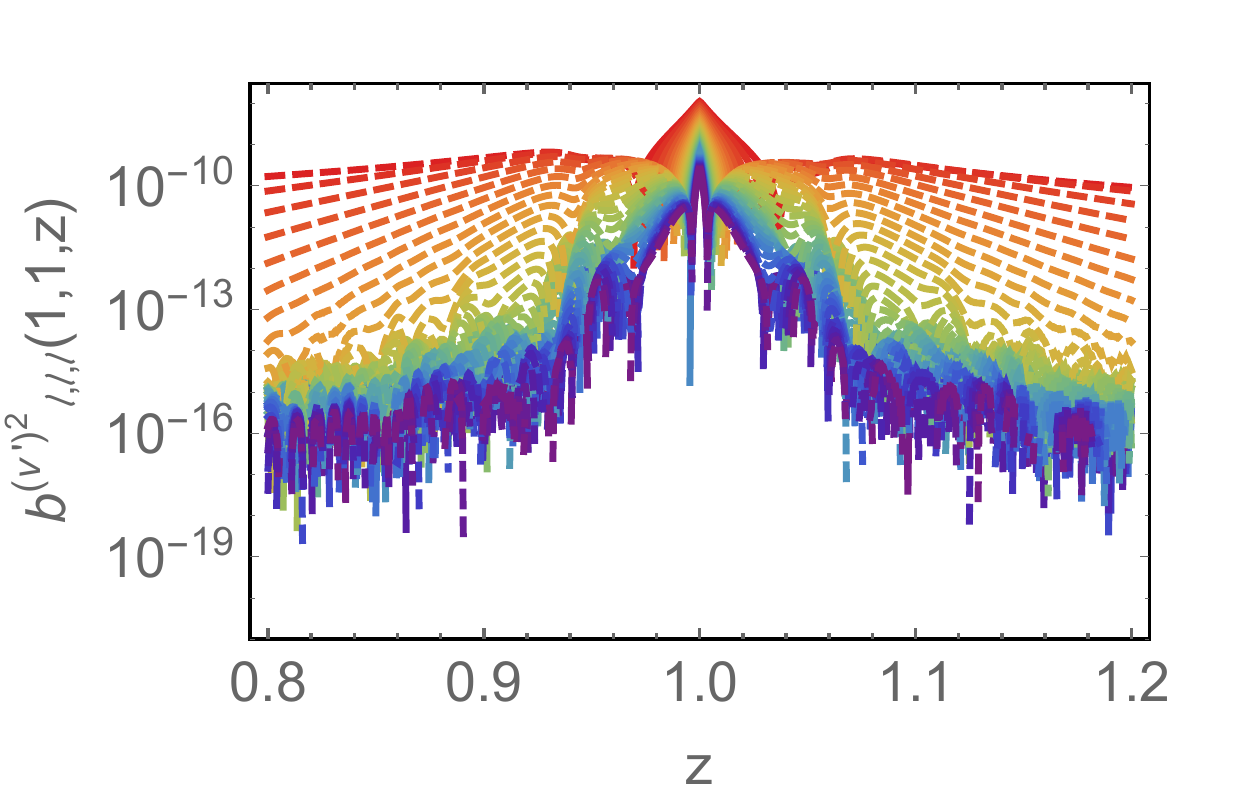}
    \includegraphics[width=0.48\textwidth]{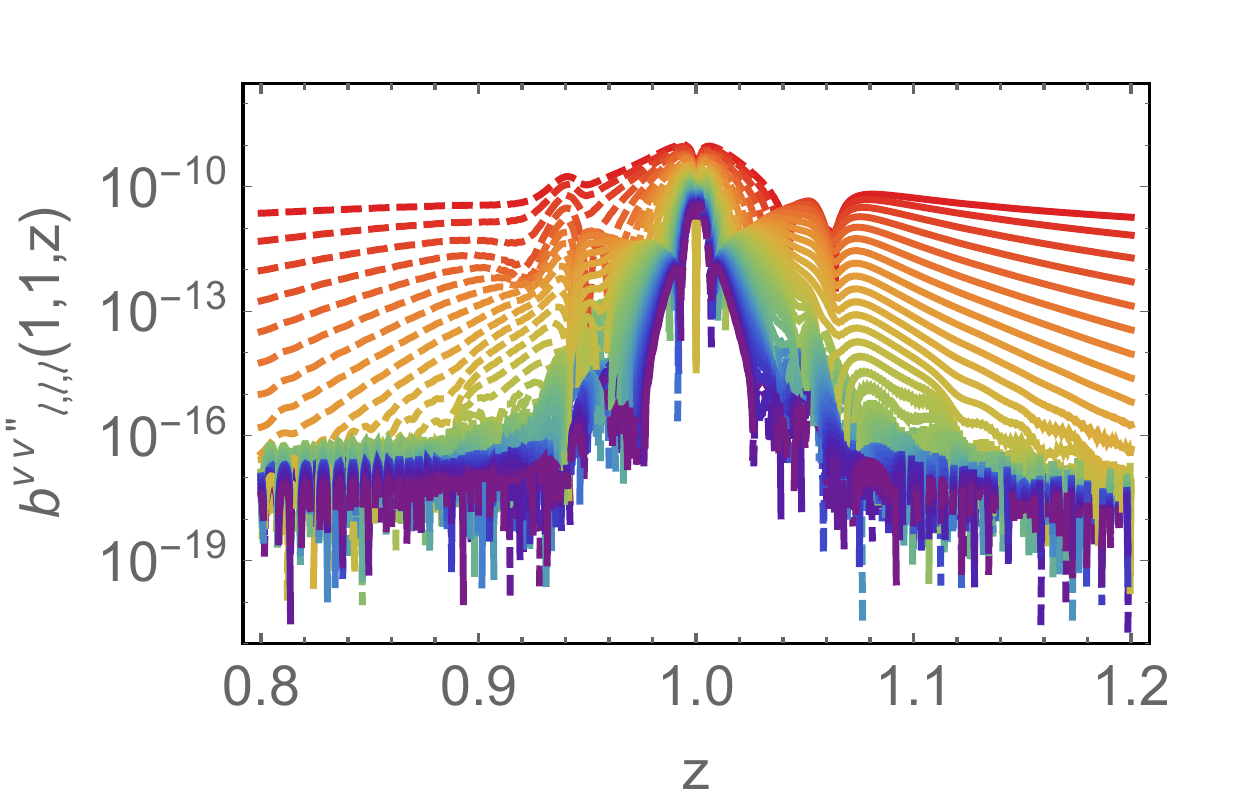}
    \includegraphics[width=0.48\textwidth]{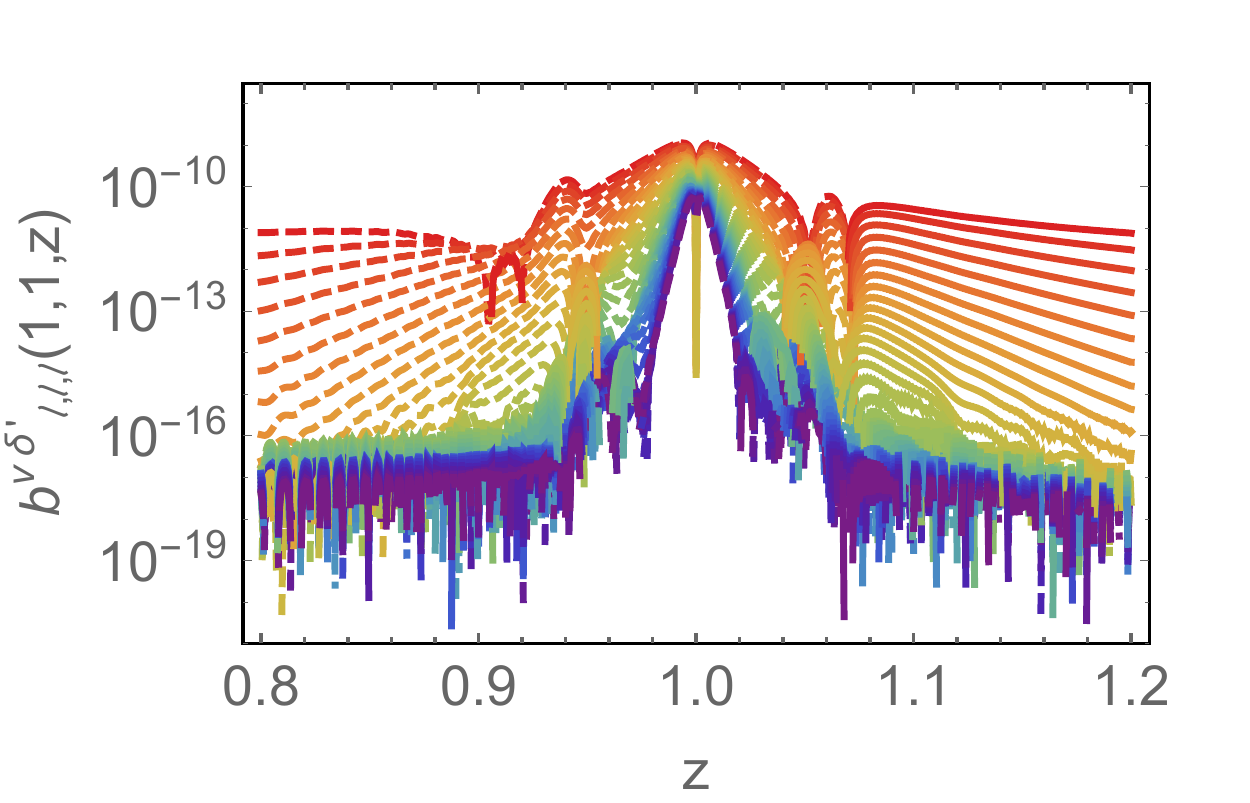}
    \includegraphics[width=0.48\textwidth]{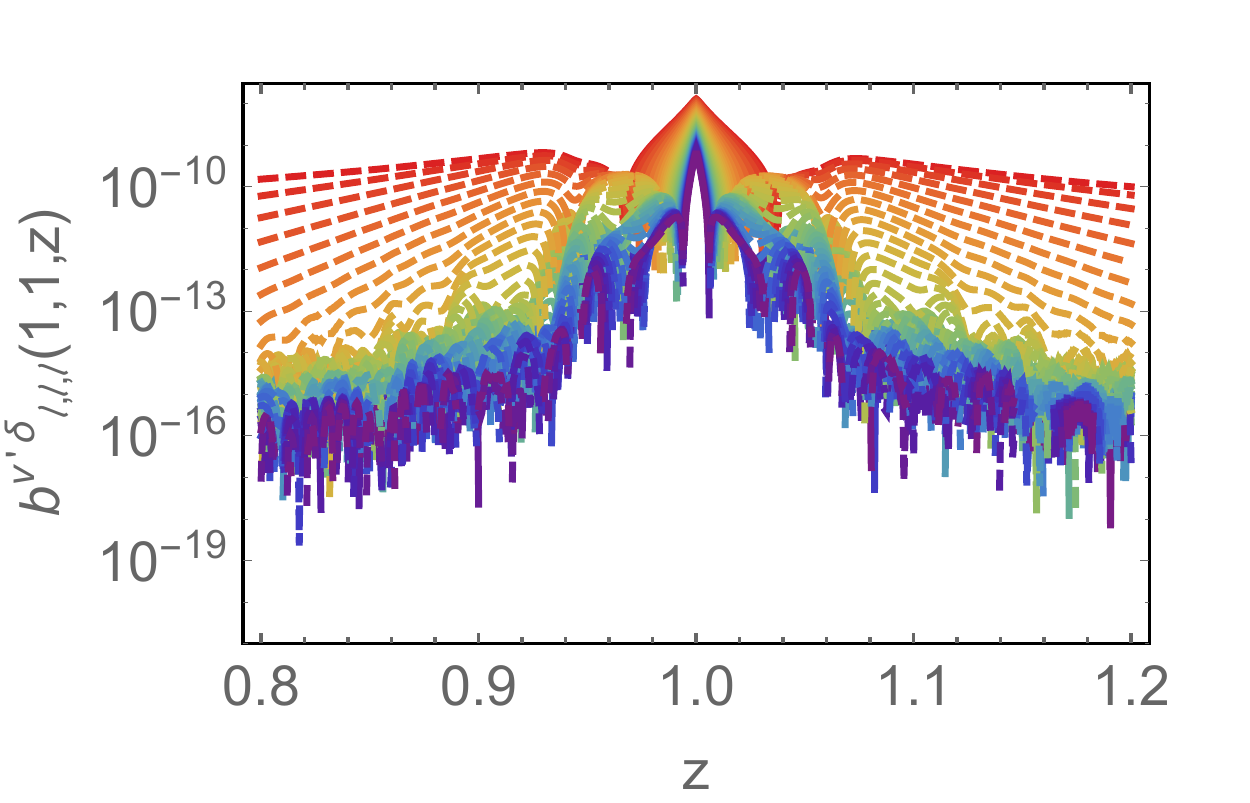}
    \caption{{\bf Newtonian terms.}
      The contributions from the Newtonian terms to the bispectrum is shown for different values of $\ell=\ell_1=\ell_2=\ell_3$, from $\ell = 4$ (red) to $\ell=400$ (purple), as a function of the third redshift $z_3=z$ for $z_1=z_2=1$. The first panel shows the first term of Eq.~(\ref{Allterms}) while the subsequent pales show the 5th to 8th terms.
    }\label{fig:newtonian}
  \end{center}
\end{figure}

In Fig.~\ref{fig:newtonian} we show the contributions from the Newtonian terms of $\Si^{(2)}$. These are the 1st and 5th to 8th terms in Eq.~(\ref{Allterms}). Their explicite expressions are given in Eqs. (\ref{e:bde0}) for the 1st panel, (\ref{e3:v'v'}) for the  2nd panel, (\ref{Newvvpp}) for the 3rd panel, (\ref{Newvdp}) for the 4th panel and (\ref{rsdxdelta}) for the 5th panel.
 We do not plot $b^{v^{(2)'}}_{\ell_1 \ell_2 \ell_3}$, which we can only determine within the Limber approximation,
see Eq.~(\ref{RSDorder2FR}), where it has a contact term (Dirac-delta) so that it can not be shown in the configuration 
of Fig.~\ref{fig:newtonian}, where no window function is included.
We fix the redshifts $z_1=z_2=1$ and plot the terms as function of $z_3=z$ for different multipoles $\ell=\ell_1=\ell_2=\ell_3$ indicated by the color coding. For better visibility we do not multiply them by $\ell^2$.
The $\de^{(2)0}$ (top panel, we consider only the monopole part of $\delta^{(2)}$ for simplicity) and the $v'\de$ (low right) terms are positive at equal redshifts and dominate the result.  They are of  order $10^{-8}$. Also the pure RSD term (middle left) has nearly the same amplitude. The $vv''$ and $v\de'$ terms actually vanish exactly at equal redshifts, $z_1=z_2=z_3$ because of the different parity of the spherical Bessel functions which appear in the expansion of these terms, see Appendices~\ref{ssa:v3v} and \ref{ssa:d1v1}.  At slightly different redshifts they are negative and of the same order of magnitude as the other terms. Most of the terms have several sign changes especially at higher $\ell$'s which are visible as the purple spikes in the figures. These oscillations arise from the fact that a fixed angular scale $\ell$ denotes different comoving scales $r(z)$ at different redshifts. All terms peak at $z_3\simeq1$ and decay rapidly with increasing redshift difference.
The same behavior has also been found for different configurations in $\ell$-space.

\begin{figure}[htbp]
  \begin{center}
    \includegraphics[width=0.48\textwidth]{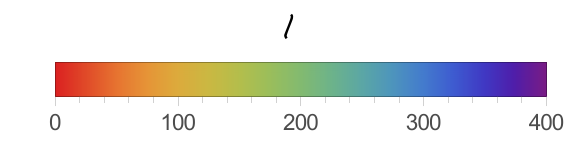} \hspace{6.8cm}\phantom{.}\\
    \includegraphics[width=0.48\textwidth]{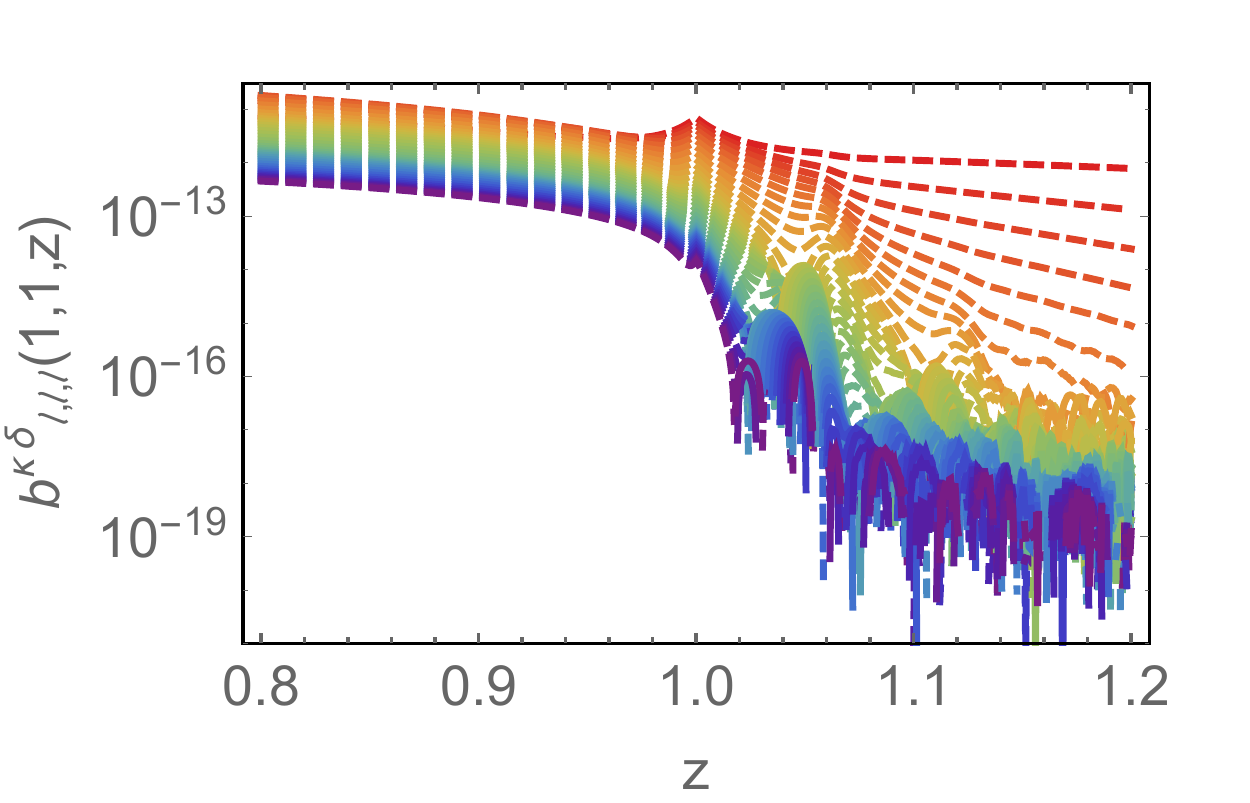}
    \includegraphics[width=0.48\textwidth]{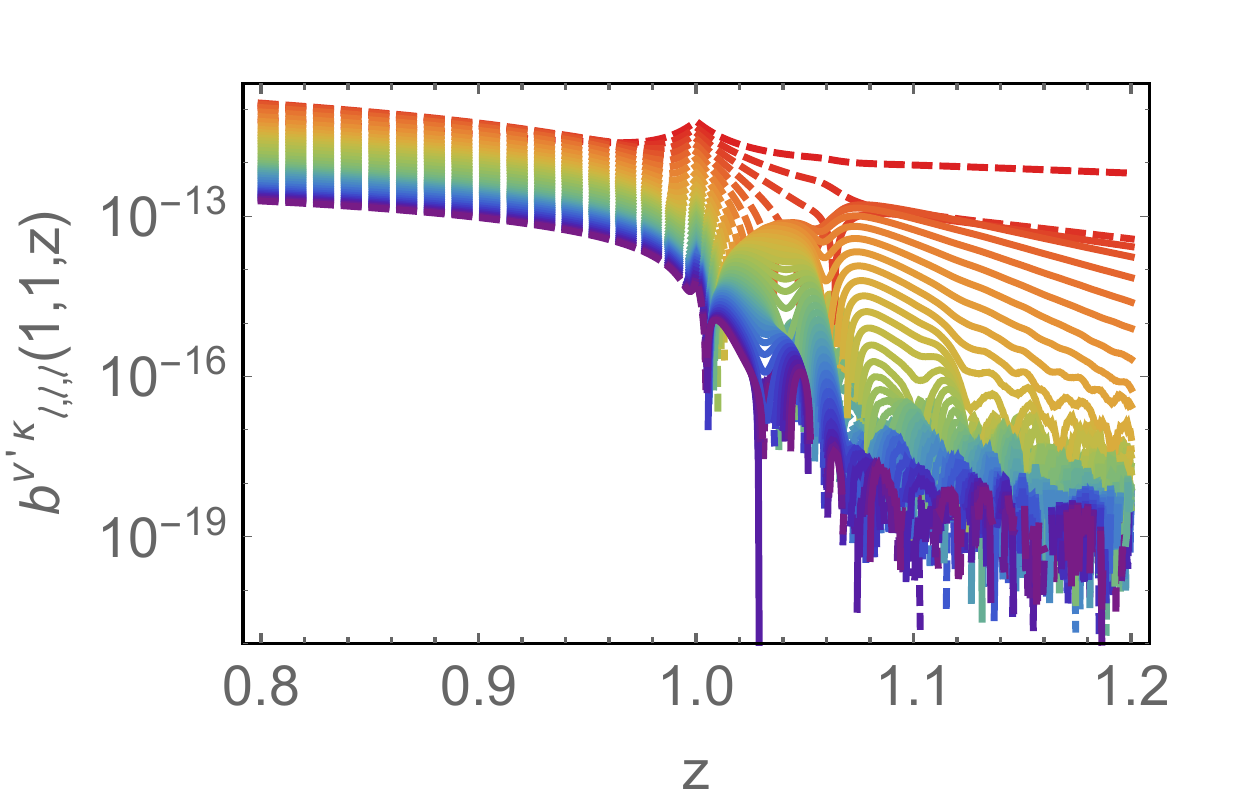}
    \includegraphics[width=0.48\textwidth]{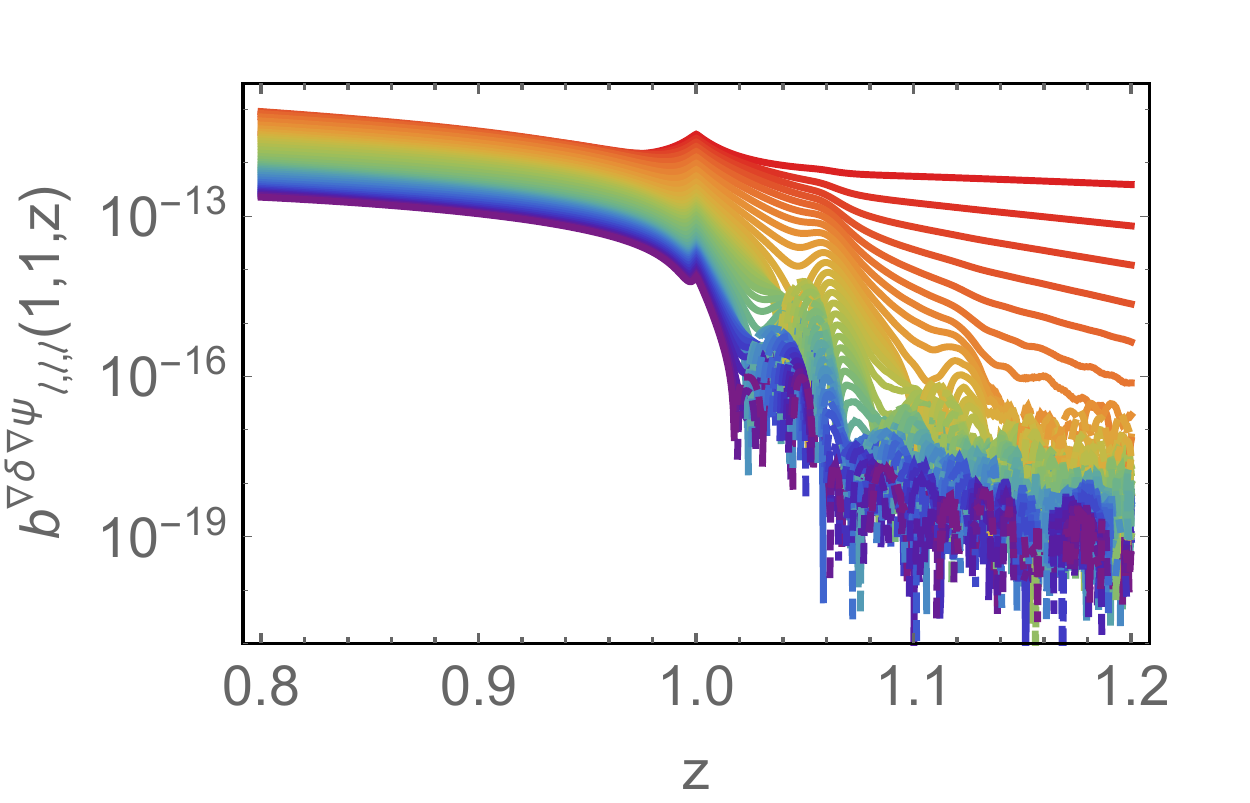}
    \includegraphics[width=0.48\textwidth]{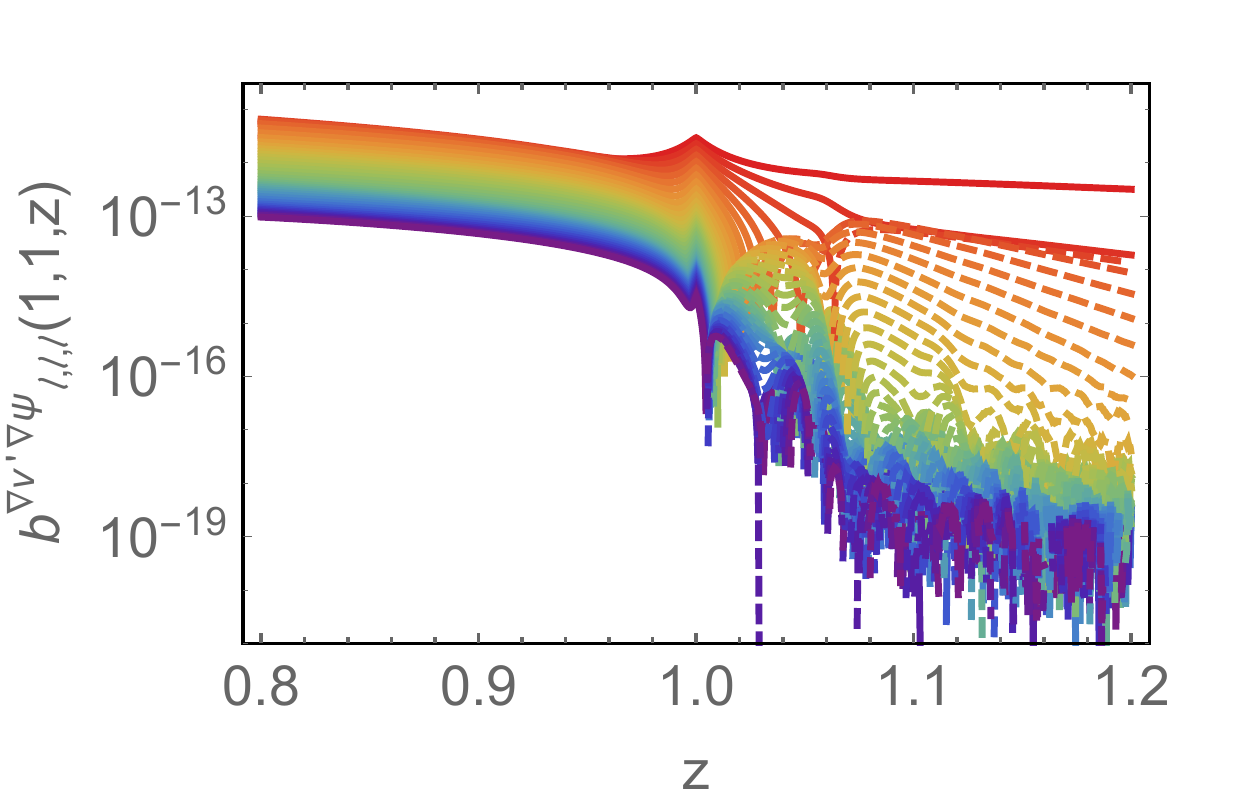}
    \caption{{\bf Newtonian~$\times$~lensing terms.}
      We plot the contributions from the  `Newtonian $\times$ lensing' terms to the bispectrum for different values of $\ell=\ell_1=\ell_2=\ell_3$, from $\ell = 4$ (red) to $\ell=400$ (purple), as a function of the third redshift $z_3=z$ for $z_1=z_2=1$. These are the 2nd (top left) and 4th (top right) and the 3rd (bottom left) and 5th (bottom right) terms in the second line of Eq.~(\ref{Allterms}). }
    \label{fig:newtXlens}
  \end{center}
\end{figure}

In Fig.~\ref{fig:newtXlens} we show terms which contain products of Newtonian and lensing terms in the same configuration $(z_i,\ell_i)$. The top left panel is the 2nd term in the second line of  Eq.~(\ref{Allterms}), explicitly given in (\ref{lenxden}).  The top right panel is the  4th term in the second line of  Eq.~(\ref{Allterms}), explicitly given in (\ref{e3:v'k}). The bottom left panel is the 3rd term in the second line of  Eq.~(\ref{Allterms}), explicitly given in (\ref{e3:nabla-d-nabla-psi}), and the bottom right panel is the 5th term in the second line of  Eq.~(\ref{Allterms}), explicitly given in (\ref{e3:nabla-v-nabla-psi}).
Also here we cannot show the $b^{\kappa^{(2)}}_{\ell_1 \ell_2 \ell_3}$ contribution  which we have computed in the Limber approximation, see Eq.~(\ref{RedBispLensOrd2}). Because of this approximation  it has contact term  (Dirac delta) and cannot be plotted in the configuration of Fig.~\ref{fig:newtXlens}, where no window function is included. These terms are typically of order $10^{-11}$, hence for equal redshifts they are about 3 orders of magnitude smaller than the Newtonian contributions and most probably not measurable.  But at different redshifts these terms do not decay significantly with increasing redshift difference as long as $z_3<1=z_{1,2}$.
At $z_3<1$, the terms containing $\ka$ (top panels) are negative while the terms with derivatives (bottom panels) are positive.
Note also that we should not trust our results for the very lowest $\ell=4$ since on this large angular scale also subdominant terms which we have not taken into account can become important.

\begin{figure}[htbp]
  \begin{center}
    \includegraphics[width=0.48\textwidth]{legend_l_row.pdf} \hspace{6.8cm}\phantom{.}\\
    \includegraphics[width=0.48\textwidth]{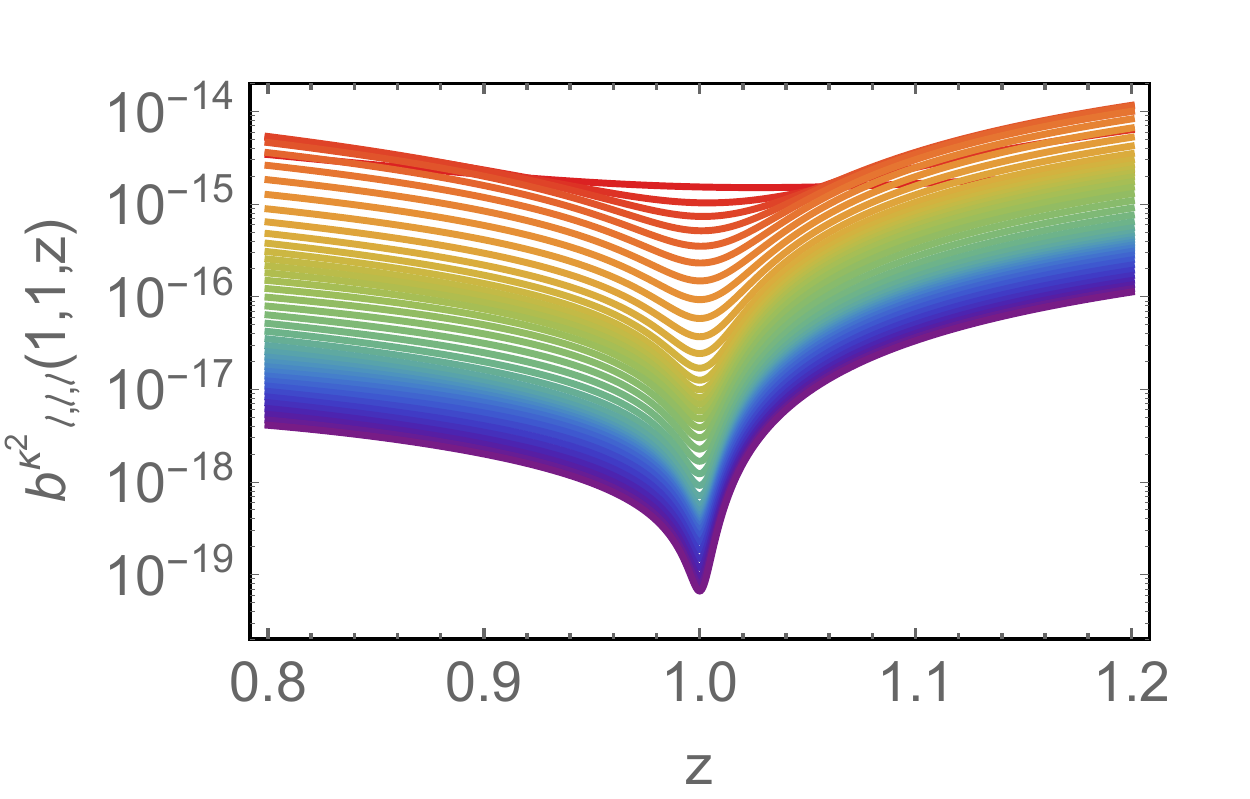}
    \includegraphics[width=0.48\textwidth]{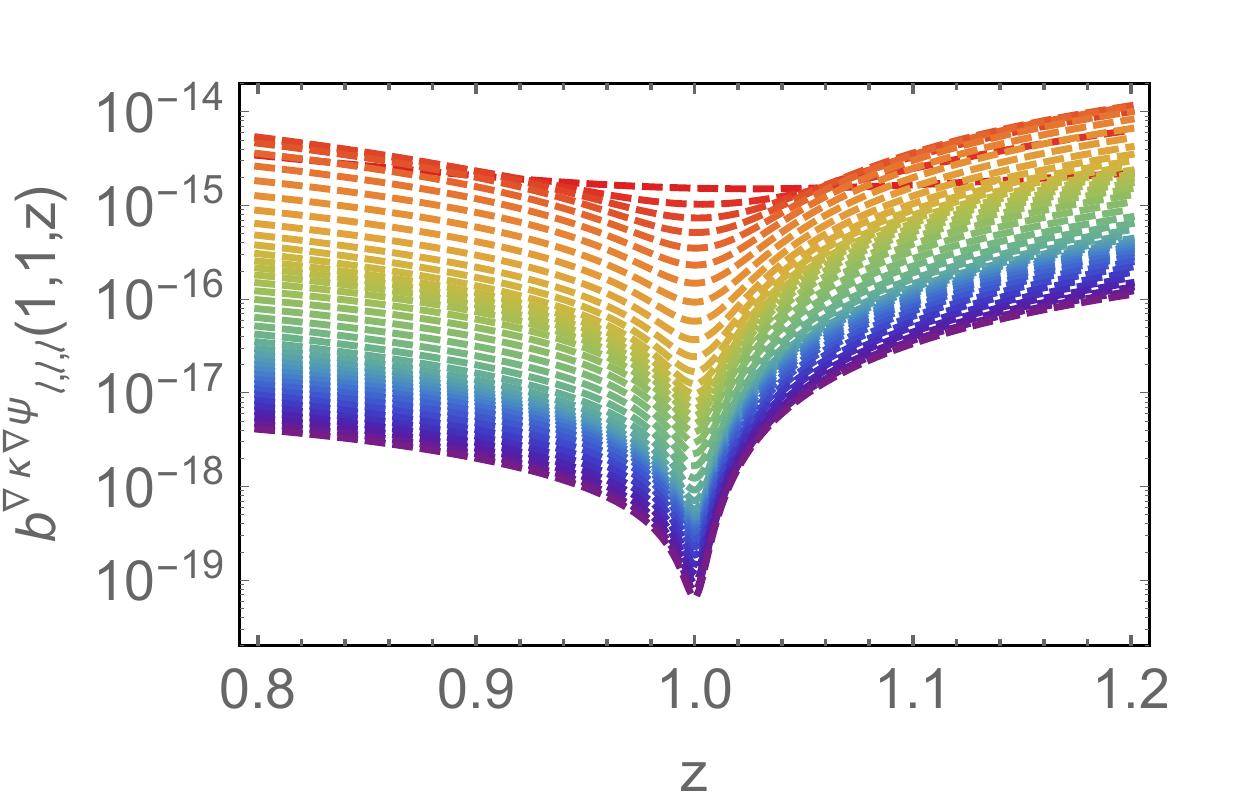} \\
    \includegraphics[width=0.48\textwidth]{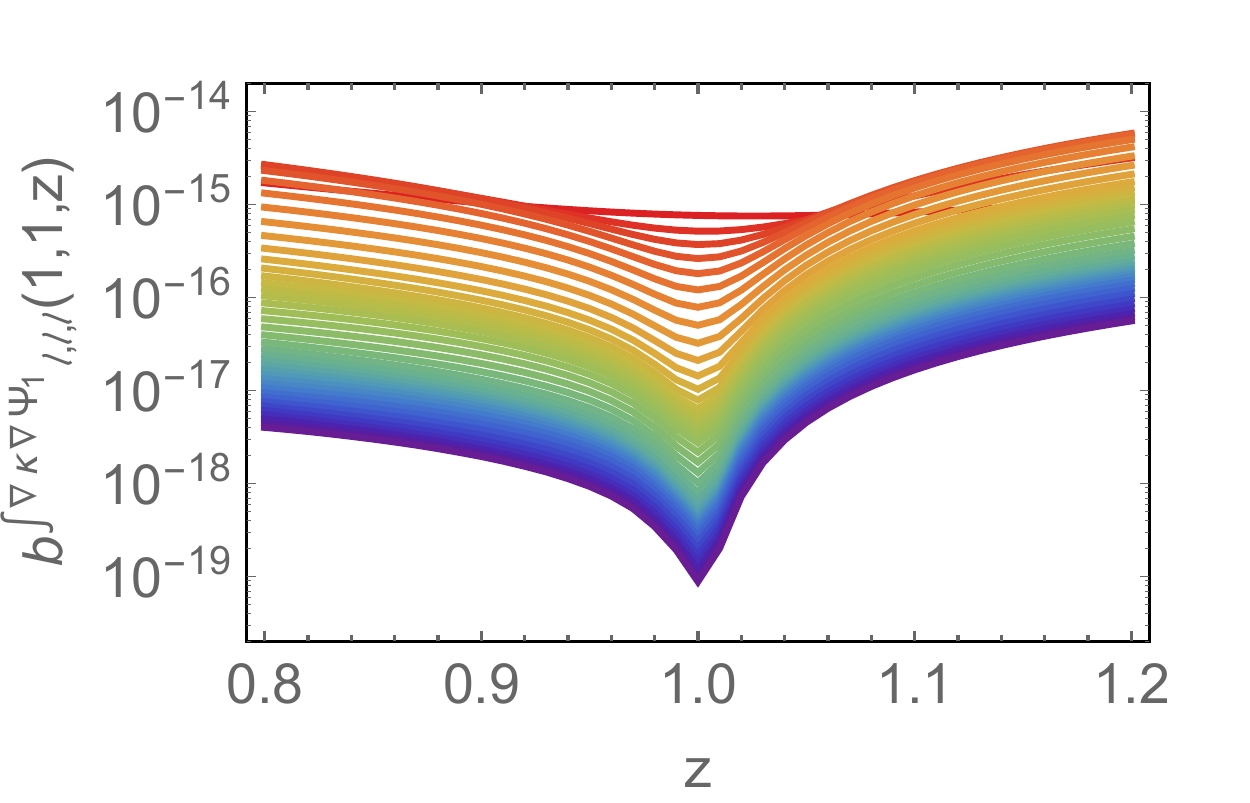}
    \includegraphics[width=0.48\textwidth]{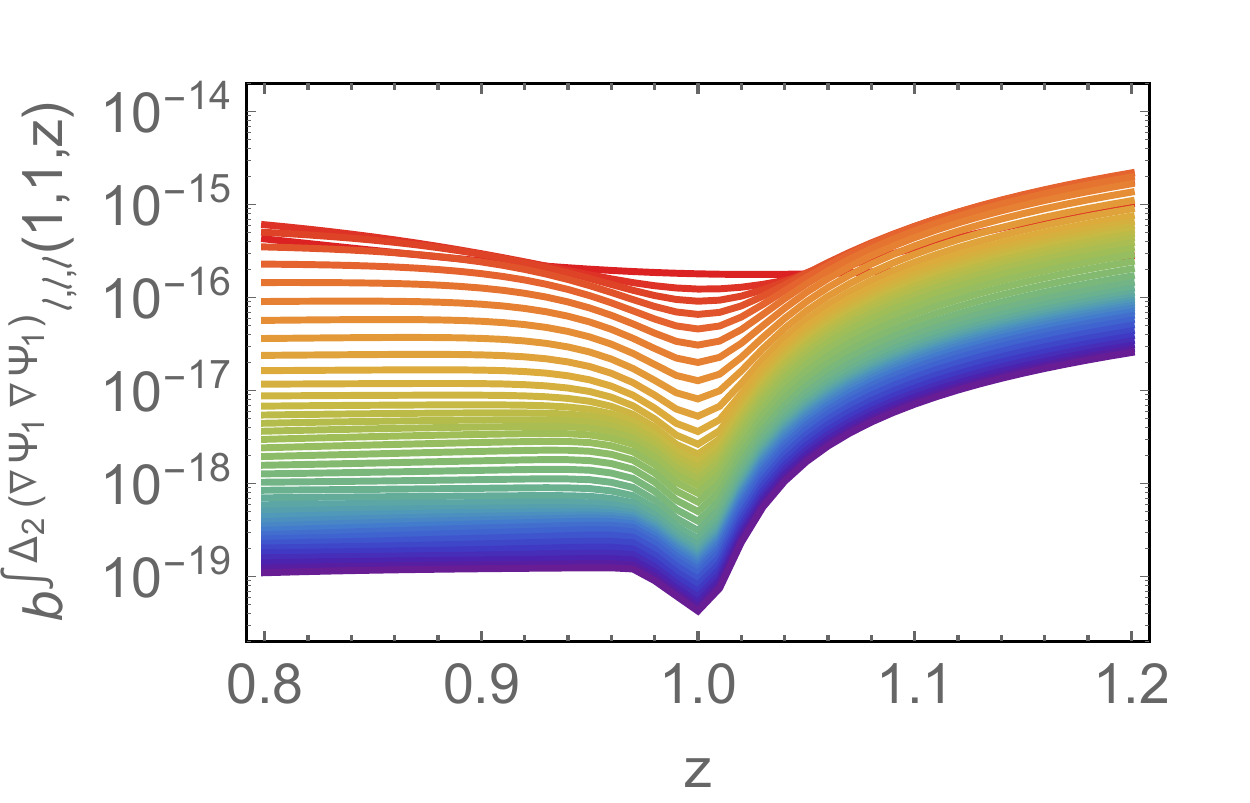}
    \caption{{\bf Lensing terms.}
      We plot the contributions from the pure lensing terms to the bispectrum for different values of $\ell=\ell_1=\ell_2=\ell_3$, from $\ell = 4$ (red) to $\ell=400$ (purple), as a function of the third redshift $z_3=z$ for $z_1=z_2=1$.
 These are the 6th (top left), 7th (top right), 8th (bottom left) and 9th (bottom right) terms on the second line of Eq.~(\ref{Allterms}).     The two terms in the top panels cancel exactly for $\ell_1=\ell_2 = \ell_3$.
    }
    \label{fig:lens}
  \end{center}
\end{figure}

In Fig.~\ref{fig:lens} we show the contribution of the pure lensing terms to the bispectrum.  The top left panel is the 6th term in the second line of  Eq.~(\ref{Allterms}), explicitly given in (\ref{e3:bkappa2}).  The top right panel is the  7th term in the second line of  Eq.~(\ref{Allterms}), explicitly given in (\ref{e3:nabla-k-nabla-psi}). The bottom left panel is the 8th term in the second line of  Eq.~(\ref{Allterms}), explicitly given in (\ref{e3:int-nabla-k-nabla-psi1}), and the bottom right panel is the 9th term in the second line of  Eq.~(\ref{Allterms}), explicitly given in (\ref{e:last}). These terms are of order $10^{-15}$ and smaller and therefore certainly  unmeasurable at equal redshifts.
The terms in the top panels, $\ka^2$ and $\nabla_a\ka\nabla^a\psi$, have opposite sign and for three  equal $\ell$'s they exactly cancel.

\begin{figure}[htbp]
\begin{center}
\subfigure{
\includegraphics[width=0.48\textwidth]{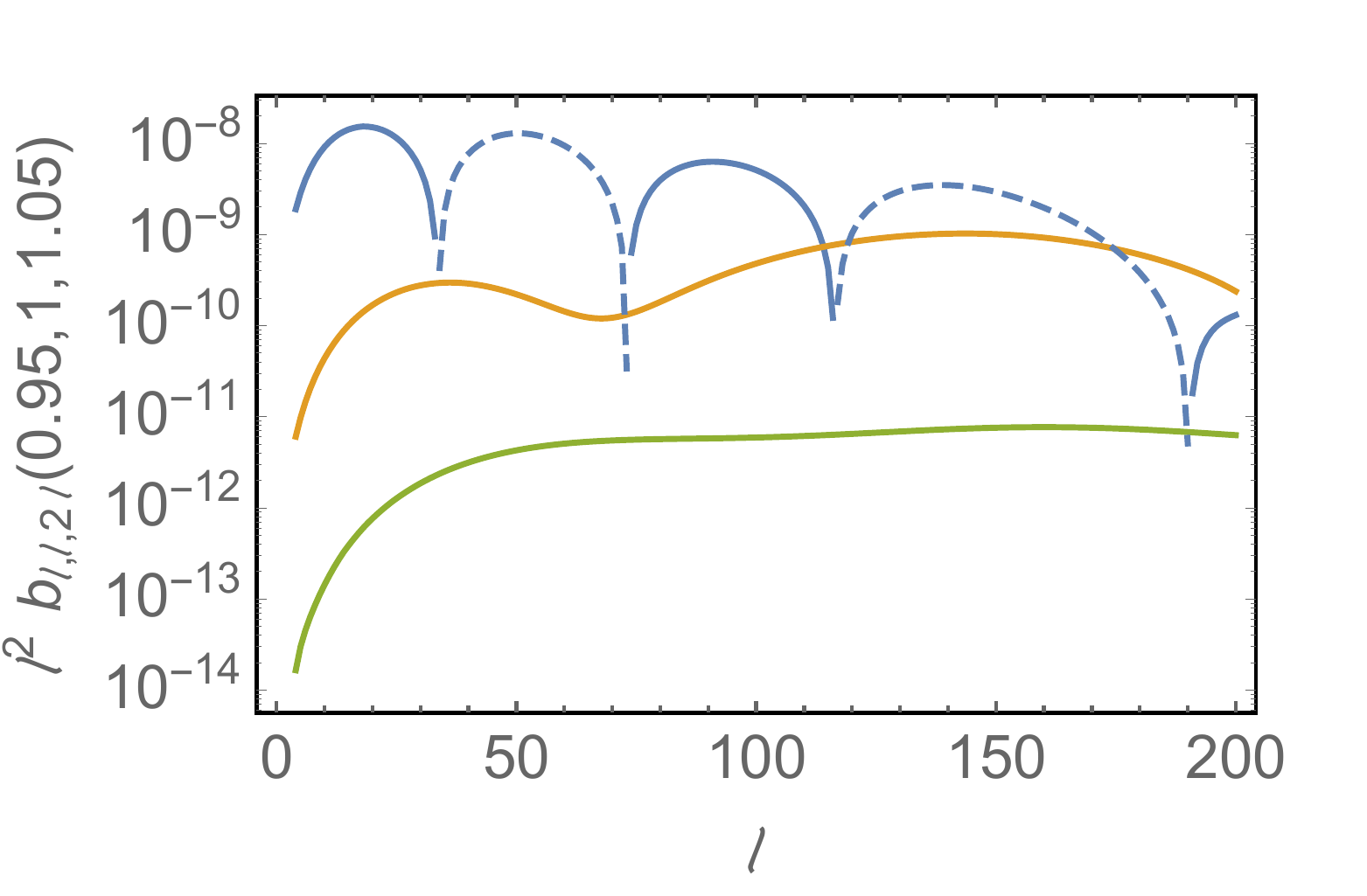}}
\subfigure{
\includegraphics[width=0.48\textwidth]{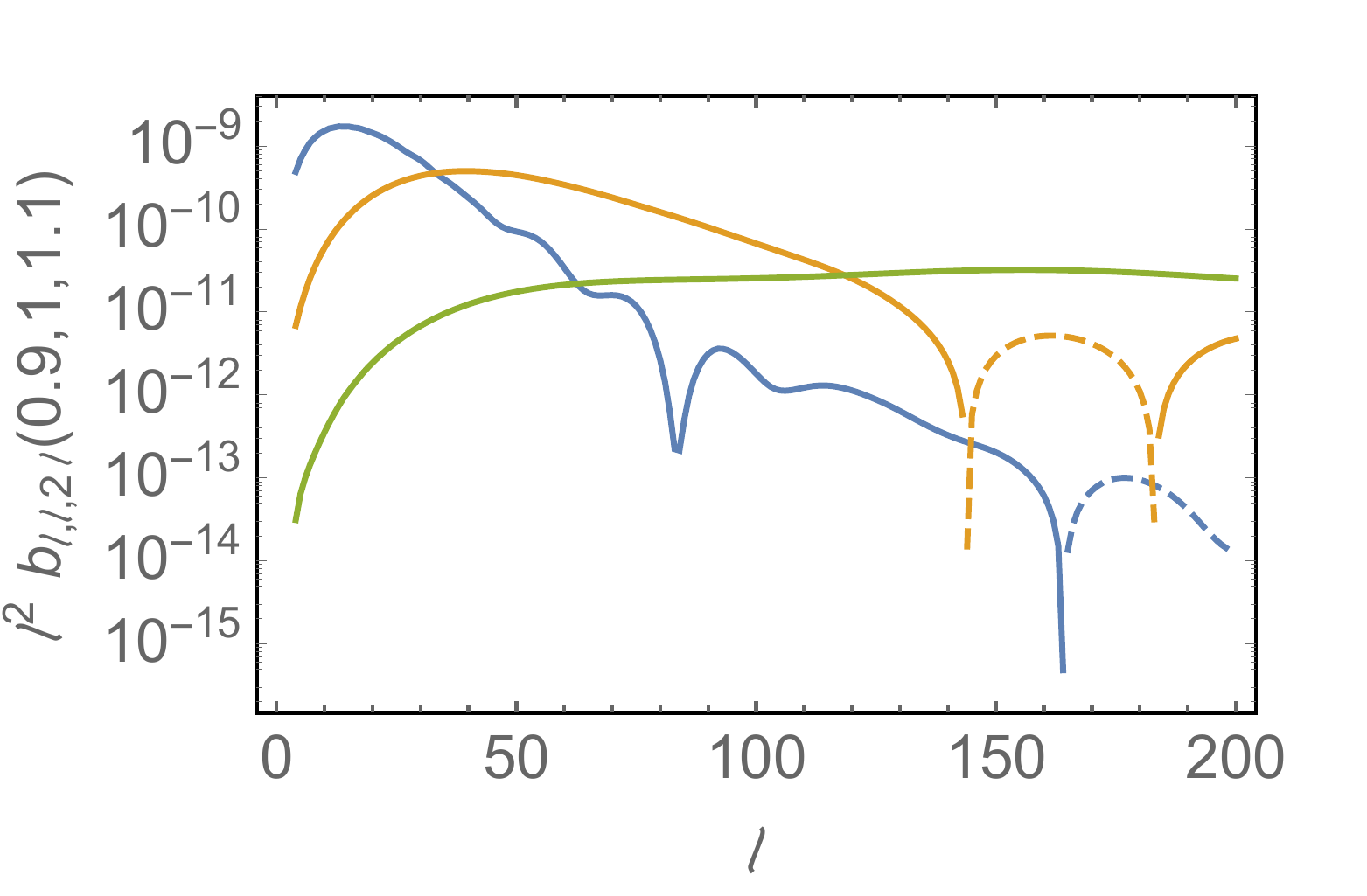}} \\
\subfigure{
\includegraphics[width=0.48\textwidth]{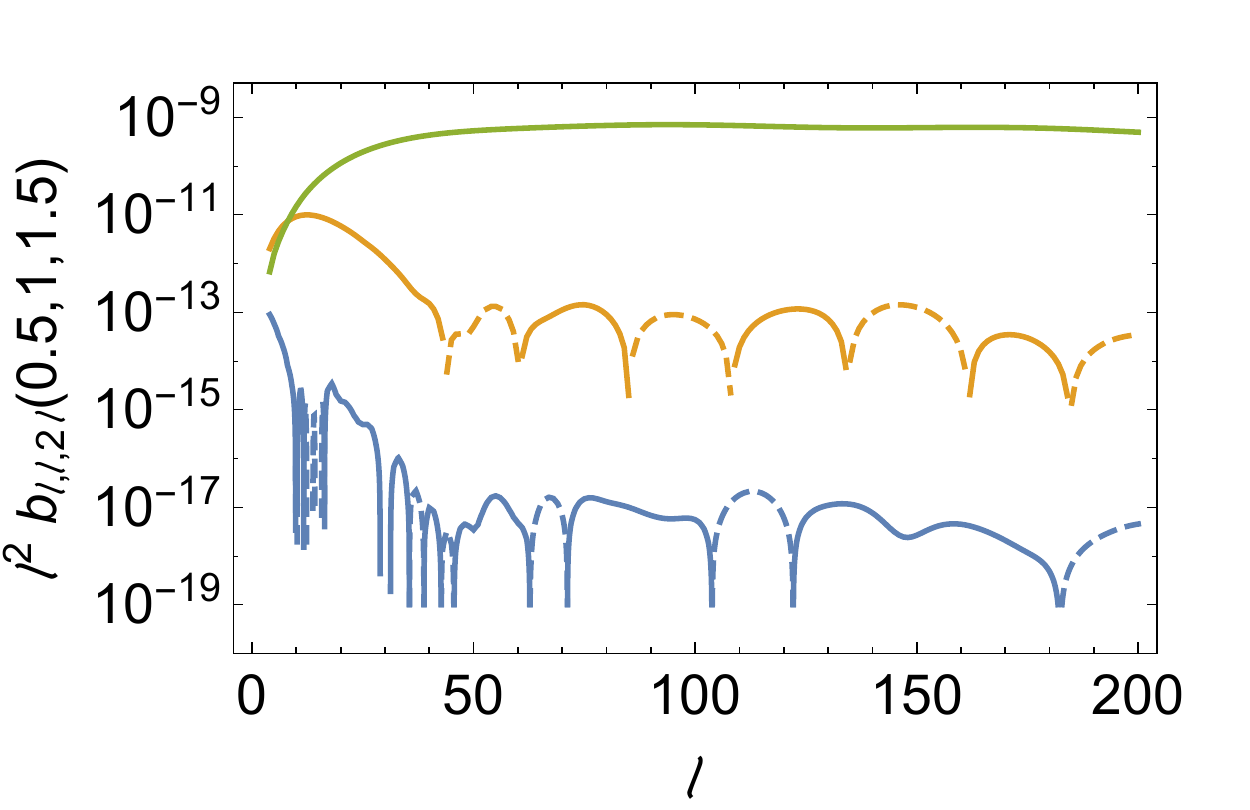}} \hspace{2.1cm}
\subfigure{
\includegraphics[trim=0mm -18mm -12mm 00mm, clip,width=5cm]{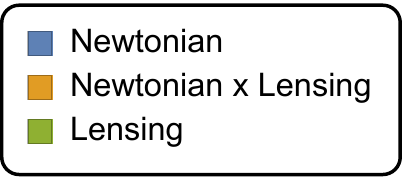}
}
\caption{{\bf Comparison.}
We plot the contributions from the Newtonian terms (blue), the Newtonian~$\times$~lensing terms (yellow) and the pure lensing terms (green)  for $z_1=0.95$, $z_2=1$ and $z_3=1.05$ (top left), for $z_1=0.9$, $z_2=1$ and $z_3=1.1$ (top right) and for $z_1=0.5$, $z_2=1$ and $z_3=1.5$ (bottom) as function of $\ell=\ell_1= \ell_2= \ell_3/2$.  Dashed lines correspond to negative values.
The lensing terms increase with increasing redshift separation from $10^{-12}$  to $10^{-9}$, while the Newtonian terms decay from $10^{-8}$ to $10^{-17}$ and the
Newtonian~$\times$~lensing terms from $10^{-9}$ to $10^{-13}$. For large redshift separations the contribution from lensing clearly dominates. Furthermore, due to its integral nature, it does not show oscillations as function of $\ell$. }
\label{fig:diff}
\end{center}
\end{figure}

On the other hand, for different, sufficiently large $\ell$'s and large enough redshift separations the total contribution from these pure lensing terms can actually dominate the result as can be seen in Fig.~\ref{fig:diff}~\footnote{In Fig.~\ref{fig:diff} the contributions from the second order RSD and $\kappa^{(2)}$ are not included. In fact, since we can evaluate them only in the Limber approximation
(see Eqs. (\ref{RSDorder2FR}) and (\ref{RedBispLensOrd2}))
their contribution vanishes in the configurations shown in Fig.~\ref{fig:diff}. However, even if their contribution could be numerically evaluated starting from the exact expressions they would be of the same order as the other contributions and not change the conclusion  drawn from Fig.~\ref{fig:diff}.}.
In particular, for the configuration $\ell=\ell_1=\ell_2=\ell_3/2$, we have found that for three different redshifts
$z_1+\Delta z=z_2=z_3-\Delta z$ with $z_2=1$ and $\Delta z=0.1$, the combinations of Newtonian~$\times$~lensing terms
dominates for $30<\ell<120$, while at larger $\ell$'s the pure lensing terms dominate, see top-right panel of Fig.~\ref{fig:diff}.
For $\Delta z=0.5$ the pure lensing terms always dominate, see bottom panel of Fig.~\ref{fig:diff}.

Furthermore, in analogy with the case of the power spectrum \cite{Montanari:2015rga}, in a fully tomographic analysis where all bin cross-correlations are taken into account, the signal coming from each triplet of redshift bins will add up.
Distant correlations (i.e., when bins are separated by more than about $150h^{-1}$Mpc, and local terms are thus suppressed) are dominated by lensing  terms. This observation shows that these new lensing terms are in principle observable in a  tomographic bispectrum.

Let us finally point out that also if we consider magnification bias in the configuration for which all the
pure lensing contributions vanish (see Appendix~\ref{app:B}), part of the  Newtonian~$\times$~lensing contribution survives and still dominates over the  Newtonian terms for large enough redshift separations.

In the present study we are not considering any specific survey including noise, and hence we cannot tell whether the signal is truly measurable in a given survey. We leave this important question for a forthcoming more detailed analysis.


\subsection{Window functions}
\label{WinFunction}
A real observation can measure the galaxy positions only with a finite resolution. This requires the introduction of window functions in redshift space in order to compare our results with observations. One also has to introduce a binning in  redshift space in order to beat shot-noise, which is inversely proportional to the number of galaxies in the bin. Once we convolve the signal with   a window function we can no longer resolve scales in radial direction which are smaller than its width. This leads to a degradation, especially,  of the velocity but also of the density terms. On the other hand integrated lensing-like terms are unaffected, as it has been already shown in the past, see e.g.~\cite{Bonvin:2011bg,DiDio:2013sea} for the angular power spectrum.

We define the observed reduced bispectrum as
\be \label{bispectrum_obs}
b^W_{\ell_1 \ell_2 \ell_3} \left( z_1, z_2, z_3 \right) = \int dz'_1 dz'_2 dz'_3 \ b_{\ell_1 \ell_2 \ell_3} \left( z'_1, z'_2, z'_3 \right) W(z_1,z'_1)W(z_2,z'_2)W(z_3,z'_3)
\ee
where $W(z,z')$ denotes a window function centered at z, with an integral over $z'$ normalized to unity. We simplify the observed reduced bispectrum~(\ref{bispectrum_obs}) by noticing that we can express all the reduced
bispectra in terms of products of angular power spectra, except for the second order terms $\de^{(2)}$, $v^{(2)}$ and $\ka^{(2)}$ which involve the kernels $F_2$ and $G_2$.
For instance, most of our contributions are of the form
\bea \label{window_product}
b^{p_\alpha p_\beta}_{\ell_1 \ell_2 \ell_3} \left( z_1, z_2, z_3 \right) &=&
c_{\ell_2}^{p_\alpha \delta} \left( z_1, z_2 \right) c_{\ell_3}^{p_\beta \delta} \left( z_1, z_3 \right)
+ c_{\ell_3}^{p_\alpha \delta} \left( z_1, z_3 \right) c_{\ell_2}^{p_\beta \delta} \left( z_1, z_2 \right)
\nonumber \\
&&
+
c_{\ell_1}^{p_\alpha \delta} \left( z_2, z_1 \right) c_{\ell_3}^{p_\beta \delta} \left( z_2, z_3 \right)
+ c_{\ell_3}^{p_\alpha \delta} \left( z_2, z_3 \right) c_{\ell_1}^{p_\beta \delta} \left( z_2, z_1 \right)
\nonumber \\
&&
+
c_{\ell_1}^{p_\alpha \delta} \left( z_3, z_1 \right) c_{\ell_2}^{p_\beta \delta} \left( z_3, z_2 \right)
+ c_{\ell_2}^{p_\alpha \delta} \left( z_3, z_2 \right) c_{\ell_1}^{p_\beta \delta} \left( z_3, z_1 \right)  \, ,
\eea
where $p_\alpha$ and $p_\beta$ denote any perturbation present in Eq.~(\ref{e:Si2}). For such terms we can reduce the above triple integral to a double integral. Indeed, defining
\be
\bar c^{p_\alpha \delta}_{\ell} \left( z'_i ,z_j \right) \equiv \int dz'  c^{p_\alpha \delta}_{\ell} \left( z'_i ,z' \right) W(z_j,z')
\ee
the observed reduced bispectrum can be written as
\bea
b^W_{\ell_1 \ell_2 \ell_3} \left( z_1, z_2, z_3 \right) &=&
\int dz'_1 W(z_1,z'_1)\left[  \bar c^{p_\alpha \delta}_{\ell_2} \left( z'_1,z_2 \right)\bar c^{p_\beta \delta}_{\ell_3} \left( z'_1,z_3 \right)
+ \bar c^{p_\alpha \delta}_{\ell_3} \left( z'_1,z_3 \right)\bar c^{p_\beta \delta}_{\ell_2} \left( z'_1,z_2 \right)
\right]
\nonumber \\
&&
+
\int dz'_2 W(z_2,z'_2)\left[  \bar c^{p_\alpha \delta}_{\ell_1} \left( z'_2,z_1 \right)\bar c^{p_\beta \delta}_{\ell_3} \left( z'_2,z_3 \right)
+ \bar c^{p_\alpha \delta}_{\ell_3} \left( z'_2,z_3 \right)\bar c^{p_\beta \delta}_{\ell_1} \left( z'_2,z_1 \right)
\right]
\nonumber \\
&&
+
\int dz'_3 W(z_3,z'_3) \left[  \bar c^{p_\alpha \delta}_{\ell_1} \left( z'_3 ,z_1\right)\bar c^{p_\beta \delta}_{\ell_2} \left( z'_3,z_2 \right)
+ \bar c^{p_\alpha \delta}_{\ell_2} \left( z'_3,z_2 \right)\bar c^{p_\beta \delta}_{\ell_1} \left( z'_3,z_1 \right)
\right]\,,
\nonumber
\\
\eea
This approach works for all the terms, except for the pure second order contributions and the integrals of products of power spectra. For the latter case it can be easily generalized. Nevertheless, for all the terms that cannot be written as a product of power spectra, we evaluate the contributions to the observed reduced bispectrum with the Limber approximation. We have tested the Limber approximation for the generalized power spectra $c_\ell^{AB}$ in several cases and it was always excellent (see also Fig.~\ref{fig:limber} in Appendix~\ref{app:kap2}).
\begin{figure}[htbp]
\begin{center}
\subfigure{
\includegraphics[width=0.64\textwidth]{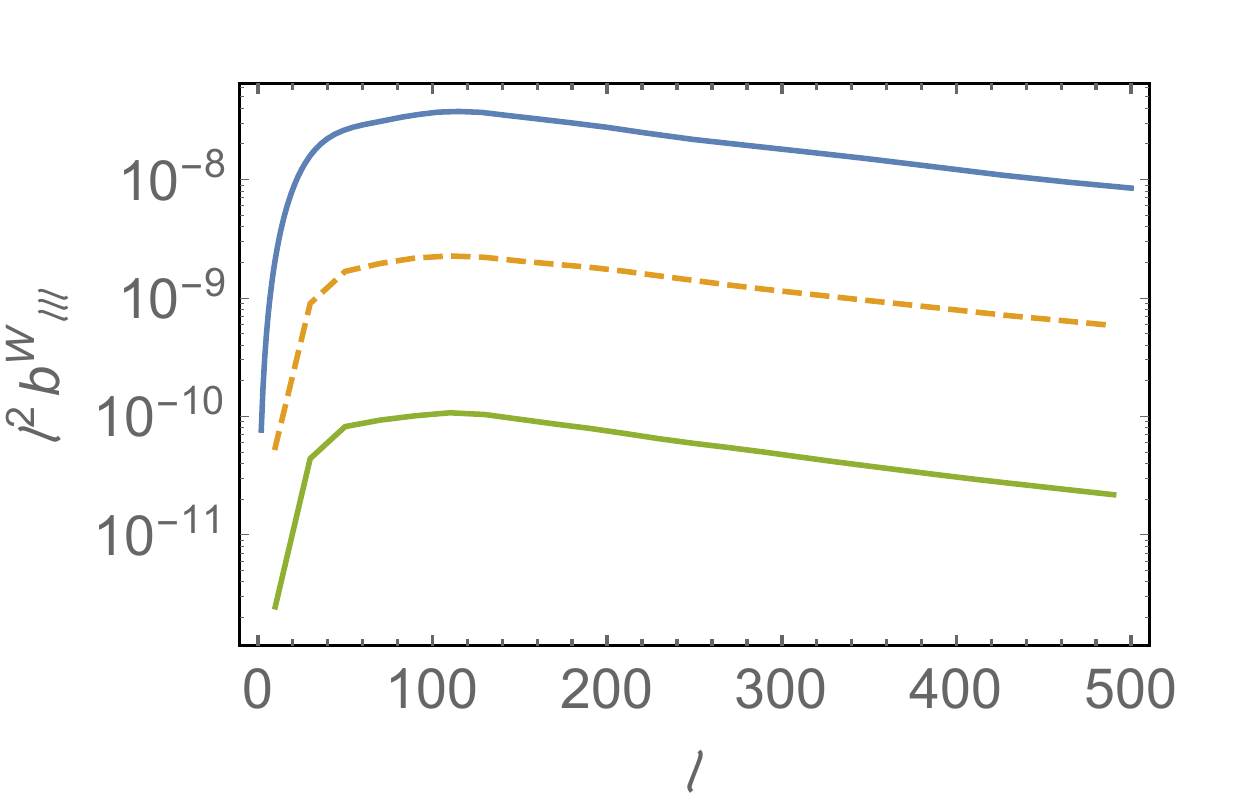}}
 \subfigure{
 \includegraphics[trim=0mm -28mm -12mm 00mm, clip,width=5cm]{legend}}
\caption{{\bf Window Function.}
We plot the contributions to the bispectrum with window function of width $\Delta z=1$ and mean redshift $z=1$, for Newtonian (blue), Newtonian~$\times$~lensing (yellow) and Lensing (green). Dashed lines correspond to negative values.
}
\label{fig:window}
\end{center}
\end{figure}

In Fig.~\ref{fig:window} we show the bispectrum with a top-hat window function of total width $\De z=1$ and mean redshift $z=1$. The density and especially the redshift space distortion terms are significantly reduced by this broad window, while the amplitudes of lensing terms are practically unchanged.
Nevertheless, also in this case, the Newtonian terms are still nearly 2 or more orders of magnitude larger than the lensing contributions, but only one order of magnitude larger of the Newtonian~$\times$~lensing terms, which are less affected by the broad window. Therefore, if the data is too sparse for a full tomographic determination of the bispectrum, integration over redshift may help to enhance the lensing contributions.

\section{Conclusions}
\label{s:con}
In this work we have calculated all the 15  terms in the bispectrum of the number counts coming from second order perturbation theory which dominate on sub-Hubble scales and at intermediate to large redshift.  We have reproduced the 6 well known Newtonian terms and formulated them in the directly observable spherical-harmonics-redshift space. In addition we have found 9 new terms which are due to lensing by foreground structures. All these terms have to be modeled precisely and subtracted if one wants to determine a primordial non-Gaussianity from inflation by measuring  the bispectrum.

On the other hand, we have seen that these lensing-like terms contain significant interesting information and it will be fascinating to measure them as they are sensitive to non-linear aspects of gravity.  In this paper we provide a first inventory of all the dominant terms. We express the bispectrum in terms of 12 contributions which can be given in terms of products of power spectra and three intrinsically second order contributions which are more involved.  We also show and discuss some numerical examples where we find that, like for the power spectrum, the bispectrum at equal redshifts is dominated by the density and redshift space distortion terms, which we call the Newtonian terms, while at well separated redshifts the lensing contributions dominate.
We have found that for the case of three different redshifts separated by $\Delta z=0.1$, the combinations of lensing terms with density and redshift space distortion dominates for $30<\ell<120$, while at larger $\ell$'s or for a larger redshift separation the pure lensing terms dominate, see Fig.~\ref{fig:diff}.

Inspired from the expressions with the Limber approximation, we can summarise the results by saying that for three equal redshifts the Newtonian terms dominate, while for two equal redshifts the Newtonian~$\times$~lensing terms dominate and for three different redshifts the pure lensing terms dominate.

Finally, we have presented some configurations of the bispectrum including a top-hat window function of total width $\De z=1$.
While the density and, especially, the redshift space distortion terms are significantly reduced by this broad window, the amplitude of lensing terms is unchanged.
Despite this reduction, the Newtonian terms are still more than two orders of magnitude larger than the pure lensing contributions, but  only one order of magnitude larger than the combinations of lensing terms with density and redshift space distortion. As they are negative they lead to a decrease by about 10\% of the redshift integrated bispectrum.  Either this, or the especially promising method of redshift separation present  opportunities to measure the Newtonian~$\times$~lensing terms or even the pure lensing terms in future surveys.

As a next step we will study how to measure this bispectrum in upcoming surveys. It will be especially interesting to investigate whether planned surveys are sensitive to the lensing contributions. Even though these are always small, we have seen that they  dominate the signal for sufficiently large redshift separations.

 It will also be interesting to test our perturbative results with N-body simulations. For the terms presented here 
Newtonian N-body simulations are sufficient, but one will have to go beyond the Born approximation in ray-tracing to see all our new lensing-like terms. 
Present ray-tracing codes use the Born approximation which calculates the lensing potential~\cite{Fosalba:2013wxa,Fosalba:2013mra}, but do not include this in the description of the density fluctuations on the light-cone. To go beyond the Born approximation, one will have to modify existing N-body codes either by including effects of General Relativity, see~\cite{Adamek:2015eda,Adamek:2014xba}, or by going to higher order in the photon propagation in a Newtonian N-body code.

\section*{Acknowledgement}
We thank David Goldberg, David Spergel and Filippo Vernizzi for useful discussions.
We are also grateful to Julien Lesgourgues and Thomas Tram for suggestions about the implementation of some transfer function in the \class{} code.
ED is supported by the ERC grant `cosmoIGM' and by INFN/PD51 INDARK grant.
GM is supported by the Marie Curie IEF, Project NeBRiC - ``Non-linear effects and backreaction in classical and quantum cosmology".
This work is supported by the Swiss National Science Foundation.

\appendix

\section{Derivation of bispectrum terms}
\label{a:deri}

In this appendix we derive in some detail the expressions for the different contributions to the bispectrum using the formalism outlined in the main text, Section~\ref{ss:bispec}.

\subsection{Density term}
In this section we follow closely \cite{Spergel:1999xn}.
We denote the $n$-order density perturbation and its Fourier transform by
\be
\delta^{(n)}(\bnx,z) = \frac{1}{(2\pi)^3} \int d^3k \delta^{(n)}(\bk,z) e^{-i\bk\cdot\bx} \;.
\ee
In  an Einstein-de Sitter universe, the second order term is given by Eq.~(\ref{e:de2}), see~\cite{Bernardeau:2001qr}.
We expand the mode-coupling term $F_2(\bk_1,\bk_2)$ in Legendre polynomials:
\be
F_2(\bk_1,\bk_2) = \sum_{\ell=0}^2 f_{\ell}(k_1,k_2) P_{\ell}(\bnk_1\cdot\bnk_2) = \sum_{\ell m} \frac{4\pi}{2\ell+1} f_{\ell}(k_1,k_2) Y_{\ell m}(\bnk_1)  Y_{\ell m}^*(\bnk_2) \;,
\ee
where the only non-vanishing coefficients are the monopole $f_0(k_1,k_2)=\frac{17}{21}$, the dipole $f_1(k_1,k_2)=\frac{1}{2}\left( \frac{k_1}{k_2} + \frac{k_2}{k_1}  \right)$ and the quadrupole $f_2(k_1,k_2)=\frac{4}{21}$, given in Eq.~(\ref{e:F2}).
The linear harmonic coefficients
of $\de$ are\footnote{We recall that in our conventions $\bn$ is the direction of propagation of photons, opposed to the direction of observation $\bnx=-\bn$.}
\be
a_{\ell_1 m_1}^{(1)} = 4\pi i^{\ell_1} \int \frac{d^3k}{(2\pi)^3} \delta^{(1)}(\bk,z) Y_{\ell_1 m_1}^*(\bnk) j_{\ell_1}(kr) \;,
\ee
while at second order we have
\bea
a_{\ell_3 m_3}^{(2)} &=& \frac{(4\pi)^3}{(2\pi)^6} \int  dk_1  dk_2 k_1^2 k_2^2
\sum_{\substack{\ell\ell'\ell''\\ mm'm''}} i^{\ell'+\ell''} \frac{1}{2\ell+1} f_{\ell}(k_1,k_2) j_{\ell'}(k_1r) j_{\ell''}(k_2r)
\nonumber \\
&& \times \int d\bn d\bnk_1 d\bnk_2 Y_{\ell_3 m_3}^*(\bn) Y_{\ell' m'}(\bn) Y_{\ell'' m''}(\bn) Y_{\ell' m'}^*(\bnk_1) Y_{\ell m}(\bnk_1) Y_{\ell'' m''}^*(\bnk_2) Y_{\ell m}^*(\bnk_2)
\nonumber \\
&&\times \delta^{(1)}(\bk_1,z) \delta^{(1)}(\bk_2,z)
\eea
The coefficients of the bispectrum are given by\footnote{
This result is in agreement with Eq.~(13) of \cite{Spergel:1999xn}, except for the  factors $i^{\ell'+\ell''} i^{\ell_1+\ell_2}$ which are missing in their equation.
 As we show in the next sections, by symmetry these terms are real and contribute as $\pm1$. But the powers of $i$ lead to important sign differences in some of the summands and cannot be ignored.
}
\bea
B_{\ell_1\ell_2\ell_3}^{m_1m_2m_3} &=& \langle a_{\ell_1 m_1}^{(1)}(z_1) a_{\ell_2 m_2}^{(1)}(z_2) a_{\ell_3 m_3}^{(2)}(z_3) \rangle_c +  \langle a_{\ell_1 m_1}^{(1)}(z_1) a_{\ell_2 m_2}^{(2)}(z_2) a_{\ell_3 m_3}^{(1)}(z_3) \rangle_c
\nonumber \\
  &&+  \langle a_{\ell_1 m_1}^{(2)}(z_1) a_{\ell_2 m_2}^{(1)}(z_2) a_{\ell_3 m_3}^{(1)}(z_3) \rangle_c
\nonumber \\
&=& \frac{(4\pi)^5}{(2\pi)^{12}} \int  \int  dk_1 dk_2 dk dk' k^2k^{'2}k_1^2  k_2^2
\nonumber \\
  && \times \sum_{\substack{\ell\ell'\ell''\\ mm'm''}} \frac{1}{2\ell+1} f_{\ell}(k',k) j_{\ell'}(k'r_3) j_{\ell''}(kr_3) j_{\ell_1}(k_1r_1) j_{\ell_2}(k_2r_2)
\nonumber \\
  &&\times \int d\bn d\bnk' d\bnk Y_{\ell m}(\bnk') Y_{\ell m}^*(\bnk) Y_{\ell' m'}(\bn) Y_{\ell' m'}^*(\bnk_1) Y_{\ell'' m''}(\bn) Y_{\ell'' m''}^*(\bnk)
\nonumber \\
  && \times Y_{\ell_1 m_1}^*(\bnk_1) Y_{\ell_2 m_2}^*(\bnk_2) Y_{\ell_3 m_3}^*(\bn)
\nonumber \\
  &&\times i^{\ell'+\ell''} i^{\ell_1+\ell_2} \langle \delta^{(1)}(\bk_1,z_1) \delta^{(1)}(\bk_2,z_2)\delta^{(1)}(\bk',z_3) \delta^{(1)}(\bk,z_3)  \rangle_c
\nonumber \\
  && +\text{perms.} \;,
\eea
where we compute explicitly only the first of the three permutations.
Wick's theorem gives three permutations for the density correlators, one of which contributes to the disconnected part.
The two remaining permutations are equal as one sees by applying the Dirac delta from Eq.~(\ref{eq:Pk_def}) to the integrals over $\bk$ and $\bk'$, and then exchanging $\bk_1 \leftrightarrow \bk_2$ together with $\ell' \leftrightarrow \ell''$.

Let us introduce the geometrical factor $g_{\ell_1\ell_2\ell_3}$ relating the reduced bispectrum $b_{\ell_1\ell_2\ell_3}$, Eq.~(\ref{bi_harm_red}), to the angle-averaged bispectrum
\bea
\label{eq:gb_red}
g_{\ell_1\ell_2\ell_3} \, b_{\ell_1\ell_2\ell_3} &\equiv& \sqrt{\frac{(2\ell_1+1)(2\ell_2+1)(2\ell_3+1)}{4\pi}} \tj{\ell_1}{\ell_2}{\ell_3}{0}{0}{0} b_{\ell_1\ell_2\ell_3}
\nonumber \\
&=& \sum_{m_1m_2m_3}  \tj{\ell_1}{\ell_2}{\ell_3}{m_1}{m_2}{m_3} B_{\ell_1\ell_2\ell_3}^{m_1m_2m_3} \;,
\eea
With this we obtain:
\bea
\label{eq:B_multi}
g_{\ell_1\ell_2\ell_3} b_{\ell_1\ell_2\ell_3} &=&
  2\frac{(4\pi)^2}{(2\pi)^6} \int  dk_1  dk_2 k_1^2  k_2^2
     P_{\mathcal{R}}(k_1) P_{\mathcal{R}}(k_2) T_{\delta}(k_1,z_1)T_{\delta}(k_1,z_3) T_{\delta}(k_2,z_2)T_{\delta}(k_2,z_3)
\nonumber \\
&& \times \sum_{\ell'\ell''} (2\ell'+1)(2\ell''+1)  j_{\ell'}(k_2r_3) j_{\ell''}(k_1r_3) j_{\ell_1}(k_1r_1) j_{\ell_2}(k_2r_2)
\nonumber \\
&&\times i^{\ell'+\ell''} (-i)^{\ell_1+\ell_2}
\left[ \frac{17}{21}Q_{0\ \ell' \ell''}^{\ell_1 \ell_2 \ell_3}
  + \frac{1}{2}\left( \frac{k_1}{k_2} + \frac{k_2}{k_1}  \right) Q_{1\ \ell' \ell''}^{\ell_1 \ell_2 \ell_3}
  + \frac{4}{21}Q_{2\ \ell' \ell''}^{\ell_1 \ell_2 \ell_3}
\right]
\nonumber \\
&&+\text{perms.} ,
\eea
where the geometrical factors $Q_{\ell\ell^{\prime }\ell^{\prime \prime }}^{\ell_1\ell_2\ell_3}$ are defined below, in Eq.~(\ref{eq:Q_predef}).
In sections \ref{sec:dens_mono}, \ref{sec:dens_dip} and \ref{sec:dens_quad} we write the three multipoles in a form suited for numerical evaluation, and prove the form of the density bispectrum given in the main text, Eqs.~(\ref{e:bde0}), (\ref{e:bde1}), (\ref{e:bde2}).
The following definition will be useful:
\be
\label{eq:c_ll}
\prescript{n}{}{c}^{AB}_{\ell\ \ell'}(z_1,z_2)
\equiv
i^{\ell-\ell'} \frac{2}{\pi}
\int dk k^{2+n} P_{\mathcal{R}}(k) T_{A}(k,z_1) T_{B}(k,z_2) j_{\ell}(k r_1) j_{\ell'}(k r_2) \;,
\ee
with the convention that, on the left hand side, the first redshift ($z_1$) and multipole ($\ell$) refer to the first transfer in the superscript ($A$), and correspondingly for the second labels. We also indicate $c^{AB}_{\ell}(z_1,z_2) \equiv \prescript{0}{}{c}^{AB}_{\ell \ell}(z_1,z_2)$.
For efficient and accurate integration of the transfer functions obtained with the \class{} code\footnote{\url{http://class-code.net/}} we rely on the GNU Scientific Library.\footnote{\url{https://www.gnu.org/software/gsl/}}

In Eq.~(\ref{eq:B_multi}) we have introduced the following integral describing the geometry of the bispectrum:
\begin{eqnarray}
\label{eq:Q_predef}
Q_{\ell\ell^{\prime }\ell^{\prime \prime }}^{\ell_1\ell_2\ell_3} &\equiv& \int d\hat{{\bf l}} d%
\hat{{\bf m}} d\hat{{\bf n}} P_{\ell}(\hat l \cdot \hat m )P_{\ell^{\prime }}(\hat m \cdot \hat n)P_{\ell^{\prime \prime }}(\hat l \cdot \hat n)  \nonumber \\
&&\quad\times \sum_{m_1,m_2m_3}\left(
\begin{array}{ccc}
\ell_1 & \ell_2 & \ell_3 \\
m_1 & m_2 & m_3
\end{array}
\right) Y_{\ell_1m_1}^{*}(\hat l)Y_{\ell_2m_2}^{*}(\hat m)Y_{\ell_3m_3}^{*}(\hat n)
\nonumber \\
&=&\left( \frac{4\pi }{2\ell+1}\right) \left( \frac{4\pi }{2\ell^{\prime }+1}%
\right) \left( \frac{4\pi }{2\ell^{\prime \prime }+1}\right)
\sum_{m_1,m_2m_3}\left(
\begin{array}{ccc}
\ell_1 & \ell_2 & \ell_3 \\
m_1 & m_2 & m_3
\end{array}
\right)
\nonumber \\
&& \times \sum_{m,m^{\prime },m^{\prime \prime }} \int d\hat{{\bf l}} d%
\hat{{\bf m}} d\hat{{\bf n}} Y_{\ell m}(\hat l) Y_{\ell m}^{*}(\hat m)
Y_{\ell^{\prime}m^{\prime}}(\hat m) Y_{\ell^{\prime}m^{\prime}}^{*}(\hat n)
Y_{\ell^{\prime \prime }m^{\prime \prime }}(\hat n)Y_{\ell^{\prime \prime
  }m^{\prime \prime }}^{*}(\hat l)
\nonumber \\
&& \times Y_{\ell_1m_1}^{*}(\hat
l)Y_{\ell_2m_2}^{*}(\hat m)Y_{\ell_3m_3}^{*}(\hat n)  \nonumber \\
&=&I_{\ell\ell^{\prime }\ell^{\prime \prime }}^{\ell_1\ell_2\ell_3}\sum_{m_1,m_2m_3}\left(
\begin{array}{ccc}
\ell_1 & \ell_2 & \ell_3 \\
m_1 & m_2 & m_3
\end{array}
\right) \sum_{m,m^{\prime },m^{\prime \prime }}\left(
\begin{array}{ccc}
\ell & \ell^{\prime \prime } & \ell_1 \\
m & -m^{\prime \prime } & -m_1
\end{array}
\right)  \nonumber \\
&&\times \left(
\begin{array}{ccc}
\ell^{\prime } & \ell & \ell_2 \\
m^{\prime } & -m & -m_2
\end{array}
\right) \left(
\begin{array}{ccc}
\ell^{\prime\prime} & \ell^{\prime} & \ell_3 \\
m^{\prime\prime} & -m^{\prime} & -m_3
\end{array}
\right) (-1)^{(m+m^{\prime}+m^{\prime\prime})} \;,
\end{eqnarray}
where we have used the Gaunt integral and defined
\bea
I_{\ell\ell^{\prime }\ell^{\prime \prime }}^{\ell_1\ell_2\ell_3} &\equiv&
\sqrt{(4\pi
  )^3(2\ell_1+1)(2\ell_2+1)(2\ell_3+1)}
\nonumber \\
&& \times
\left(
\begin{array}{ccc}
\ell & \ell^{\prime \prime } & \ell_1 \\
0 & 0 & 0
\end{array}
\right) \left(
\begin{array}{ccc}
\ell^{\prime } & \ell & \ell_2 \\
0 & 0 & 0
\end{array}
\right) \left(
\begin{array}{ccc}
\ell^{\prime \prime } & \ell^{\prime} & \ell_3 \\
0 & 0 & 0
\end{array}
\right) \;.
\eea
Furthermore, Eq.~(\ref{eq:Q_predef}) can be written in terms of the Wigner 6$j$ symbol as
\begin{eqnarray}
\label{eq:Q_def}
Q_{\ell\ell^{\prime }\ell^{\prime \prime }}^{\ell_1\ell_2\ell_3}
=
I_{\ell\ell^{\prime }\ell^{\prime \prime }}^{\ell_1\ell_2\ell_3}\left\{
\begin{array}{ccc}
\ell_1 & \ell_2 & \ell_3 \\
\ell^{\prime} & \ell^{\prime\prime} & \ell
\end{array}
\right\} \left( -1\right) ^{\ell+\ell^{\prime }+\ell^{\prime \prime }} \,.
\end{eqnarray}
A useful property is that the $6j$ symbol is non-vanishing only if the triangle condition is satisfied by the triplets
\be
(\ell_1,\ell_2,\ell_3),\ (\ell_1,\ell'',\ell),\ (\ell',\ell_2,\ell) \mbox{ and } (\ell',\ell'',\ell_3) \;.
\ee
For example, for the first triplet this implies:
\be
|\ell_i-\ell_j| \leq \ell_k \leq \ell_i+\ell_j \;,
\ee
where $(i,j,k)$ is an arbitrary permutation of $(1,2,3)$.
The non-vanishing coefficients of $Q_{\ell\ell^{\prime }\ell^{\prime \prime }}^{\ell_1\ell_2\ell_3}$ can be determined by using these conditions, together with those related to $I_{\ell\ell^{\prime }\ell^{\prime \prime }}^{\ell_1\ell_2\ell_3}$, whose $3j$ symbols further require the following sums to be even:
\be
\ell_1+\ell''+\ell,\ \ell'+\ell_2+\ell,\ \ell'+\ell''+\ell_3 \;,
\ee
which also implies
\be
\ell_1+\ell_2+\ell_3 = \text{even} \;.
\ee
For other properties in relation with Wigner 3$j$ symbols see, e.g., \cite{Varshalovich,Komatsu:2002db}.

The geometrical factors can be efficiently computed numerically in terms of Wigner symbols with publicly available libraries, e.g., WIGXJPF\footnote{\url{http://fy.chalmers.se/subatom/wigxjpf/}} \cite{Johansson:2015cca}.

\subsubsection{Monopole}
\label{sec:dens_mono}
For  $\ell=0$, the only non-vanishing term is:
\be
Q_{0\ \ell_2 \ell_1}^{\ell_1\ell_2\ell_3} =  \sqrt{\frac{\left( 2\ell_3+1\right) (4\pi )^3}{%
    (2\ell_2+1)(2\ell_1+1)}}
\tj{\ell_1}{\ell_2}{\ell_3}{0}{0}{0}\,.
\ee
The reduced bispectrum of the monopole reads
\bea
b_{\ell_1\ell_2\ell_3}^{\delta0}(z_1,z_2,z_3) &=& \sqrt{\frac{4\pi}{(2\ell_1+1)(2\ell_2+1)(2\ell_3+1)}} \tj{\ell_1}{\ell_2}{\ell_3}{0}{0}{0}^{-1} B_{\ell_1\ell_2\ell_3}^{\delta0}
\nonumber \\
&=& 2 \frac{17}{21} \frac{2}{\pi} \int  dk_1 k_1^2 P_{\mathcal{R}}(k_1) T_{\delta}(k_1,z_1)T_{\delta}(k_1,z_3) j_{\ell_1}(k_1r_3) j_{\ell_1}(k_1r_1)
\nonumber \\
&& \times \frac{2}{\pi} \int dk_2 k_2^2 P_{\mathcal{R}}(k_2) T_{\delta}(k_2,z_2)T_{\delta}(k_2,z_3) j_{\ell_2}(k_2r_3) j_{\ell_2}(k_2r_2)
\nonumber \\
&&+\text{perms.}
\eea
Introducing the angular-redshift power spectra, we finally obtain:
\be\label{ea:d0}
b_{\ell_1\ell_2\ell_3}^{\delta0}(z_1,z_2,z_3) = \frac{34}{21} c_{\ell_1}^{\delta\delta}(z_1,z_3) c_{\ell_2}^{\delta\delta}(z_2,z_3) + \text{perms.}
\,.
\ee
With this we recover the result already obtained in Eq.~(\ref{e:bde0}).

\subsubsection{Dipole}
\label{sec:dens_dip}
For $\ell=1,$ $Q_{1\ \ell'\ell''}^{\ell_1\ell_2\ell_3}$ is zero unless
\be
\ell'=\ell_2 \pm 1 \quad \text{and}\quad \ell''=\ell_1 \pm 1 \;.
\ee
This guarantees that $i^{\ell'+\ell''} (-i)^{\ell_1+\ell_2} = \pm 1$.
The dipole reduced bispectrum reads
\bea
b_{\ell_1\ell_2\ell_3}^{\delta 1}(z_1,z_2,z_3) &=& \left(g_{\ell_1\ell_2\ell_3}\right)^{-1} \frac{1}{16\pi^2} \sum_{\ell'\ell''} (2\ell'+1)(2\ell''+1) Q_{1\ \ell' \ell''}^{\ell_1 \ell_2 \ell_3}
\nonumber \\
&& \times \left[
  \prescript{1}{}{c}^{\delta \delta}_{\ell'' \ell_1}(z_3,z_1)
  \prescript{-1}{}{c}^{\delta \delta}_{\ell' \ell_2}(z_3,z_2)
  +
  \prescript{-1}{}{c}^{\delta \delta}_{\ell'' \ell_1}(z_3,z_1)
  \prescript{1}{}{c}^{\delta \delta}_{\ell' \ell_2}(z_3,z_2)
  \right]
\nonumber \\
&&+\text{perms.}\,. \label{ea:d1}
\eea

\subsubsection{Quadrupole}
\label{sec:dens_quad}
If $\ell=2,$ $Q_{2\ \ell'\ell''}^{\ell_1\ell_2\ell_3}$ is zero unless
\be
\ell'=\ell_2\pm 2, \ell_2
\quad \text{and} \quad
\ell''=\ell_1\pm 2, \ell_1 \;.
\ee
This guarantees that $i^{\ell'+\ell''} (-i)^{\ell_1+\ell_2} = \pm 1$.
The quadrupole bispectrum reads
\bea
b_{\ell_1\ell_2\ell_3}^{\delta 2}(z_1,z_2,z_3) &=&
\left(g_{\ell_1\ell_2\ell_3}\right)^{-1}
\frac{1}{42\pi^2}
\sum_{\ell'\ell''} (2\ell'+1)(2\ell''+1) Q_{2\ \ell' \ell''}^{\ell_1 \ell_2 \ell_3}
\;\prescript{0}{}{c}^{\delta \delta}_{\ell'' \ell_1}(z_3,z_1) \;\prescript{0}{}{c}^{\delta \delta}_{\ell' \ell_2}(z_3,z_2)
\nonumber \\
&&+\text{perms.}\,. \label{ea:d2}
\eea

\subsection{Term $\HH^{-2}  \dd_rv\partial_r^3v$}
\label{ssa:v3v}
We consider
\be\label{ea:12}
 \left\langle \left(\HH^{-2}  \dd_rv\partial_r^3v \right)\left( \bn_1, z_1 \right) \delta\left( \bn_2, z_2 \right) \delta \left( \bn_3, z_3 \right) \right\rangle_c + \text{perm.} \, .
\ee
We write the first factor as a product of integrals in Fourier space,
\bea
\Delta^{v  v''} \left( \bn, z \right) &=& \frac{\HH^{-2}(z)}{\left( 2 \pi \right)^6} \int d^3k d^3k_1 k_1^2 V \left( \bk , \eta \right) V \left( \bk_1 , \eta \right) \partial_{\left( k r \right)} e^{i \bk \cdot \bn r} \partial_{\left( k_1 r \right)}^3  e^{i \bk_1 \cdot \bn r}\,,
\eea
where  $V$ is  the velocity potential (in Newtonian gauge) defined through ${\bf v} = i \hat \bk V$. Here $\eta=\eta(z)$ and $r=r(z)=\eta_0-\eta(z)$.
We then compute~(\ref{ea:12})
\bea
&&\hspace{-0.6cm}\big\langle \cdots \big\rangle  =
 \frac{1}{\HH \left(\!z_1\! \right)^2\left( 2 \pi \right)^{12}} \! \int \!d^3k  d^3k_1 d^3 k_2 d^3 k_3 k_1^2 \!\left(\!\partial_{\left( k r_1\!\right)} e^{i \bk \cdot \bn_1 r_1} \!\right)\!
 \left(\partial_{\left( k_1 r_1 \right)}^3 e^{i \bk_1 \cdot \bn_1 r_1} \right)
e^{i \left( \bk_2 \cdot \bn_2r_2 +  \bk_3 \cdot \bn_3r_3 \right)}
\nonumber \\
&&\times  \ T_v \left( k, \eta_1 \right) T_v\left( k_1, \eta_1 \right) T_\delta \left( k_2, \eta_2 \right) T_\delta \left( k_3 , \eta_3 \right)
\langle R \left( \bk \right) R \left( \bk_1 \right) R \left( \bk_2 \right) R \left( \bk_3 \right)\rangle
\nonumber \\
&=& \frac{1}{\HH \left( z_1 \right)^2\left( 2 \pi \right)^6} \int d^3k_2 d^3k_3 k_3^2 \left(\partial_{\left( k_2 r_1\right)} e^{- i \bk_2 \cdot \bn_1 r_1} \right)
 \left(\partial_{\left( k_3 r_1 \right)}^3 e^{-i \bk_3 \cdot \bn_1 r_1} \right)
e^{i \left( \bk_2 \cdot \bn_2r_2 +  \bk_3 \cdot \bn_3r_3 \right)}
\nonumber \\
&& \times T_v \left( k_2, \eta_1 \right) T_v\left( k_3, \eta_1 \right) T_\delta \left( k_2, \eta_2 \right) T_\delta \left( k_3 , \eta_3 \right)P_R  \left( k_2 \right)P_R  \left( k_3 \right)
\nonumber \\
&&+ \frac{1}{\HH \left( z_1 \right)^2\left( 2 \pi \right)^6} \int d^3k_2 d^3k_3 k_2^2 \left(\partial_{\left( k_3 r_1\right)} e^{- i \bk_3 \cdot \bn_1 r_1} \right)
 \left(\partial_{\left( k_2 r_1 \right)}^3 e^{-i \bk_2 \cdot \bn_1 r_1} \right)
e^{i \left( \bk_2 \cdot \bn_2r_2 +  \bk_3 \cdot \bn_3r_3 \right)}
\nonumber \\
&& \times \ T_v \left( k_2, \eta_1 \right) T_v\left( k_3, \eta_1 \right) T_\delta \left( k_2, \eta_2 \right) T_\delta \left( k_3 , \eta_3 \right)P_R  \left( k_2 \right)P_R  \left( k_3 \right)\,.
\eea
Expanding the exponentials in spherical harmonics and spherical Bessel functions we obtain
\bea
B^{v v''} \left( \bn_1, \bn_2 ,\bn_3, z_1,z_2 , z_3 \right)&=& \frac{4 }{\pi^2} \sum_{\substack{\ell, \ell'\\ m, m'}} Y_{\ell m} \left( \bn_1 \right) Y_{\ell' m'} \left( \bn_1 \right) Y^*_{\ell m } \left( \bn_2 \right) Y^*_{\ell' m'}\left( \bn_3 \right) Z^{vv''}_{\ell \ell'} \left( z_1 , z_2 ,z_3 \right)
\nonumber \\
&&
 + \text{perm.}\,,
\eea
where
\bea
&&\hspace{-1.2cm}Z^{v v''}_{\ell \ell'} \left( z_1 , z_2 ,z_3 \right)
\nonumber \\
&=& \frac{1}{\HH \left( z_1 \right)^2} \int dk_2 dk_3 k_2^2 k_3^4P_R  \left( k_2 \right)P_R  \left( k_3 \right) T_v \left( k_2 ,\eta_1 \right) T_v \left( k_3 ,\eta_1 \right) T_\delta \left( k_2, \eta_2 \right) T_\delta \left( k_3 , \eta_3 \right)
\nonumber \\
&& \qquad\qquad \times \ j'_{\ell} \left( k_2 r_1 \right) j'''_{\ell'} \left( k_3 r_1 \right) j_{\ell}\left( k_2 r_2 \right) j_{\ell'} \left( k_3 r_3 \right)
\nonumber \\
&& + \frac{1}{\HH \left( z_1 \right)^2} \int dk_2 dk_3 k_2^4 k_3^2P_R  \left( k_2 \right)P_R  \left( k_3 \right) T_v \left( k_2 ,\eta_1 \right) T_v \left( k_3 ,\eta_1 \right) T_\delta \left( k_2, \eta_2 \right) T_\delta \left( k_3 , \eta_3 \right)
\nonumber \\
&& \qquad\qquad \times \ j'''_{\ell} \left( k_2 r_1 \right) j'_{\ell'} \left( k_3 r_1 \right) j_{\ell}\left( k_2 r_2 \right) j_{\ell'} \left( k_3 r_3 \right) \, .
\eea
Note that this expression vanishes if $r_1=r_2$ or $r_1=r_3$ i.e. $z_1=z_2$ or $z_1=z_3$ due to the opposite parity of $j_\ell$ and $j_\ell'$ as well as  $j_\ell$
and $j'''_\ell$.
The reduced bispectrum is therefore given by
\be \label{31}
b_{\ell_1 \ell_2 \ell_3}^{ vv''} = \frac{4}{\pi^2} Z^{vv''}_{\ell_2 \ell_3} \left( z_1 , z_2 ,z_3 \right) + \text{perm..}
\ee
Due to the sum this vanishes only if all three redshifts coincide.

Making use of the definition~(\ref{e:cAB}) and of the transfer functions in multipole space~(\ref{trans_v}) and (\ref{trans_v2}), we can express the bispectrum~(\ref{31}) as
\bea\label{ea:vv''}
b_{\ell_1 \ell_2 \ell_3}^{vv''}&=&
c_{\ell_2}^{v \delta} \left( z_1, z_2 \right)c_{\ell_3}^{v'' \delta} \left( z_1, z_3 \right)
+ c_{\ell_2}^{v'' \delta} \left( z_1, z_2 \right)c_{\ell_3}^{v \delta} \left( z_1, z_3 \right)
\nonumber \\
&&
 + c_{\ell_1}^{v \delta} \left( z_2, z_1 \right)c_{\ell_3}^{v'' \delta} \left( z_2, z_3 \right)
 + c_{\ell_1}^{v'' \delta} \left( z_2, z_1 \right)c_{\ell_3}^{v \delta} \left( z_2, z_3 \right)
 \nonumber \\
&&
 + c_{\ell_1}^{v \delta} \left( z_3, z_1 \right)c_{\ell_2}^{v'' \delta} \left( z_3, z_2 \right)
 + c_{\ell_1}^{v'' \delta} \left( z_3, z_1 \right)c_{\ell_2}^{v \delta} \left( z_3, z_2 \right) \, .
\eea

\subsection{Term $\HH^{-2}  \dd_r^2v\partial_r^2v$}
\label{ssa:v2v2}
We compute the contribution of the following term.
\be
 \left\langle \left(\HH^{-2}  \dd_r^2 v\partial_r^2v \right)\left( \bn_1, z_1 \right) \delta \left( \bn_2, z_2 \right) \delta\left( \bn_3, z_3 \right) \right\rangle_c + \text{perm.}\,.
\ee
In Fourier space we can rewrite the product of the redshift space distortions as
\bea
\Delta^{{v'} ^2} \left( \bn, z \right) &=& \frac{\HH^{-2}(z)}{\left( 2 \pi \right)^6} \int d^3k d^3k_1 k k_1 V \left( \bk , \eta \right) V \left( \bk_1 , \eta \right) \partial_{\left( k r \right)}^2 e^{i \bk \cdot \bn r} \partial_{\left( k_1 r \right)}^2  e^{i \bk_1 \cdot \bn r} \, ,
\eea
such that
\bea
\big\langle \cdots \big\rangle&=& \frac{1}{\HH \left( z_1 \right)^2\left( 2 \pi \right)^{12}} \int d^3k  d^3k_1 d^3 k_2 d^3 k_3 k  k_1\left(\partial_{\left( k r_1\right)}^2 e^{i \bk \cdot \bn_1 r_1} \right)
 \left(\partial_{\left( k_1 r_1 \right)}^2 e^{i \bk_1 \cdot \bn_1 r_1} \right)
\nonumber \\
&&\hspace{-0.7cm}
\times \
e^{i \left( \bk_2 \cdot \bn_2r_2 +  \bk_3 \cdot \bn_3r_3 \right)} T_v \left( k, \eta_1 \right) T_v\left( k_1, \eta_1 \right) T_\delta \left( k_2, \eta_2 \right) T_\delta \left( k_3 , \eta_3 \right)
\langle R \left( \bk \right) R \left( \bk_1 \right) R \left( \bk_2 \right) R \left( \bk_3 \right)\rangle
\nonumber \\
&=& \frac{2}{\HH \left( z_1 \right)^2\left( 2 \pi \right)^6} \int d^3k_2 d^3k_3 k_2 k_3 \left(\partial_{\left( k_2 r_1\right)}^2 e^{- i \bk_2 \cdot \bn_1 r_1} \right)
 \left(\partial_{\left( k_3 r_1 \right)}^2 e^{-i \bk_3 \cdot \bn_1 r_1} \right)
\nonumber \\
&& \hspace{-0.7cm}
 \times \ e^{i \left( \bk_2 \cdot \bn_2r_2 +  \bk_3 \cdot \bn_3r_3 \right)} T_v \left( k_2, \eta_1 \right) T_v\left( k_3, \eta_1 \right) T_\delta \left( k_2, \eta_2 \right) T_\delta \left( k_3 , \eta_3 \right)P_R  \left( k_2 \right)P_R  \left( k_3 \right) \, .
\eea
Again, by expanding the exponentials in spherical harmonics and spherical Bessel functions we find
\bea
B^{{v'}^2} \left( \bn_1, \bn_2 ,\bn_3, z_1,z_2 , z_3 \right)&=&\frac{4 }{\pi^2} \sum_{\substack{\ell, \ell'\\ m, m'}} Y_{\ell m} \left( \bn_1 \right) Y_{\ell' m'} \left( \bn_1 \right) Y^*_{\ell m } \left( \bn_2 \right) Y^*_{\ell' m'}\left( \bn_3 \right) Z^{{v'}^2}_{\ell \ell'} \left( z_1 , z_2 ,z_3 \right)
\nonumber \\
&&
+ \text{perm.}\,,
\eea
where
\bea
Z^{{v'}^2}_{\ell \ell'} \left( z_1 , z_2 ,z_3 \right) &=& \frac{2}{\HH \left( z_1 \right)^2} \int dk_2 dk_3 k_2^3 k_3^3 T_v \left( k_2 ,\eta_1 \right) T_v \left( k_3 ,\eta_1 \right) T_\delta \left( k_2, \eta_2 \right) T_\delta \left( k_3 , \eta_3 \right)
\nonumber \\
&& \qquad \times \ P_R  \left( k_2 \right)P_R  \left( k_3 \right) j''_{\ell} \left( k_2 r_1 \right) j''_{\ell'} \left( k_3 r_1 \right) j_{\ell}\left( k_2 r_2 \right) j_{\ell'} \left( k_3 r_3 \right)
\,.
\eea
These leads to the reduced bispectrum
\be \label{31a}
b_{\ell_1 \ell_2 \ell_3}^{ {v'}^2} = \frac{4}{\pi^2} Z^{{v'}^2}_{\ell_2 \ell_3} \left( z_1 , z_2 ,z_3 \right) + \text{perm.}\, ,
\ee
which can be rewritten in terms of product of angular power spectra
\be\label{ea:v'v'}
b_{\ell_1 \ell_2 \ell_3}^{v'^2}=
2\left( c_{\ell_2}^{v'\delta} \left( z_1, z_2 \right)c_{\ell_3}^{v' \delta} \left( z_1, z_3 \right)
 + c_{\ell_1}^{v' \delta} \left( z_2, z_1 \right)c_{\ell_3}^{v' \delta} \left( z_2, z_3 \right)
 +  c_{\ell_1}^{v' \delta} \left( z_3, z_1 \right)c_{\ell_2}^{v' \delta} \left( z_3, z_2 \right) \right)\, .
\ee

\subsection{Term $ \HH^{-1}  \dd_rv \partial_r \delta $}
\label{ssa:d1v1}
We consider the term
\be\label{ea:vd}
 \left\langle \left( \HH^{-1}  \dd_rv \partial_r \delta\right)\left( \bn_1, z_1 \right) \delta\left( \bn_2, z_2 \right) \delta \left( \bn_3, z_3 \right) \right\rangle_c + \text{perm.}\,.
\ee
Again we write the first factor as a product of Fourier integrals,
\bea
\Delta^{v \delta'} \left( \bn , z \right) &=& \frac{\HH^{-1}}{\left( 2 \pi \right)^6} \int d^3 k d^3k_1 k_1 V \left( \bk, \eta\right) \delta \left( \bk_1, \eta \right) \partial_{\left( k r\right)}e^{i \bk \cdot \bn r } \partial_{\left( k_1r \right)} e^{i \bk_1 \cdot \bn r } \,.
\eea
We now compute~(\ref{ea:vd}) as above by using its Fourier representation
\bea
\langle\cdots\rangle &=& \frac{1}{\HH\left( z_1 \right) \left( 2 \pi \right)^{12}} \int d^3k d^3k_1 d^3k_2 d^3k_3 k_1 \partial_{\left( k r_1 \right)} e^{i \bk \cdot \bn_1 r_1 }
\partial_{\left( k_1 r_1 \right)} e^{i \bk_1 \cdot \bn_1 r_1 }
e^{i\left(   \bk_2 \cdot \bn_2 r_2+\bk_3 \cdot \bn_3 r_3\right)}
\nonumber \\
&& \times \
T_v \left( k, \eta_1 \right) T_\delta \left( k_1 , \eta_1 \right) T_\delta \left( k_2 , \eta_2 \right) T_\delta \left( k_3 , \eta_3 \right)\langle R \left( \bk \right) R \left( \bk_1 \right) R \left( \bk_2 \right) R \left( \bk_3 \right) \rangle
\nonumber \\
&=&
 \frac{1}{\HH \left( z_1 \right)\left( 2 \pi \right)^6} \int d^3k_2 d^3k_3 k_3
 \partial_{\left(k_2 r_1\right)}e^{- i \bk_2 \cdot \bn_1 r_1}
\partial_{\left( k_3 r_1 \right)} e^{-i \bk_3 \cdot \bn_1 r_1}
e^{i \left( \bk_2 \cdot \bn_2r_2 +  \bk_3 \cdot \bn_3r_3 \right)}
\nonumber \\
&& \times \ T_v \left( k_2, \eta_1 \right) T_\delta \left( k_3, \eta_1 \right) T_\delta \left( k_2, \eta_2 \right) T_\delta \left( k_3 , \eta_3 \right)P_R  \left( k_2 \right)P_R  \left( k_3 \right)
\nonumber \\
&&+ \frac{1}{\HH \left( z_1 \right)\left( 2 \pi \right)^6} \int d^3k_2 d^3k_3 k_2  \partial_{\left( k_3 r_1 \right)} e^{- i \bk_3 \cdot \bn_1 r_1}
\partial_{\left( k_2 r_1 \right)} e^{-i \bk_2 \cdot \bn_1 r_1}
e^{i \left( \bk_2 \cdot \bn_2r_2 +  \bk_3 \cdot \bn_3r_3 \right)}
\nonumber \\
&& \times \ T_v \left( k_3, \eta_1 \right) T_\delta\left( k_2, \eta_1 \right) T_\delta \left( k_2, \eta_2 \right) T_\delta \left( k_3 , \eta_3 \right)P_R  \left( k_2 \right)P_R  \left( k_3 \right)
\eea
from which follows
\bea
B^{v \delta'} \left( \bn_1, \bn_2 ,\bn_3, z_1,z_2 , z_3 \right)&=& \frac{4 }{\pi^2} \sum_{\substack{\ell, \ell'\\ m, m'}} Y_{\ell m} \left( \bn_1 \right) Y_{\ell' m'} \left( \bn_1 \right) Y^*_{\ell m } \left( \bn_2 \right) Y^*_{\ell' m'}\left( \bn_3 \right) Z^{v \delta'}_{\ell \ell'} \left( z_1 , z_2 ,z_3 \right)
\nonumber \\
&&
+ \text{perm.}\,,
\eea
where
\bea
Z^{v \delta'}_{\ell \ell'} \left( z_1 , z_2 ,z_3 \right) &=& \frac{1}{\HH \left( z_1 \right)} \int dk_2 dk_3 k_2^2 k_3^3 T_v \left( k_2 ,\eta_1 \right) T_\delta \left( k_3 ,\eta_1 \right) T_\delta \left( k_2, \eta_2 \right) T_\delta \left( k_3 , \eta_3 \right)
\nonumber \\
&& \qquad \times \ P_R  \left( k_2 \right)P_R  \left( k_3 \right) j'_{\ell} \left( k_2 r_1 \right) j'_{\ell'} \left( k_3 r_1 \right) j_{\ell}\left( k_2 r_2 \right) j_{\ell'} \left( k_3 r_3 \right)
\nonumber \\
&& + \frac{1}{\HH \left( z_1 \right)} \int dk_2 dk_3 k_2^3 k_3^2 T_v \left( k_3 ,\eta_1 \right) T_\delta \left( k_2 ,\eta_1 \right) T_\delta \left( k_2, \eta_2 \right) T_\delta \left( k_3 , \eta_3 \right)
\nonumber \\
&& \qquad \times \  P_R  \left( k_2 \right)P_R  \left( k_3 \right) j'_{\ell} \left( k_2 r_1 \right) j'_{\ell'} \left( k_3 r_1 \right) j_{\ell}\left( k_2 r_2 \right) j_{\ell'} \left( k_3 r_3 \right) \, .
\eea
The reduced bispectrum is then given by
\be
b_{\ell_1 \ell_2 \ell_3}^{v \delta'} = \frac{4}{\pi^2} Z^{v\delta'}_{\ell_2 \ell_3} \left( z_1 , z_2 ,z_3 \right) + \text{perm..}
\ee
Also this contribution vanishes if all three redshifts coincide, due to the different parities of the spherical Bessel functions involved.
Using the definition~(\ref{e:cAB}) and the transfer function~(\ref{trans_delta1}) we can express the bispectrum as
\bea
b_{\ell_1 \ell_2 \ell_3}^{ v\delta'}&=&
c_{\ell_2}^{v \delta} \left( z_1, z_2 \right)c_{\ell_3}^{\delta' \delta} \left( z_1, z_3 \right)
+ c_{\ell_2}^{\delta' \delta} \left( z_1, z_2 \right)c_{\ell_3}^{v \delta} \left( z_1, z_3 \right)
\nonumber \\
&&
 + c_{\ell_1}^{v \delta} \left( z_2, z_1 \right)c_{\ell_3}^{\delta' \delta} \left( z_2, z_3 \right)
 + c_{\ell_1}^{\delta' \delta} \left( z_2, z_1 \right)c_{\ell_3}^{v \delta} \left( z_2, z_3 \right)
 \nonumber \\
&&
 + c_{\ell_1}^{v \delta} \left( z_3, z_1 \right)c_{\ell_2}^{\delta' \delta} \left( z_3, z_2 \right)
 + c_{\ell_1}^{\delta' \delta} \left( z_3, z_1 \right)c_{\ell_2}^{v \delta} \left( z_3, z_2 \right)\,. \label{ea:vd'}
\eea

\subsection{Term $\HH^{-1}\dd_r^2 v \  \de$}
\label{ssa:v2d}
We do not compute this term here since this is done in Ref.~\cite{DiDio:2014lka} but we repeat the result for completeness.
\bea
b^{v'\de}_{\ell_1 \ell_2 \ell_3} &=& c_{\ell_2}^{v'\delta} ( z_1, z_2 ) c_{\ell_3}^{\delta\delta} ( z_1,z_3 ) +  c_{\ell_2}^{\delta\delta} (z_1,z_2) c_{\ell_3}^{v'\delta} (z_1, z_3) \nonumber \\
&&\hspace{-6pt}+  c_{\ell_1}^{v'\delta} ( z_2, z_1 ) c_{\ell_3}^{\delta\delta} ( z_2,z_3 ) +  c_{\ell_1}^{\delta\delta} (z_1,z_2) c_{\ell_3}^{v'\delta} (z_2, z_3) \nonumber \\
&& \hspace{-6pt}+  c_{\ell_1}^{v'\delta} ( z_3, z_1 ) c_{\ell_2}^{\delta\delta} ( z_2,z_3 ) +  c_{\ell_1}^{\delta\delta} (z_1,z_3) c_{\ell_2}^{v'\delta} (z_3, z_2) \, ,
\label{ea:v'd}
\eea

\subsection{Term $\HH^{-1}\dd_r^2 v \  \kappa$}
\label{ssa:v2ka}
We compute the term
\be
 \left\langle \left( -2 \HH^{-1} \dd_r^2 v \  \kappa \right)\left( \bn_1, z_1 \right) \delta \left( \bn_2, z_2 \right) \delta\left( \bn_3, z_3 \right) \right\rangle_c + \text{perm.}\,.
\ee
Following the same approach of previous sections we obtain
\bea
\langle ... \rangle &=& - \frac{1}{\HH\left( z_1 \right) \left( 2 \pi \right)^{12}} \int d^3k d^3k_1 d^3k_2 d^3k_3
 \int_0^{r_1} dr' \frac{r_1-r'}{r_1r'}
k  \partial_{\left( k r_1 \right)}^2 e^{i \bk \cdot \bn_1 r_1 }
\nonumber \\
& &\times
\left( \Delta_2 e^{i \bk_1 \cdot \bn_1 r' } \right)
e^{i\left(   \bk_2 \cdot \bn_2 r_2+\bk_3 \cdot \bn_3 r_3\right)}
\nonumber \\
&&\times \
T_v \left( k, \eta_1 \right) T_{\Phi + \Psi} \left( k_1 , \eta' \right) T_\delta \left( k_2 , \eta_2 \right) T_\delta \left( k_3 , \eta_3 \right)\langle R \left( \bk \right) R \left( \bk_1 \right) R \left( \bk_2 \right) R \left( \bk_3 \right) \rangle
\nonumber \\
&=&
- \frac{1}{\HH\left( z_1 \right) \left( 2 \pi \right)^{6}} \int  d^3k_2 d^3k_3
 \int_0^{r_1} dr' \frac{r_1-r'}{r_1r'}
  k_2  \partial_{\left( k_2 r_1 \right)}^2 e^{-i \bk_2 \cdot \bn_1 r_1 }
\left( \Delta_2 e^{-i \bk_3 \cdot \bn_1 r' } \right)
\nonumber \\
&&\times \
e^{i\left(   \bk_2 \cdot \bn_2 r_2+\bk_3 \cdot \bn_3 r_3\right)}
T_v \left( k_2, \eta_1 \right) T_{\Phi + \Psi} \left( k_3 , \eta' \right) T_\delta \left( k_2 , \eta_2 \right) T_\delta \left( k_3 , \eta_3 \right)P_R  \left( k_2 \right)P_R  \left(k_3 \right)
\nonumber \\
&&
- \frac{1}{\HH\left( z_1 \right) \left( 2 \pi \right)^{6}} \int  d^3k_2 d^3k_3
 \int_0^{r_1} dr' \frac{r_1-r'}{r_1r'}
k_3  \partial_{\left( k_3 r_1 \right)}^2 e^{-i \bk_3 \cdot \bn_1 r_1 }
\left( \Delta_2 e^{-i \bk_2 \cdot \bn_1 r' } \right)
\nonumber \\
&&\times \
e^{i\left(   \bk_2 \cdot \bn_2 r_2+\bk_3 \cdot \bn_3 r_3\right)}
T_v \left( k_3, \eta_1 \right) T_{\Phi + \Psi} \left( k_2 , \eta' \right) T_\delta \left( k_2 , \eta_2 \right) T_\delta \left( k_3 , \eta_3 \right)P_R  \left( k_2 \right)P_R  \left(k_3 \right)
\nonumber \\
& &
\eea
which leads to
\bea
B^{v' \kappa} \left( \bn_1, \bn_2 ,\bn_3, z_1,z_2 , z_3 \right)&=& \frac{4 }{\pi^2} \sum_{\substack{\ell, \ell'\\ m, m'}} Y_{\ell m} \left( \bn_1 \right) Y_{\ell' m'} \left( \bn_1 \right) Y^*_{\ell m } \left( \bn_2 \right) Y^*_{\ell' m'}\left( \bn_3 \right) Z^{v' \kappa}_{\ell \ell'} \left( z_1 , z_2 ,z_3 \right)
\nonumber \\
&&
+ \text{perm.}\,,
\eea
where
\bea
&&\hspace{-.7cm}Z^{v' \kappa}_{\ell \ell'} \left( z_1 , z_2 ,z_3 \right)
\nonumber \\
&=&
\frac{1}{\HH \left( z_1 \right)} \int dk_2 dk_3 k_2^3 k_3^2P_R  \left( k_2 \right)P_R  \left( k_3 \right)
 \int_0^{r_1} dr' \frac{r_1-r'}{r_1r'}
T_v \left( k_2 ,\eta_1 \right) T_{\Psi+\Phi} \left( k_3 ,\eta' \right) T_\delta \left( k_2, \eta_2 \right)
\nonumber \\
&& \qquad\qquad \times \  T_\delta \left( k_3 , \eta_3 \right)  \ell' \left( \ell' +1 \right) j'_{\ell} \left( k_2 r_1 \right) j_{\ell'} \left( k_3 r' \right) j_{\ell}\left( k_2 r_2 \right) j_{\ell'} \left( k_3 r_3 \right)
\nonumber \\
&& + \frac{1}{\HH \left( z_1 \right)} \int dk_2 dk_3 k_2^2 k_3^2P_R  \left( k_2 \right)P_R  \left( k_3 \right)
 \int_0^{r_1} dr' \frac{r_1-r'}{r_1r'}
T_v \left( k_3 ,\eta_1 \right) T_{\Psi +\Phi} \left( k_2 ,\eta' \right) T_\delta \left( k_2, \eta_2 \right)
\nonumber \\
&& \qquad\qquad \times \ T_\delta \left( k_3 , \eta_3 \right)  \ell \left( \ell +1 \right) j_{\ell} \left( k_2 r' \right) j'_{\ell'} \left( k_3 r_1 \right) j_{\ell}\left( k_2 r_2 \right) j_{\ell'} \left( k_3 r_3 \right) \, .
\eea
Hence, the reduced bispectrum is
\be
b_{\ell_1 \ell_2 \ell_3}^{v' \kappa } = \frac{4}{\pi^2} Z^{v' \kappa}_{\ell_2 \ell_3} \left( z_1 , z_2 ,z_3 \right) + \text{perm.}\,,
\ee
and it can be rewritten in terms of products of power spectra, using the transfer function in multipole space~(\ref{trans_L}), as
\bea
b_{\ell_1 \ell_2 \ell_3}^{ v' \kappa}&=&
c_{\ell_2}^{v' \delta} \left( z_1, z_2 \right)c_{\ell_3}^{\kappa \delta} \left( z_1, z_3 \right)
+ c_{\ell_2}^{\kappa \delta} \left( z_1, z_2 \right)c_{\ell_3}^{v' \delta} \left( z_1, z_3 \right)
\nonumber \\
&&
 + c_{\ell_1}^{v' \delta} \left( z_2, z_1 \right)c_{\ell_3}^{\kappa \delta} \left( z_2, z_3 \right)
 + c_{\ell_1}^{\kappa \delta} \left( z_2, z_1 \right)c_{\ell_3}^{v' \delta} \left( z_2, z_3 \right)
 \nonumber \\
&&
 + c_{\ell_1}^{v' \delta} \left( z_3, z_1 \right)c_{\ell_2}^{\kappa \delta} \left( z_3, z_2 \right)
 + c_{\ell_1}^{\kappa \delta} \left( z_3, z_1 \right)c_{\ell_2}^{v' \delta} \left( z_3, z_2 \right)\,. \label{ea:v'k}
\eea

\subsection{Term $\kappa\de$}
This term has been computed in detail in Ref.~\cite{DiDio:2014lka}. We repeat only the result here for completeness.
\bea
b_{\ell_1 \ell_2 \ell_3}^{ \kappa\de}&=&
c^{\ka\delta}_{\ell_2} ( z_1, z_2 ) c_{\ell_3}^{\delta\delta} (z_1,z_3) +  c^{\ka\delta}_{\ell_3} ( z_1, z_3 ) c_{\ell_2}^{\delta\delta} (z_1,z_2) \nonumber \\
&& \hspace{-6pt}+ c^{\ka\delta}_{\ell_1} ( z_2, z_1 ) c_{\ell_3}^{\delta\delta} (z_2,z_3) +  c^{\ka\delta}_{\ell_3} ( z_2, z_3 ) c_{\ell_1}^{\delta\delta} (z_1,z_2) \nonumber \\
&&\hspace{-6pt}+ c^{\ka\delta}_{\ell_1} ( z_3, z_1 ) c_{\ell_2}^{\delta\delta} (z_2,z_3) +  c^{\ka\delta}_{\ell_2} ( z_3, z_2 ) c_{\ell_1}^{\delta\delta} (z_1,z_3)\, .
\label{ea:kd}
\eea

\subsection{Term $\kappa^2$}\label{app:kap2}

We consider
\be
 \left\langle \left( 2 \kappa^2 \right)\left( \bn_1, z_1 \right) \delta \left( \bn_2, z_2 \right) \delta\left( \bn_3, z_3 \right) \right\rangle_c
 + \text{perm.} \,,
\ee
from which we compute
\bea
\langle ... \rangle &=&  \frac{1}{2 \left( 2 \pi \right)^{12}} \int d^3k d^3k_1 d^3k_2 d^3k_3
 \int_0^{r_1} dr \frac{r_1-r}{r_1r}
 \int_0^{r_1} dr' \frac{r_1-r'}{r_1r'}
\left( \Delta_2 e^{i \bk \cdot \bn_1 r } \right)
\left( \Delta_2 e^{i \bk_1 \cdot \bn_1 r'} \right)
\nonumber \\
& &\times \
e^{i\left(   \bk_2 \cdot \bn_2 r_2+\bk_3 \cdot \bn_3 r_3\right)}
T_{\Psi+ \Phi} \left( k, \eta \right) T_{\Phi + \Psi} \left( k_1 , \eta' \right)
\nonumber \\
&&\times \
T_\delta \left( k_2 , \eta_2 \right) T_\delta \left( k_3 , \eta_3 \right)\langle R \left( \bk \right) R \left( \bk_1 \right) R \left( \bk_2 \right) R \left( \bk_3 \right) \rangle
\nonumber \\
&=&
\frac{1}{\left( 2 \pi \right)^{6}} \int  d^3k_2 d^3k_3
 \int_0^{r_1} dr \frac{r_1-r}{r_1r}
 \int_0^{r_1} dr' \frac{r_1-r'}{r_1r'}
\left( \Delta_2 e^{-i \bk_2 \cdot \bn_1 r} \right)
\left( \Delta_2 e^{-i \bk_3 \cdot \bn_1 r' } \right)
\nonumber \\
&&\times \
e^{i\left(   \bk_2 \cdot \bn_2 r_2+\bk_3 \cdot \bn_3 r_3\right)}
T_{\Psi + \Phi}\! \left( k_2, \eta \right) T_{\Phi + \Psi}\! \left( k_3 , \eta' \right)
T_\delta \!\left( k_2 , \eta_2 \right) T_\delta\! \left( k_3 , \eta_3 \right)P_R  \!\left( k_2 \right)P_R \! \left(k_3 \right)
\,. \nonumber \\
& &
\eea
Analogously to previous section, we get
\bea
B^{ \kappa^2} \left( \bn_1, \bn_2 ,\bn_3, z_1,z_2 , z_3 \right)&=& \frac{4 }{\pi^2} \sum_{\substack{\ell, \ell'\\ m, m'}} Y_{\ell m} \left( \bn_1 \right) Y_{\ell' m'} \left( \bn_1 \right) Y^*_{\ell m } \left( \bn_2 \right) Y^*_{\ell' m'}\left( \bn_3 \right) Z^{\kappa^2}_{\ell \ell'} \left( z_1 , z_2 ,z_3 \right)
\nonumber \\
&&
+ \text{perm.}\,,
\eea
where
\bea
Z^{\kappa^2}_{\ell \ell'} \left( z_1 , z_2 ,z_3 \right)\! &=& \!\! \int\! dk_2 dk_3 k_2^2 k_3^2P_R  \left( k_2 \right)P_R  \left( k_3 \right)
 \int_0^{r_1} dr \frac{r_1-r}{r_1r}
 \int_0^{r_1} dr' \frac{r_1-r'}{r_1r'}
\nonumber \\
&&\times \
T_{\Psi + \Phi} \left( k_2 ,\eta \right) T_{\Psi+\Phi} \left( k_3 ,\eta' \right) T_\delta \left( k_2, \eta_2 \right) T_\delta \left( k_3 , \eta_3 \right)
\nonumber \\
&& \times \
\ell \left( \ell +1 \right)\ell' \left( \ell' +1 \right) j_{\ell} \left( k_2 r \right) j_{\ell'} \left( k_3 r' \right) j_{\ell}\left( k_2 r_2 \right) j_{\ell'} \left( k_3 r_3 \right) \, .
\label{Zk2}
\eea
Therefore, the reduced bispectrum is
\be
b_{\ell_1 \ell_2 \ell_3}^{ \kappa^2 } = \frac{4}{\pi^2} Z^{\kappa^2}_{\ell_2 \ell_3} \left( z_1 , z_2 ,z_3 \right) + \text{perm.} \,,
\ee
and in terms of products of power spectra it becomes
\be \label{e:bkappa2}
b_{\ell_1 \ell_2 \ell_3}^{ \kappa^2}=
 c_{\ell_2}^{\kappa \delta} \left( z_1, z_2 \right)c_{\ell_3}^{\kappa \delta} \left( z_1, z_3 \right)
 + c_{\ell_1}^{\kappa \delta} \left( z_2, z_1 \right)c_{\ell_3}^{\kappa \delta} \left( z_2, z_3 \right)
 +  c_{\ell_1}^{\kappa \delta} \left( z_3, z_1 \right)c_{\ell_2}^{\kappa \delta} \left( z_3, z_2 \right)
 \,.
\ee
We remark that in Eq.~(\ref{e:Si2}) only the two terms, $\HH^{-2}  \dd_r^2v\partial_r^2v$ and $2 \kappa^2$, enter as squares of first order perturbations, but with a different numerical pre-factor. Indeed in terms of the first order lensing, i.e.~$-2 \kappa$, at second order we have $2 \kappa^2= \left( - 2\kappa \right)^2 /2$.
For this reason Eq.~(\ref{e:bkappa2}) has no pre-factor 2.

\begin{figure}[htbp]
\begin{center}
\includegraphics[width=0.9\textwidth]{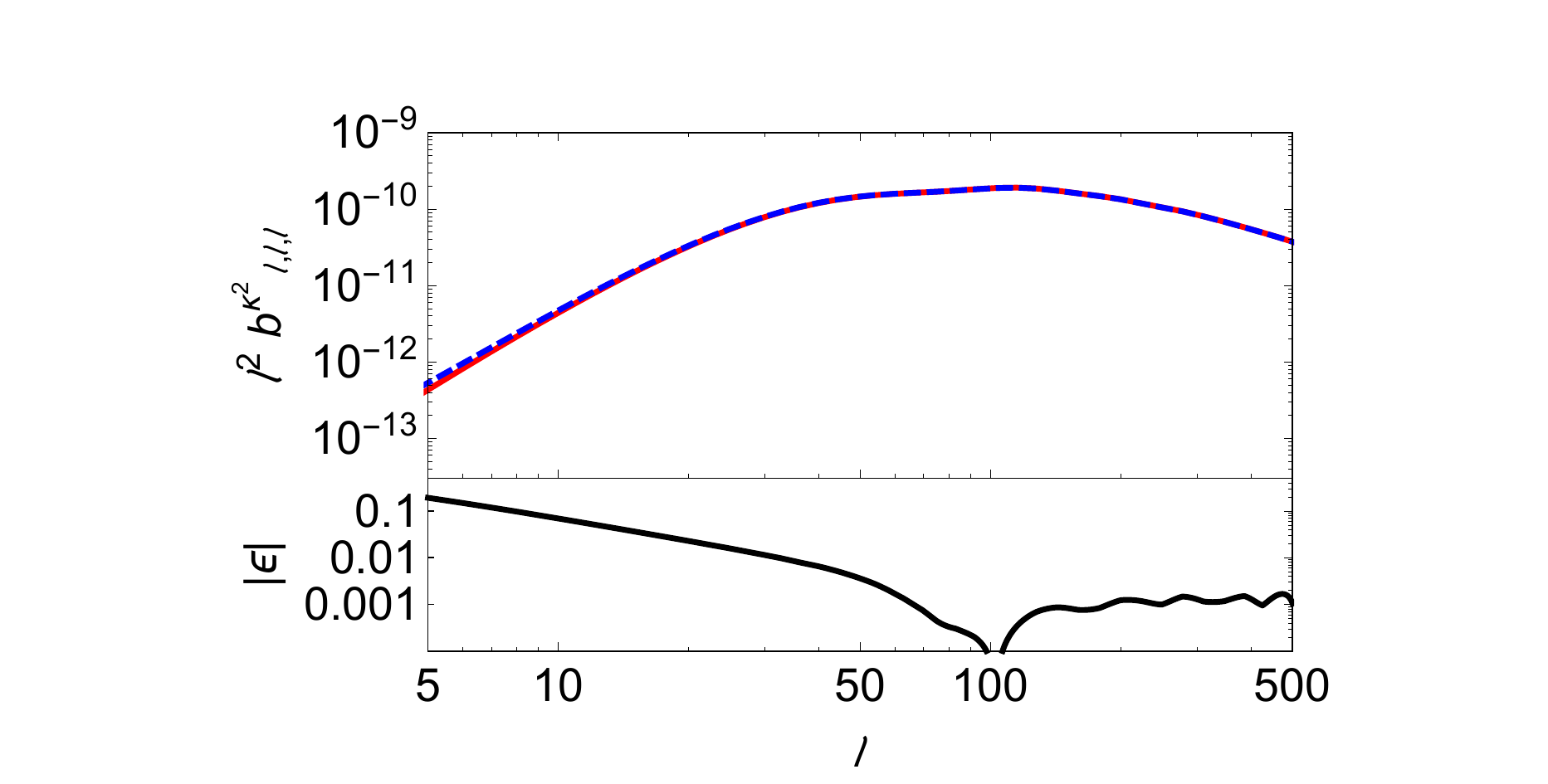}
\caption{{\bf Limber approximation.}  
The Limber approximation is excellent for all lensing like terms. Here we plot (top panel) the Limber approximation (solid red) and the full result (dashed blue) for the term $\ka^2$ evaluated at the mean redshift $z=1$ and with a window function of width $\Delta z=1$, and the relative difference (bottom panel) $\epsilon = 1 - b^\text{Limber}_{\ell,\ell,\ell} / b^\text{exact}_{\ell,\ell,\ell}$.}
\label{fig:limber}
\end{center}
\end{figure}

In order to test the accuracy of the Limber approximation, that we use to evaluate the contribution of $b^{\int \nabla \kappa \nabla \Psi_1}_{\ell_1 \ell_2 \ell_3}$ and $b^{\int \Delta_2 \left( \nabla \Psi_1 \nabla \Psi_1 \right)}_{\ell_1 \ell_2 \ell_3}$ to the total bispectrum with window functions (Fig.~\ref{fig:window}), we have checked it  on the term $b^{ \kappa^2}_{\ell_1 \ell_2 \ell_3}$. Our tests show that the Limber approximation for like-lensing terms in configurations with window functions is very accurate for $\ell\ge20$. The difference between the fully integrated term and the Limber approximation is not visible by eye, see Fig.~\ref{fig:limber}.
Using Eq.~(\ref{e:limber}) we easily obtain the following result for the Limber approximation of (\ref{Zk2})

\bea
\left[Z^{\kappa^2}_{\ell \ell'} \left( z_1 , z_2 ,z_3 \right)\right]^{(\text{Limber})}&=& \Theta(z_1-z_2) \Theta(z_1-z_3)  \frac{\pi^2}{4}
\ell \left( \ell +1 \right)\ell' \left( \ell' +1 \right)\frac{1}{r_2^2 r_3^2} \frac{r_1-r_2}{r_1 r_2}\frac{r_1-r_3}{r_1 r_3}
\nonumber \\
&&\hspace{-2cm}\times
P_R\left(\frac{\ell+1/2}{r_2}\right)
P_R\left(\frac{\ell'+1/2}{r_3}\right)
T_{\Psi + \Phi} \left(\frac{\ell+1/2}{r_2}, \eta_2 \right)
T_{\Psi + \Phi} \left(\frac{\ell'+1/2}{r_3}, \eta_3 \right)
\nonumber \\
&&\hspace{-2cm} \times \
T_{\delta} \left(\frac{\ell+1/2}{r_2}, \eta_2 \right)
T_{\delta} \left(\frac{\ell'+1/2}{r_3}, \eta_3 \right)\,.
\label{ZA8Limber}
\eea
It is interesting to note that here we have no Dirac-$\de$ function, but two Heaviside-$\Theta$ functions due to the double integral along the line of sight.  Therefore this term remains large even if all three redshifts are different, while in the $\ka^{(2)}$-term in the Limber approximation, Eq.~(\ref{RedBispLensOrd2}), at least two redshifts have to be equal for the result to remain substantial.

\subsection{Term $\nabla_a \delta\nabla^a \psi$}\label{ssa:lens}
Next we consider the lensing term,
\be
 \left\langle \left( \nabla_a \delta \nabla^a \psi \right)\left( \bn_1, z_1 \right) \delta \left( \bn_2, z_2 \right) \delta\left( \bn_3, z_3 \right) \right\rangle_c + \text{perm.} \, .
\ee
To profit from the known identities of the spin raising and spin lowering operators, see e.g.~\cite{book}, we
 introduce the helicity basis $\left\{ {\bf e_+}, {\bf e_-} \right\}$, where
\be
 {\bf e_+}= \frac{{\bf e_1} - i {\bf e_2}}{\sqrt{2}}, \quad  {\bf e_-}= \frac{{\bf e_1} + i {\bf e_2}}{\sqrt{2}}
\ee
with $\left\{ {\bf e_1} , {\bf e_2} \right\} = \left\{ \partial_\theta, \frac{1}{\sin \left( \theta \right) } \partial_\phi \right\}$. In terms of the helicity basis the metric on the sphere is determined by $ \gamma_0^{++}=\gamma_0^{--} =0$ and $ \gamma_0^{+-}=\gamma_0^{-+} =1$ so that
\be
\left( \nabla_a \delta\nabla^a\psi \right) = \nabla_+ \delta \nabla_-\psi  + \nabla_- \delta\nabla_+ \psi \, .
\ee
Using  Eq.~(A4.69) in Ref.~\cite{book},
\be
\spart = - \sqrt{2} \nabla_- \quad \text{and} \quad \spart^* = - \sqrt{2} \nabla_+ \,,
\ee
we can rewrite the previous expression in terms of lowering and rising spin operators,
\be
\left( \nabla_a \delta\nabla^a \psi \right) = \frac{1}{2} \spart^* \delta\spart \psi + \frac{1}{2}\spart \delta\spart^*\psi  \, .
\ee
We then compute
\bea
\langle\cdots\rangle&=&
-\frac{1}{\left( 2 \pi \right)^{12}}\int dk^3 dk_1^3 dk_2^3 dk_3^3 \int_0^{r_1} dr \frac{r_1-r}{r_1r}
\nonumber \\
&&\times \
 \frac{1}{2} \left( \spart^* e^{i \bk_1 \cdot \bn_1 r_1}  \spart e^{i \bk \cdot \bn_1 r} + \spart e^{i \bk_1 \cdot \bn_1 r_1} \spart^* e^{i \bk \cdot \bn_1 r} \right) e^{i \left( \bk_2 \cdot \bn_2 r_2 + \bk_3 \cdot \bn_3 r_3\right)}\nonumber \\
&&  \times
T_{\Psi+ \Phi} \left( k, \eta\right) T_\delta \left( k_1 , \eta_1 \right) T_\delta \left( k_2 , \eta_2 \right) T_\delta \left( k_3 , \eta_3 \right)
\langle R \left( \bk \right) R \left( \bk_1 \right) R \left( \bk_2 \right) R \left( \bk_3 \right) \rangle
\nonumber \\
&=&
-\frac{1}{\left( 2 \pi \right)^{6}} \frac{1}{2}\int  dk_2^3 dk_3^3 \int_0^{r_1} dr \frac{r_1-r}{r_1r}
\nonumber \\
&&\times
\left( \spart^* e^{-i \bk_3 \cdot \bn_1 r_1}  \spart e^{-i \bk_2 \cdot \bn_1 r} + \spart e^{-i \bk_3 \cdot \bn_1 r_1} \spart^* e^{-i \bk_2 \cdot \bn_1 r} \right) e^{i \left( \bk_2 \cdot \bn_2 r_2 + \bk_3 \cdot \bn_3 r_3\right)} \nonumber \\
&&\times \
T_{\Psi+ \Phi} \left( k_2, \eta \right) T_\delta \left( k_3 , \eta_1 \right) T_\delta \left( k_2 , \eta_2 \right) T_\delta \left( k_3 , \eta_3 \right)P \left( k_2 \right)P_R  \left( k_3 \right) \nonumber \\
&&
 -\frac{1}{\left( 2 \pi \right)^{6}} \frac{1}{2 }\int  dk_2^3 dk_3^3 \int_0^{r_1} dr \frac{r_1-r}{r_1r}
 \nonumber \\
&&\times
  \left( \spart^* e^{-i \bk_2 \cdot \bn_1 r_1}  \spart e^{-i \bk_3 \cdot \bn_1 r} + \spart e^{-i \bk_2 \cdot \bn_1 r_1} \spart^* e^{-i \bk_3 \cdot \bn_1 r} \right) e^{i \left( \bk_2 \cdot \bn_2 r_2 + \bk_3 \cdot \bn_3 r_3\right)} \nonumber \\
&& \times \
T_{\Psi+ \Phi} \left( k_3, \eta \right) T_\delta \left( k_2 , \eta_1 \right) T_\delta \left( k_2 , \eta_2 \right) T_\delta \left( k_3 , \eta_3 \right)P_R  \left( k_2 \right)P_R  \left( k_3 \right)   \label{ea:depsi} \,.
\eea
We now use the identities
\bea \label{spin-up}
\spart \ _s\!Y_{\ell m } &=& \sqrt{\left( \ell -s \right)  \left( \ell + s + 1 \right) } \ _{s+1}\!Y_{\ell m }\, , \\
\spart^* \ _s\!Y_{\ell m } &=& - \sqrt{\left( \ell +s \right)  \left( \ell - s + 1 \right) } \ _{s-1}\!Y_{\ell m } \, , \label{spin-down}
\eea
which lead to
\bea
\spart e^{i \bk \cdot \bn r} &=& 4 \pi \sum_{\ell m } i^\ell j_\ell \left( k r \right) \sqrt{\ell \left( \ell +1 \right)}  \ _{1}\!Y_{\ell m } \left( \bn \right) Y^*_{\ell m } \left( \hat \bk \right) \, , \\
\spart^* e^{i \bk \cdot \bn r} &=& -4 \pi \sum_{\ell m } i^\ell  j_\ell \left( k r \right) \sqrt{\ell \left( \ell +1 \right)}  \ _{-1}\!Y_{\ell m } \left( \bn \right) Y^*_{\ell m } \left( \hat \bk \right) \, .
\eea
Inserting these identities in~(\ref{ea:depsi}) we find the bispectrum
\bea
B^{\nabla \delta \nabla\psi} \left( \bn_1, \bn_2 ,\bn_3, z_1,z_2 , z_3 \right)&=&
 \frac{4 }{\pi^2} \sum_{\substack{\ell, \ell'\\ m, m'}} \frac{1}{2}\left[_{-1}\!Y_{\ell m} \left( \bn_1 \right) _{1}\!Y_{\ell' m'} \left( \bn_1 \right)
+ _{1}\!Y_{\ell m} \left( \bn_1 \right) _{-1}\!Y_{\ell' m'} \left( \bn_1 \right) \right]
\nonumber \\
&&
\times \
Y^*_{\ell m } \left( \bn_2 \right) Y^*_{\ell' m'}\left( \bn_3 \right) Z^{\nabla \delta \nabla \psi}_{\ell \ell'} \left( z_1 , z_2 ,z_3 \right) + \text{perm.}\,,
\eea
where
\bea
&& \hspace{-0.5cm} Z^{\nabla \delta \nabla \psi}_{\ell \ell'} \left( z_1 , z_2 ,z_3 \right)
\nonumber \\
&=& \quad
  \int dk_2 k_2^2 dk_3 k_3^2 \int_0^{r_1} \hspace{-3mm}dr \frac{r_1-r}{r_1r}
 \sqrt{\ell \left( \ell +1 \right)} \sqrt{\ell' \left( \ell' +1 \right)}
   j_{\ell} \left( k_2r \right) j_{\ell'} \left( k_3 r_1 \right) j_{\ell} \left( k_2 r_2 \right) j_{\ell'}\left( k_3 r_3 \right)
 \nonumber \\
&& \qquad
\times \
T_{\Psi+ \Phi} \left( k_2, \eta \right)
  T_\delta \left( k_3 , \eta_1 \right) T_\delta \left( k_2 , \eta_2 \right) T_\delta \left( k_3 , \eta_3 \right)P_R  \left( k_2 \right)P_R  \left( k_3 \right)
   \nonumber \\
   && +
     \int dk_2 k_2^2 dk_3 k_3^2 \int_0^{r_1} dr \frac{r_1-r}{r_1r}
 \sqrt{\ell \left( \ell +1 \right)} \sqrt{\ell' \left( \ell' +1 \right)}
j_{\ell'} \left( k_3r \right) j_{\ell} \left( k_2 r_1 \right) j_{\ell} \left( k_2 r_2 \right) j_{\ell'}\left( k_3 r_3 \right)
 \nonumber \\
&& \qquad
\times \
  T_{\Psi+ \Phi} \left( k_3, \eta \right) T_\delta \left( k_2 , \eta_1 \right) T_\delta \left( k_2 , \eta_2 \right) T_\delta \left( k_3 , \eta_3 \right)P_R  \left( k_2 \right)P_R  \left( k_3 \right)   \,.
\eea
We now introduce the generalized Gaunt integral
\bea
&&\hspace{-2cm}\int d \Omega  \sY{s_1}{\ell_1} {m_1} \left( \bn \right) \sY{s_2}{\ell_2}{m_2} \left( \bn\right) \sY{s_3}{\ell_3}{m_3} \left( \bn \right)
\nonumber\\
&=& \sqrt{\frac{\left( 2 \ell_1 +1 \right) \left( 2 \ell_2 + 1 \right) \left( 2 \ell_3 +1 \right)}{4 \pi}}
\left( \begin{array}{ccc} \ell_1 & \ell_2 & \ell_3 \\ -s_1 & -s_2 & -s_3 \end{array} \right)
\left( \begin{array}{ccc} \ell_1 & \ell_2 & \ell_3 \\ m_1 & m_2 & m_3 \end{array} \right)\,,
\eea
and we define
\bea\label{def:Afactor}
A_{\ell_1 \ell_2 \ell_3}\equiv\frac{1}{2}\frac{\left[
\left( \begin{array}{ccc} \ell_1 & \ell_2 & \ell_3 \\ 0 & 1 & -1 \end{array} \right) +  \left( \begin{array}{ccc} \ell_1 & \ell_2 & \ell_3 \\ 0 & -1 & 1 \end{array} \right)
\right] }{\left( \begin{array}{ccc} \ell_1 & \ell_2 & \ell_3 \\ 0 & 0 & 0 \end{array} \right)} \, .
\eea
With this we can express the reduced bispectrum as
\bea
b^{\nabla \delta \nabla \psi }_{\ell_1 \ell_2 \ell_3}= \frac{4}{\pi^2} A_{\ell_1 \ell_2 \ell_3 } Z^{\nabla \delta \nabla \psi}_{\ell_2 \ell_3} \left( z_1 , z_2 ,z_3 \right) + \text{perm.} \,.
\eea
This can also be written in terms of the following angular power spectra
\bea
b^{\nabla \delta \nabla \psi}_{\ell_1 \ell_2 \ell_3}&=&A_{\ell_1 \ell_2 \ell_3} \left(\sqrt{\frac{\ell_3 \left( \ell_3 + 1 \right)}{\ell_2 \left( \ell_2 + 1 \right)}} c_{\ell_2}^{\ka \delta} \left( z_1, z_2\right) c_{\ell_3}^{\delta \delta} \left( z_1 , z_3 \right)
+ \sqrt{\frac{\ell_2 \left( \ell_2 + 1 \right)}{\ell_3 \left( \ell_3 + 1 \right)}} c_{\ell_3}^{\ka \delta} \left( z_1, z_3\right) c_{\ell_2}^{\delta \delta} \left( z_1 , z_2 \right) \right)
\nonumber \\
&&
+ A_{\ell_2 \ell_1 \ell_3} \left(\sqrt{\frac{\ell_3 \left( \ell_3 + 1 \right)}{\ell_1 \left( \ell_1 + 1 \right)}} c_{\ell_1}^{\ka \delta} \left( z_2, z_1\right) c_{\ell_3}^{\delta \delta} \left( z_2 , z_3 \right)
+ \sqrt{\frac{\ell_1 \left( \ell_1 + 1 \right)}{\ell_3 \left( \ell_3 + 1 \right)}} c_{\ell_3}^{\ka \delta} \left( z_2, z_3\right) c_{\ell_1}^{\delta \delta} \left( z_2 , z_1 \right) \right)
\nonumber \\
&&
+ A_{\ell_3 \ell_1 \ell_2} \left(\sqrt{\frac{\ell_2 \left( \ell_2 + 1 \right)}{\ell_1 \left( \ell_1 + 1 \right)}} c_{\ell_1}^{\ka \delta} \left( z_3, z_1\right) c_{\ell_2}^{\delta \delta} \left( z_3 , z_2 \right)
+ \sqrt{\frac{\ell_1 \left( \ell_1 + 1 \right)}{\ell_2 \left( \ell_2 + 1 \right)}} c_{\ell_2}^{\ka \delta} \left( z_3, z_2\right) c_{\ell_1}^{\delta \delta} \left( z_3 , z_1 \right) \right)\, .
\nonumber \\  \label{e0:nabla-d-nabla-psi}
&&
\eea

\subsection{Term $\HH^{-1} \nabla_a \partial^2_r v \nabla^a \psi$}
We consider
\be\label{ea:46}
 \langle \left( \HH^{-1} \nabla_a \partial^2_r v \nabla^a \psi\right)\left( \bn_1, z_1 \right) \delta\left( \bn_2, z_2 \right) \delta \left( \bn_3, z_3 \right) \rangle_c + \text{perm.} \, .
\ee
Analogously to the previous case we have
\be
   \HH^{-1} \nabla_a \partial^2_rv \nabla^a \psi = \frac{1}{2} \spart^* \partial^2_rv \spart \psi+ \frac{1}{2}\spart \partial^2_rv \spart^* \psi \, .
\ee
Let us compute~(\ref{ea:46}),
\bea
\langle\cdots\rangle&=&
-\frac{1}{\left( 2 \pi \right)^{12}} \frac{1}{\HH \left( z_1 \right)}\int dk^3 dk_1^3 dk_2^3 dk_3^3 k_1 \int_0^{r_1} dr \frac{r_1-r}{r_1r}
\nonumber \\
&&\times \
 \frac{1}{2} \left( \spart^* \partial^2_{\left( k_1 r_1 \right)} e^{i \bk_1 \cdot \bn_1 r_1}  \spart e^{i \bk \cdot \bn_1 r} + \spart \partial^2_{\left( k_1 r_1 \right)}e^{i \bk_1 \cdot \bn_1 r_1} \spart^* e^{i \bk \cdot \bn_1 r} \right) e^{i \left( \bk_2 \cdot \bn_2 r_2 + \bk_3 \cdot \bn_3 r_3\right)}\nonumber \\
&&  \times \
T_{\Psi+ \Phi} \left( k, \eta\right) T_v \left( k_1 , \eta_1 \right) T_\delta \left( k_2 , \eta_2 \right) T_\delta \left( k_3 , \eta_3 \right)
\langle R \left( \bk \right) R \left( \bk_1 \right) R \left( \bk_2 \right) R \left( \bk_3 \right) \rangle
\nonumber \\
&=&
-\frac{1}{\left( 2 \pi \right)^{6}} \frac{1}{2\HH \left( z_1 \right)}\int  dk_2^3 dk_3^3 k_3 \int_0^{r_1} dr \frac{r_1-r}{r_1r}
\nonumber \\
&&\times
\left( \spart^* \partial^2_{\left( k_3 r_1 \right)}e^{-i \bk_3 \cdot \bn_1 r_1}  \spart e^{-i \bk_2 \cdot \bn_1 r} + \spart \partial^2_{\left( k_3 r_1 \right)}e^{-i \bk_3 \cdot \bn_1 r_1} \spart^* e^{-i \bk_2 \cdot \bn_1 r} \right) e^{i \left( \bk_2 \cdot \bn_2 r_2 + \bk_3 \cdot \bn_3 r_3\right)} \nonumber \\
&&\times \
T_{\Psi+ \Phi} \left( k_2, \eta \right) T_\delta \left( k_3 , \eta_1 \right) T_\delta \left( k_2 , \eta_2 \right) T_\delta \left( k_3 , \eta_3 \right)P \left( k_2 \right)P_R  \left( k_3 \right) \nonumber \\
&&
 -\frac{1}{\left( 2 \pi \right)^{6}} \frac{1}{2 \HH \left( z_1 \right)}\int  dk_2^3 dk_3^3 k_2 \int_0^{r_1} dr \frac{r_1-r}{r_1r}
 \nonumber \\
 && \times
  \left( \spart^* \partial^2_{\left( k_2 r_1 \right)}e^{-i \bk_2 \cdot \bn_1 r_1}  \spart e^{-i \bk_3 \cdot \bn_1 r} + \spart \partial^2_{\left( k_2 r_1 \right)}e^{-i \bk_2 \cdot \bn_1 r_1} \spart^* e^{-i \bk_3 \cdot \bn_1 r} \right) e^{i \left( \bk_2 \cdot \bn_2 r_2 + \bk_3 \cdot \bn_3 r_3\right)} \nonumber \\
&& \times \
T_{\Psi+ \Phi} \left( k_3, \eta \right) T_\delta \left( k_2 , \eta_1 \right) T_\delta \left( k_2 , \eta_2 \right) T_\delta \left( k_3 , \eta_3 \right)P_R  \left( k_2 \right)P_R  \left( k_3 \right) \,.
\eea
Again, $\eta=\eta_0-r$, $\eta_1=\eta_0-r_1=\eta(z_1)$ etc.
With this, the bispectrum is given by
\bea
&&\hspace{-3cm}B^{\nabla v' \nabla \psi} \left( \bn_1, \bn_2 ,\bn_3, z_1,z_2 , z_3 \right)
\nonumber \\
&=& \frac{4 }{\pi^2} \!\sum_{\substack{\ell, \ell'\\ m, m'}} \!\frac{1}{2}\left[_{-1}\!Y_{\ell m} \!\left( \bn_1 \right) _{1}\!Y_{\ell' m'}\! \left( \bn_1 \right) + _{1}\!Y_{\ell m}\! \left( \bn_1 \right) _{-1}\!Y_{\ell' m'} \!\left( \bn_1 \right) \right]
\nonumber \\
&& \times \
Y^*_{\ell m } \!\left( \bn_2 \right) Y^*_{\ell' m'} \! \left( \bn_3 \right) Z^{\nabla v' \nabla \psi}_{\ell \ell'} \!\left( z_1 , z_2 ,z_3 \right)
 ~+ ~ \text{perm.}\,,
\eea
where
\bea
Z^{\nabla v' \nabla \psi}_{\ell \ell'} \left( z_1 , z_2 ,z_3 \right) &=&
  \frac{1}{\HH \left( z_1 \right)} \int dk_2 k_2^2 dk_3 k_3^3 \int_0^{r_1} dr \frac{r_1-r}{r_1r}     \nonumber \\
   && \times \
  T_{\Psi+ \Phi} \left( k_2, \eta\right) T_v \left( k_3 , \eta_1 \right) T_\delta \left( k_2 , \eta_2 \right) T_\delta \left( k_3 , \eta_3 \right)P_R  \left( k_2 \right)P_R  \left( k_3 \right) \nonumber \\
&&\times \
  j_{\ell} \left( k_2r \right) j''_{\ell'} \left( k_3 r_1 \right) j_{\ell} \left( k_2 r_2 \right) j_{\ell'}\left( k_3 r_3 \right)  \sqrt{\ell \left( \ell +1 \right)} \sqrt{\ell' \left( \ell' +1 \right)}
   \nonumber \\
   && +
     \frac{1}{\HH \left( z_1 \right)} \int dk_2 k_2^3 dk_3 k_3^2 \int_0^{r_1} dr \frac{r_1-r}{r_1r}        \nonumber \\
   &&\times \
     T_{\Psi+ \Phi} \left( k_3, \eta\right) T_v \left( k_2 , \eta_1 \right) T_\delta \left( k_2 , \eta_2 \right) T_\delta \left( k_3 , \eta_3 \right)P_R  \left( k_2 \right)P_R  \left( k_3 \right) \nonumber \\
&&\times \
  j_{\ell'} \left( k_3r \right) j''_{\ell} \left( k_2 r_1 \right) j_{\ell} \left( k_2 r_2 \right) j_{\ell'}\left( k_3 r_3 \right)  \sqrt{\ell \left( \ell +1 \right)} \sqrt{\ell' \left( \ell' +1 \right)}  \, .\qquad
\eea
Like in the previous case, with the use of the generalised Gaunt factors, the reduced bispectrum is then given by
\be
b^{ \nabla v' \nabla\psi}_{\ell_1 \ell_2 \ell_3 } = \frac{4}{\pi^2} A_{\ell_1 \ell_2 \ell_3 } Z^{\nabla v' \nabla \psi}_{\ell_2 \ell_3} \left( z_1 , z_2, z_3 \right) + \text{perm.} \, .
\ee
In terms of the angular power spectra this can be rewritten as
\bea
b^{\nabla v' \nabla\psi}_{\ell_1 \ell_2 \ell_3}&=&A_{\ell_1 \ell_2 \ell_3} \left(\!\sqrt{\frac{\ell_3 \left( \ell_3 + 1 \right)}{\ell_2 \left( \ell_2 + 1 \right)}} c_{\ell_2}^{\ka \delta} \left( z_1, z_2\right) c_{\ell_3}^{v' \delta} \left( z_1 , z_3 \right)
\!+\! \sqrt{\frac{\ell_2 \left( \ell_2 + 1 \right)}{\ell_3 \left( \ell_3 + 1 \right)}} c_{\ell_3}^{\ka \delta} \left( z_1, z_3\right) c_{\ell_2}^{v' \delta} \left( z_1 , z_2 \right) \!\right)
\nonumber \\
&&
+ A_{\ell_2 \ell_1 \ell_3} \left(\!\sqrt{\frac{\ell_3 \left( \ell_3 + 1 \right)}{\ell_1 \left( \ell_1 + 1 \right)}} c_{\ell_1}^{\ka \delta} \left( z_2, z_1\right) c_{\ell_3}^{v' \delta} \left( z_2 , z_3 \right)
\!+\! \sqrt{\frac{\ell_1 \left( \ell_1 + 1 \right)}{\ell_3 \left( \ell_3 + 1 \right)}} c_{\ell_3}^{\ka \delta} \left( z_2, z_3\right) c_{\ell_1}^{v' \delta} \left( z_2 , z_1 \right) \!\right)
\nonumber \\
&&
+ A_{\ell_3 \ell_1 \ell_2} \left(\!\sqrt{\frac{\ell_2 \left( \ell_2 + 1 \right)}{\ell_1 \left( \ell_1 + 1 \right)}} c_{\ell_1}^{\ka \delta} \left( z_3, z_1\right) c_{\ell_2}^{v' \delta} \left( z_3 , z_2 \right)
\!+\! \sqrt{\frac{\ell_1 \left( \ell_1 + 1 \right)}{\ell_2 \left( \ell_2 + 1 \right)}} c_{\ell_2}^{\ka \delta} \left( z_3, z_2\right) c_{\ell_1}^{v' \delta} \left( z_3 , z_1 \right) \!\right)\,. \nonumber \\
&& \label{e0:nabla-v-nabla-psi}
\eea

\subsection{Term $ \nabla_b \ka \nabla^b \psi$} \label{sec:6.1}
We consider
\be\label{ea:53}
 \left\langle \left( -2 \nabla_b \ka \nabla^b \psi \right)\left( \bn_1, z_1 \right) \delta \left( \bn_2, z_2 \right) \delta \left( \bn_3, z_3 \right) \right\rangle_c + \text{perm.} \, .
\ee
We start by rewriting
\be
 -2\nabla_b \ka \nabla^b \psi = \left(  \nabla_b \Delta_2 \psi \right) \nabla^b \psi = \frac{1}{2} \spart^* \Delta_2 \psi\spart \psi + \frac{1}{2}\spart \Delta_2 \psi
\spart^* \psi  \, .
\ee
We now compute~(\ref{ea:53})
\bea
\langle\cdots\rangle &=& \frac{1}{\left( 2 \pi \right)^{12}}\int dk^3 dk_1^3 dk_2^3 dk_3^3
\int_0^{r_1} dr \frac{r_1-r}{r_1r} \int_0^{r_1} dr' \frac{r_1-r'}{r_1r'}
\nonumber \\
&&\times \
\frac{1}{2} \left( \spart^* \Delta_2 e^{i \bk \cdot \bn_1 r} \spart e^{i \bk_1 \cdot \bn_1 r'} + \spart \Delta_2 e^{i \bk \cdot \bn_1 r} \spart^* e^{i \bk_1 \cdot \bn_1 r'} \right) e^{i \left( \bk_2 \cdot \bn_2 r_2 + \bk_3 \cdot \bn_3 r_3\right)}
\nonumber \\
&&\times \
T_{\Psi + \Phi} \left( k, \eta \right)T_{\Psi + \Phi} \left( k_1, \eta'\right)  T_\delta \left( k_2 , \eta_2 \right) T_\delta \left( k_3 , \eta_3 \right)
\langle R \left( \bk \right) R \left( \bk_1 \right) R \left( \bk_2 \right) R \left( \bk_3 \right) \rangle
\nonumber \\
&=&
\frac{1}{\left( 2 \pi \right)^6}\int dk_2^3 dk_3^3 \int_0^{r_1} dr \frac{r_1-r}{r_1r}  \int_0^{r_1} dr' \frac{r_1-r'}{r_1r'}
\nonumber \\
&&\times \
\frac{1}{2} \left( \spart^* \Delta_2 e^{-i \bk_2 \cdot \bn_1 r} \spart e^{-i \bk_3 \cdot \bn_1 r'} + \spart \Delta_2 e^{-i \bk_2 \cdot \bn_1 r} \spart^* e^{-i \bk_3 \cdot \bn_1 r'} \right) e^{i \left( \bk_2 \cdot \bn_2 r_2 + \bk_3 \cdot \bn_3 r_3\right)}
\nonumber \\
&&\times \
T_{\Psi + \Phi} \left( k_2, \eta \right)T_{\Psi + \Phi} \left( k_3, \eta' \right)  T_\delta \left( k_2 , \eta_2 \right) T_\delta \left( k_3 , \eta_3 \right)P_R  \left( k_2 \right)P_R  \left( k_3 \right)
\nonumber \\
&&
+
\frac{1}{\left( 2 \pi \right)^6} \int dk_2^3 dk_3^3 \int_0^{r_1} dr \frac{r_1-r}{r_1r}  \int_0^{r_1} dr' \frac{r_1-r'}{r_1r'}
\nonumber \\
&&\times \
\frac{1}{2} \left( \spart^* \Delta_2 e^{-i \bk_3 \cdot \bn_1 r} \spart e^{-i \bk_2 \cdot \bn_1 r'} + \spart \Delta_2 e^{-i \bk_3 \cdot \bn_1 r} \spart^* e^{-i \bk_2 \cdot \bn_1 r'} \right) e^{i \left( \bk_2 \cdot \bn_2 r_2 + \bk_3 \cdot \bn_3 r_3\right)}
\nonumber \\
&&\times \
T_{\Psi + \Phi} \left( k_3, \eta \right)T_{\Psi + \Phi} \left( k_2, \eta' \right)  T_\delta \left( k_2 , \eta_2 \right) T_\delta \left( k_3 , \eta_3 \right)P_R  \left( k_2 \right)P_R  \left( k_3 \right)\,.
\eea
With this the bispectrum becomes
\bea
&&\hspace{-0.5cm} B^{ \nabla_b \ka\nabla^b \psi} \left( \bn_1, \bn_2 ,\bn_3, z_1,z_2 , z_3 \right)
\nonumber \\
&& \qquad  = \frac{4 }{\pi^2} \sum_{\substack{\ell, \ell'\\ m, m'}} \frac{1}{2}\left[_{-1}\!Y_{\ell m} \left( \bn_1 \right) _{1}\!Y_{\ell' m'} \left( \bn_1 \right) + _{1}\!Y_{\ell m} \left( \bn_1 \right) _{-1}\!Y_{\ell' m'} \left( \bn_1 \right) \right]Y^*_{\ell m } \left( \bn_2 \right) Y^*_{\ell' m'}\left( \bn_3 \right)
\nonumber \\
&&  \qquad \quad \times \ Z^{ \nabla_b \ka\nabla^b \psi }_{\ell \ell'} \left( z_1 , z_2 ,z_3 \right) ~+ ~\text{perm.}\,,
\eea
where
\bea
Z^{  \nabla_b \ka \nabla^b \psi}_{\ell \ell'} \left( z_1 , z_2 ,z_3 \right) &=&
 \int dk_2 k_2^2 dk_3 k_3^3 \int_0^{r_1} dr \frac{r_1-r}{r_1r}\int_0^{r_1} dr' \frac{r_1-r'}{r_1r'}  \nonumber \\
 &&\times \
  T_{\Psi+ \Phi} \left( k_2, \eta\right) T_{\Psi+ \Phi} \left( k_3, \eta' \right)  T_\delta \left( k_2 , \eta_2 \right) T_\delta \left( k_3 , \eta_3 \right)P_R  \left( k_2 \right)P_R  \left( k_3 \right) \nonumber \\
&&\times \
  j_{\ell} \left( k_2r \right) j_{\ell'} \left( k_3 r' \right) j_{\ell} \left( k_2 r_2 \right) j_{\ell'}\left( k_3 r_3 \right)  \left(\ell \left( \ell +1 \right)\right)^{3/2} \sqrt{\ell' \left( \ell' +1 \right)}
   \nonumber \\
   && +
 \int dk_2 k_2^3 dk_3 k_3^2\int_0^{r_1} dr \frac{r_1-r}{r_1r} \int_0^{r_1} dr' \frac{r_1-r'}{r_1r'}     \nonumber \\
    &&\times \
    T_{\Psi+ \Phi} \left( k_3, \eta \right) T_{\Psi+ \Phi} \left( k_2, \eta' \right) T_\delta \left( k_2 , \eta_2 \right) T_\delta \left( k_3 , \eta_3 \right)P_R  \left( k_2 \right)P_R  \left( k_3 \right) \nonumber \\
&&\times \
  j_{\ell'} \left( k_3r \right) j_{\ell} \left( k_2 r' \right) j_{\ell} \left( k_2 r_2 \right) j_{\ell'}\left( k_3 r_3 \right)  \sqrt{\ell \left( \ell +1 \right)} \left(\ell' \left( \ell' +1 \right)\right)^{3/2}  . \qquad \
\eea
From which the reduced bispectrum is given by
\be
b^{ \nabla\ka  \nabla\psi }_{\ell_1 \ell_2 \ell_3 } = \frac{4}{\pi^2} A_{\ell_1 \ell_2 \ell_3 } Z^{ \nabla_b \ka \nabla^b \psi}_{\ell_2 \ell_3} \left( z_1 , z_2, z_3 \right) + \text{perm.} \, .
\ee
In terms of power spectra this leads to
\bea
b^{ \nabla\ka \nabla\psi }_{\ell_1 \ell_2 \ell_3}&=&
A_{\ell_1 \ell_2 \ell_3} \left(\sqrt{\frac{\ell_2\left( \ell_2 + 1 \right)}{\ell_3 \left( \ell_3 + 1 \right)}}c_{\ell_2}^{\ka \delta} \left( z_1, z_2\right) c_{\ell_3}^{\ka  \delta} \left( z_1 , z_3 \right)
+ \sqrt{\frac{\ell_3 \left( \ell_3 + 1 \right)}{\ell_2 \left( \ell_2 + 1 \right)}} c_{\ell_3}^{\ka \delta} \left( z_1, z_3\right) c_{\ell_2}^{\ka  \delta} \left( z_1 , z_2 \right) \right)
\nonumber \\
&&
+ A_{\ell_2 \ell_1 \ell_3} \left(\sqrt{\frac{\ell_1 \left( \ell_1 + 1 \right)}{\ell_3 \left( \ell_3 + 1 \right)}}
 c_{\ell_1}^{\ka \delta} \left( z_2, z_1\right) c_{\ell_3}^{\ka \delta} \left( z_2 , z_3 \right)
+ \sqrt{\frac{\ell_3 \left( \ell_3 + 1 \right)}{\ell_1 \left( \ell_1 + 1 \right)}}c_{\ell_3}^{\ka \delta} \left( z_2, z_3\right) c_{\ell_1}^{\ka  \delta} \left( z_2 , z_1 \right) \right)
\nonumber \\
&&
+ A_{\ell_3 \ell_1 \ell_2} \left(\sqrt{\frac{\ell_1 \left( \ell_1 + 1 \right)}{\ell_2 \left( \ell_2 + 1 \right)}} c_{\ell_1}^{\ka \delta} \left( z_3, z_1\right) c_{\ell_2}^{\ka  \delta} \left( z_3 , z_2 \right) +
\sqrt{\frac{\ell_2 \left( \ell_2 + 1 \right)}{\ell_1 \left( \ell_1 + 1 \right)}}
  c_{\ell_2}^{\ka \delta} \left( z_3, z_2\right) c_{\ell_1}^{\ka  \delta} \left( z_3 , z_1 \right) \right) \nonumber \\
  &&=
  A_{\ell_1 \ell_2 \ell_3} \frac{\ell_2\left( \ell_2 + 1 \right) + \ell_3 \left( \ell_3 + 1 \right)}{\sqrt{\ell_2 \left( \ell_2 + 1 \right) \ell_3 \left( \ell_3 + 1 \right)}} c_{\ell_2}^{\ka \delta} \left( z_1, z_2\right) c_{\ell_3}^{\ka  \delta} \left( z_1 , z_3 \right)
  \nonumber \\
  && +
   A_{\ell_2 \ell_1 \ell_3} \frac{\ell_1\left( \ell_1 + 1 \right) + \ell_3 \left( \ell_3 + 1 \right)}{\sqrt{\ell_1 \left( \ell_1 + 1 \right) \ell_3 \left( \ell_3 + 1 \right)}} c_{\ell_1}^{\ka \delta} \left( z_2, z_1\right) c_{\ell_3}^{\ka  \delta} \left( z_2 , z_3 \right) \nonumber \\
   && + A_{\ell_3 \ell_1 \ell_2} \frac{\ell_1\left( \ell_1 + 1 \right) + \ell_2 \left( \ell_2 + 1 \right)}{\sqrt{\ell_1 \left( \ell_1 + 1 \right) \ell_2 \left( \ell_2 + 1 \right)}} c_{\ell_1}^{\ka \delta} \left( z_3, z_1\right) c_{\ell_2}^{\ka  \delta} \left( z_3 , z_2 \right)\, .
\label{ea:nabla-k-nabla-psi}
\eea

\subsection{Term $ \int_0^{r(z)}\frac{dr}{   r} \nabla^b\ka \nabla_b \Psi_1$}
We want to compute
\be
\left\langle \left(-2 \int_0^{r_1} \frac{dr}{   r}\nabla^b \ka \nabla_b \Psi_1 \right) \left( \bn_1, z_1 \right) \delta\left( \bn_2, z_2 \right) \delta\left( \bn_3, z_3 \right) \right\rangle_c + \text{perm.} \, .
\ee
Expressing the perturbation variables in Fourier space, we find
\bea
\langle\cdots\rangle &=&-  \frac{1}{\left( 2 \pi \right)^{12}} \int_0^{r_1} \frac{dr}{ {  r^2}} \int d^3k d^3k_1 d^3k_2 d^3k_3 \int_0^r dr'\frac{r-r'}{{ r}r'} \int_0^r dr''\nonumber \\
&&\times \
\frac{1}{2} \left( \spart^* \Delta_2 e^{i \bk \cdot \bn_1 r'} \spart e^{i \bk_1 \cdot \bn_1 r''} + \spart \Delta_2 e^{i \bk \cdot \bn_1 r'} \spart^* e^{i \bk_1 \cdot \bn_1 r''} \right) e^{i \left( \bk_2 \cdot \bn_2 r_2 + \bk_3 \cdot \bn_3 r_3\right)}
\nonumber \\
&&\times \
T_{\Psi + \Phi} \left( k, \eta' \right)T_{\Psi + \Phi} \left( k_1, \eta'' \right)  T_\delta \left( k_2 , \eta_2 \right) T_\delta \left( k_3 , \eta_3 \right)
\langle R \left( \bk \right) R \left( \bk_1 \right) R \left( \bk_2 \right) R \left( \bk_3 \right) \rangle
\nonumber \\
&=&-  \frac{1}{\left( 2 \pi \right)^6}  \int_0^{r_1} \frac{dr}{{  r^2}} \int  d^3k_2 d^3k_3  \int_0^r dr'\frac{r-r'}{{ r} r'} \int_0^r dr''
\nonumber \\
&&\times \
\frac{1}{2} \left( \spart^* \Delta_2 e^{-i \bk_2 \cdot \bn_1 r'} \spart e^{-i \bk_3 \cdot \bn_1 r''} + \spart \Delta_2 e^{-i \bk_2 \cdot \bn_1 r'} \spart^* e^{-i \bk_3 \cdot \bn_1 r''} \right) e^{i \left( \bk_2 \cdot \bn_2 r_2 + \bk_3 \cdot \bn_3 r_3\right)}
\nonumber \\
&&\times \
T_{\Psi + \Phi} \left( k_2, \eta' \right)T_{\Psi + \Phi} \left( k_3, \eta'' \right)  T_\delta \left( k_2 , \eta_2 \right) T_\delta \left( k_3 , \eta_3 \right)
P \left( k_2 \right)P_R  \left( k_3 \right)
\nonumber \\
&& - \frac{1}{\left( 2 \pi \right)^6}  \int_0^{r_1} \frac{dr}{{  r^2}} \int  d^3k_2 d^3k_3 \int_0^r dr'\frac{r-r'}{ { r} r'} \int_0^r dr''
\nonumber \\
&&\times \
\frac{1}{2} \left( \spart^* \Delta_2 e^{-i \bk_3 \cdot \bn_1 r'} \spart e^{-i \bk_2 \cdot \bn_1 r''} + \spart \Delta_2 e^{-i \bk_3 \cdot \bn_1 r'} \spart^* e^{-i \bk_2 \cdot \bn_1 r''} \right) e^{i \left( \bk_2 \cdot \bn_2 r_2 + \bk_3 \cdot \bn_3 r_3\right)}
\nonumber \\
&&\times \
T_{\Psi + \Phi} \left( k_3, \eta' \right)T_{\Psi + \Phi} \left( k_2, \eta'' \right)  T_\delta \left( k_2 , \eta_2 \right) T_\delta \left( k_3 , \eta_3 \right)
P \left( k_2 \right)P_R  \left( k_3 \right) \,.
\eea
With this we obtain the following expression for  the bispectrum
\bea
&& \hspace{-1cm}B^{\int_0^{r_1} \frac{d r}{r} \nabla^b \ka \nabla_b \Psi_1} \left( \bn_1, \bn_2 ,\bn_3, z_1,z_2 , z_3 \right)
\nonumber \\
&& =\frac{4 }{\pi^2} \sum_{\substack{\ell, \ell'\\ m, m'}} \frac{1}{2}\left[_{-1}\!Y_{\ell m} \left( \bn_1 \right) _{1}\!Y_{\ell' m'} \left( \bn_1 \right) + _{1}\!Y_{\ell m} \left( \bn_1 \right) _{-1}\!Y_{\ell' m'} \left( \bn_1 \right) \right]Y^*_{\ell m } \left( \bn_2 \right) Y^*_{\ell' m'}\left( \bn_3 \right) \nonumber \\
&& \qquad \times  \ Z^{\int_0^{r_1}  {\frac{dr}{r}} \nabla^b \ka \nabla_b \Psi_1}_{\ell \ell'} \left( z_1 , z_2 ,z_3 \right)   ~ + ~ \text{perm.}
\,,
\eea
where
\bea
Z^{\int_0^{r_1} {\frac{dr}{r}} \nabla^b \ka \nabla_b \Psi_1}_{\ell \ell'} \left( z_1 , z_2 ,z_3 \right) &=&
-\int_0^{r_1} \frac{dr}{  {  r^2}} \int dk_2 k_2^2 dk_3 k_3^2 \int_0^r dr'\frac{r-r'}{r r'} \int_0^r dr''\nonumber \\
&&
\hspace{-1.5cm}\times \
T_{\Psi + \Phi}\left( k_2, \eta' \right) T_{\Psi + \Phi}\left( k_3, \eta'' \right) T_\delta \left( k_2, \eta_2 \right) T_\delta \left( k_3 , \eta_3 \right)P_R  \left( k_2 \right)P_R  \left( k_3 \right)
\nonumber \\
&& \hspace{-1.5cm}
\times \left( \ell \left( \ell + 1 \right)\right)^{3/2} \sqrt{\ell' \left( \ell' +1 \right)}  j_\ell \left( k_2 r' \right) j_{\ell'}\left( k_3 r'' \right) j_\ell \left( k_2 r_2 \right) j_{\ell'} \left( k_3 r_3 \right)
\nonumber \\
&& \hspace{-1.5cm}-\int_0^{r_1} \frac{dr}{   {  r^2}} \int dk_2 k_2^2 dk_3 k_3^2 \int_0^r dr'\frac{r-r'}{{r}r'} \int_0^r dr''  \nonumber \\
&&  \hspace{-1.5cm} \times \
T_{\Psi + \Phi}\left( k_3, \eta' \right) T_{\Psi + \Phi}\left( k_2, \eta'' \right) T_\delta \left( k_2, \eta_2 \right) T_\delta \left( k_3 , \eta_3 \right)P_R  \left( k_2 \right)P_R  \left( k_3 \right)
\nonumber \\
&&  \hspace{-2.5cm} \times
\sqrt{\ell \left( \ell + 1 \right)} \left(\ell' \left( \ell' + 1 \right)\right)^{3/2}  j_\ell \left( k_2 r'' \right) j_{\ell'}\left( k_3 r' \right) j_\ell \left( k_2 r_2 \right) j_{\ell'} \left( k_3 r_3 \right) \,.
\label{ZA12}
\eea
Inserting this above, we obtain the reduced bispectrum
\be
b^{\int\nabla\ka\nabla\Psi_1}_{\ell_1 \ell_2 \ell_3 } = \frac{4}{\pi^2} A_{\ell_1 \ell_2 \ell_3 } Z^{\int_0^{r_1}  {\frac{dr}{r}} \nabla^b \ka \nabla_b \Psi_1}_{\ell_2 \ell_3} \left( z_1 , z_2, z_3 \right) + \text{perm.} \, .
\ee
To express this in terms of angular power spectra, we define an additional transfer function in multipole space~(\ref{trans_psi1}).
With this we can rewrite the reduced bispectrum as follows
\bea \label{eq:A10}
b^{\int\nabla\ka\nabla\Psi_1}_{\ell_1 \ell_2 \ell_3 } &=&-
A_{\ell_1 \ell_2 \ell_3 } { \sqrt{ \ell_2 \left( \ell_2+1\right)} \sqrt{\ell_3 \left( \ell_3 + 1 \right)}}
 \nonumber \\
 && \times
  \int_0^{r_1} \frac{dr}{{  r}}
\left[
c_{\ell_2}^{\ka  \delta} \left( z, z_2 \right) c_{\ell_3}^{\Psi_1 \delta}\left( z, z_3\right)
 +
 c_{\ell_3}^{\ka  \delta} \left( z, z_3 \right) c_{\ell_2}^{\Psi_1  \delta}\left( z, z_2\right) \right]
 \nonumber \\
 &&
  -A_{\ell_2 \ell_1 \ell_3 } {  \sqrt{ \ell_1 \left( \ell_1+1\right)} \sqrt{\ell_3 \left( \ell_3 + 1 \right)}}
  \nonumber \\
  &&\times
  \int_{0}^{r_2} \frac{d r}{{  r}}
 \left[
   c_{\ell_1}^{\ka  \delta} \left( z, z_1 \right) c_{\ell_3}^{\Psi_1  \delta}\left( z, z_3\right)
 +  c_{\ell_3}^{\ka  \delta} \left( z, z_3 \right) c_{\ell_1}^{\Psi_1  \delta}\left( z, z_1\right) \right]
 \nonumber \\
 &&
 -A_{\ell_3 \ell_1 \ell_2 } {  \sqrt{ \ell_1 \left( \ell_1+1\right)} \sqrt{\ell_2 \left( \ell_2 + 1 \right)}}
 \nonumber \\
 &&\times
  \int_0^{r_3} \frac{d r}{{  r}}
  \left[
  c_{\ell_1}^{\ka  \delta} \left( z, z_1 \right) c_{\ell_2}^{\Psi_1  \delta}\left( z, z_2\right)
 c_{\ell_2}^{\ka  \delta} \left( z, z_2 \right) c_{\ell_1}^{\Psi_1  \delta}\left( z, z_1\right) \right]\,, \label{ea:int-nabla-k-nabla-psi1}
\eea
where $z=z(r)$ everywhere.

To include this and the following terms in the configuration with window functions we apply Limber approximation. Using the Limber approximation for the $k_2$ and $k_3$ integrals of Eq.~(\ref{ZA12}) we obtain
\bea
&& \hspace{-1cm} \left[ Z^{\int_0^{r_1}  {\frac{dr}{r}} \nabla^b \ka \nabla_b \Psi_1}_{\ell \ell'} \left( z_1 , z_2 ,z_3 \right) \right]^{(\text{Limber})}=
- \frac{\pi^2}{4} \int_0^{r_1} \frac{dr}{  {  r^2}} \frac{r-r_2}{r r_2} \Theta(r-r_2) \Theta(r-r_3)
\nonumber \\
& &\times
\frac{1}{r_2^2 r_3^2}
P_R\left(\frac{\ell+1/2}{r_2}\right)
P_R\left(\frac{\ell'+1/2}{r_3}\right)
T_{\Psi + \Phi} \left(\frac{\ell+1/2}{r_2}, \eta_2 \right)
T_{\Psi + \Phi} \left(\frac{\ell'+1/2}{r_3}, \eta_3 \right)
\nonumber \\
&&\times \
T_{\delta} \left(\frac{\ell+1/2}{r_2}, \eta_2 \right)
T_{\delta} \left(\frac{\ell'+1/2}{r_3}, \eta_3 \right) \left[\ell (\ell+1)\right]^{3/2}
\sqrt{\ell' (\ell'+1)}
\nonumber \\
& &
- \frac{\pi^2}{4} \int_0^{r_1} \frac{dr}{  {  r^2}} \frac{r-r_3}{r r_3} \Theta(r-r_2) \Theta(r-r_3)
\frac{1}{r_2^2 r_3^2}
P_R\left(\frac{\ell+1/2}{r_2}\right)
P_R\left(\frac{\ell'+1/2}{r_3}\right)
\nonumber \\
& &\times \
T_{\Psi + \Phi} \left(\frac{\ell+1/2}{r_2}, \eta_2 \right)
T_{\Psi + \Phi} \left(\frac{\ell'+1/2}{r_3}, \eta_3 \right)
\nonumber \\
&&\times \
T_{\delta} \left(\frac{\ell+1/2}{r_2}, \eta_2 \right)
T_{\delta} \left(\frac{\ell'+1/2}{r_3}, \eta_3 \right) \left[\ell' (\ell'+1)\right]^{3/2}
\sqrt{\ell (\ell+1)} \,.
\eea
The integral over $r$ can now be performed analytically, and we are finally left with
\bea
& & \left[ Z^{\int_0^{r_1}  {\frac{dr}{r}} \nabla^b \ka \nabla_b \Psi_1}_{\ell \ell'} \left( z_1 , z_2 ,z_3 \right) \right]^{(\text{Limber})}=
- \Theta(z_1-z_2) \Theta(z_2-z_3)  \frac{\pi^2}{4}
\frac{1}{r_2^2 r_3^2}
\nonumber \\
&&\times \
P_R\left(\frac{\ell+1/2}{r_2}\right)
P_R\left(\frac{\ell'+1/2}{r_3}\right)
T_{\Psi + \Phi} \left(\frac{\ell+1/2}{r_2}, \eta_2 \right)
T_{\Psi + \Phi} \left(\frac{\ell'+1/2}{r_3}, \eta_3 \right)
\nonumber \\
&&\times \
T_{\delta} \left(\frac{\ell+1/2}{r_2}, \eta_2 \right)
T_{\delta} \left(\frac{\ell'+1/2}{r_3}, \eta_3 \right)\frac{(r_1-r_2)}{2 r_1^2 r_2^2}
\nonumber \\
&&\times
\left[\left(\ell \left( \ell +1 \right)\right)^{3/2} \sqrt{\ell' \left( \ell' +1 \right)}
 (r_1-r_2)+
 \left(\ell' \left( \ell' +1 \right)\right)^{3/2} \sqrt{\ell \left( \ell +1 \right)}
 \frac{2 r_1 r_2-r_3 (r_1+r_2)}{r_3}
   \right]
   \nonumber \\
&&
- \Theta(z_1-z_3) \Theta(z_3-z_2)  \frac{\pi^2}{4}
\frac{1}{r_2^2 r_3^2}
\nonumber \\
&&\times \
P_R\left(\frac{\ell+1/2}{r_2}\right)
P_R\left(\frac{\ell'+1/2}{r_3}\right)
T_{\Psi + \Phi} \left(\frac{\ell+1/2}{r_2}, \eta_2 \right)
T_{\Psi + \Phi} \left(\frac{\ell'+1/2}{r_3}, \eta_3 \right)
\nonumber \\
&&\times \
T_{\delta} \left(\frac{\ell+1/2}{r_2}, \eta_2 \right)
T_{\delta} \left(\frac{\ell'+1/2}{r_3}, \eta_3 \right)\frac{(r_1-r_3)}{2 r_1^2 r_3^2}
\nonumber \\
&&\times \
\left[\left(\ell \left( \ell +1 \right)\right)^{3/2} \sqrt{\ell' \left( \ell' +1 \right)}
\frac{2 r_1 r_3-r_2 (r_1+r_3)}{r_2}+
 \left(\ell' \left( \ell' +1 \right)\right)^{3/2} \sqrt{\ell \left( \ell +1 \right)}
(r_1-r_3)   \right]\,.
   \nonumber \\
&&
\label{ZA12Limber}
 \eea

\subsection{Term $\int_0^{r_s} dr \frac{r_s-r}{{ r_s r}}  \Delta_2 \left(\nabla^b  \Psi_1 \nabla_b \Psi_1 \right) $}\label{ssa:int}
It is convenient to split this term as follows
\bea
&&- \frac{1}{2}\int_0^{r_s} dr \frac{r_s-r}{{ r_s r}}  \Delta_2 \left(\nabla^b  \Psi_1 \nabla_b \Psi_1 \right)
\nonumber \\
&& \hspace{1.8cm} =
-\frac{1}{2}\int_0^{r_s} dr \frac{r_s-r}{{r_s  r}}
\left[ 2 \left(\nabla^b\Delta_2  \Psi_1 \right)\nabla_b \Psi_1 + 2 \nabla^a\nabla^b \Psi_1 \nabla_a \nabla_b \Psi_1 \right]
\nonumber \\
&& \hspace{1.8cm} =
-\frac{1}{2}\int_0^{r_s} dr \frac{r_s-r}{{ r_s r}}
\left[ 2 \left(\nabla^b\Delta_2   \Psi_1 \right)\nabla_b \Psi_1 +  \spart^{*2}  \Psi_1 \spart^2  \Psi_1  + \left( \Delta_2  \Psi_1 \right)^2 \right]\,.  \label{ea:66}
\eea
We want to compute the expectation value of the first of these terms
\be
\left\langle \left( -\int_0^rdr' \frac{r-r'}{{ r r'}}
\left(\nabla^b\Delta_2  \Psi_1 \right)\nabla_b  \Psi_1 \right) \left( \bn_1, z_1 \right) \delta \left( \bn_2, z_2 \right) \delta\left( \bn_3, z_3 \right) \right\rangle_c + \text{perm.} \, .
\ee
We can use the results of section~(\ref{sec:6.1}) finding the bispectrum
\bea
&& \hspace{-1cm}B^{ \int_0^{r_s} dr \frac{r_s-r}{r_s r}  \left((\nabla^b \Delta_2  \Psi_1) \nabla_b  \Psi_1 \right)} \left( \bn_1, \bn_2 ,\bn_3, z_1,z_2 , z_3 \right)
\nonumber \\
&& =\frac{4 }{\pi^2} \sum_{\substack{\ell, \ell'\\ m, m'}} \frac{1}{2}\left[_{-1}\!Y_{\ell m} \left( \bn_1 \right) _{1}\!Y_{\ell' m'} \left( \bn_1 \right) + _{1}\!Y_{\ell m} \left( \bn_1 \right) _{-1}\!Y_{\ell' m'} \left( \bn_1 \right) \right]Y^*_{\ell m } \left( \bn_2 \right) Y^*_{\ell' m'}\left( \bn_3 \right)
\nonumber \\
&&
 \qquad \times  \ Z^{\int_0^{r_s} dr\frac{r_s-r}{r_s r} \left(\nabla^b\Delta_2    \Psi_1  \right)\nabla_b \Psi_1}_{\ell \ell'} \left( z_1 , z_2 ,z_3 \right)
+ \text{perm.}\,,
\eea
where
\bea
Z^{\int_0^{r_s} dr\frac{r_s-r}{r_s r}\left(\nabla^b  \Delta_2   \Psi_1\right) \nabla_b \Psi_1 }_{\ell \ell'} \left( z_1 , z_2 ,z_3 \right) &&
\nonumber \\
&&  \hspace{-4cm}=
-
 \int_0^{r_1} dr\frac{(r_1-r)}{r_1r^3} \int dk_2 k_2^2 dk_3 k_3^2 \int_0^{r} dr'  \int_0^{r} dr''
\nonumber \\
&&  \hspace{-4cm}
\times\Big[T_{\Psi + \Phi}\left( k_2, \eta' \right) T_{\Psi + \Phi}\left( k_3, \eta'' \right) T_\delta \left( k_2, \eta_2 \right) T_\delta \left( k_3 , \eta_3 \right)P_R  \left( k_2 \right)P_R  \left( k_3 \right)
\nonumber \\
&& \hspace{-4cm}  \times  \left( \ell \left( \ell + 1 \right) \right)^{3/2} \sqrt{\ell' \left( \ell' + 1 \right)} j_\ell \left( k_2 r' \right) j_{\ell'}\left( k_3 r'' \right) j_\ell \left( k_2 r_2 \right) j_{\ell'} \left( k_3 r_3 \right) \Big]
\nonumber \\
&&\hspace{-4cm}  -  \int_0^{r_1}dr \frac{(r_1-r)}{r_1r^3}  \int dk_2 k_2^2 dk_3 k_3^2  \int_0^{r} dr'  \int_0^{r} dr''
\nonumber \\
&& \hspace{-4cm}
\times\Big[T_{\Psi + \Phi}\left( k_2, \eta'' \right) T_{\Psi + \Phi}\left( k_3, \eta' \right) T_\delta \left( k_2, \eta_2 \right) T_\delta \left( k_3 , \eta_3 \right)P_R  \left( k_2 \right)P_R  \left( k_3 \right)
\nonumber \\
&& \hspace{-4cm} \times \sqrt{\ell \left( \ell +1 \right)} \left( \ell' \left( \ell' + 1 \right)\right)^{3/2}  j_\ell \left( k_2 r'' \right) j_{\ell'}\left( k_3 r' \right) j_\ell \left( k_2 r_2 \right) j_{\ell'} \left( k_3 r_3 \right)\!\Big].
\label{ZA13p1}
\eea
Again, the reduced bispectrum is
\be
b^{\int(\nabla\De_2\Psi_1)\nabla\Psi_1}_{\ell_1 \ell_2 \ell_3 } = \frac{4}{\pi^2} A_{\ell_1 \ell_2 \ell_3 }
Z^{\int_0^{r_s} dr\frac{r_s-r}{r_s r}\left(\nabla^b  \Delta_2   \Psi_1\right) \nabla_b \Psi_1 }_{\ell_2 \ell_3} \left( z_1 , z_2, z_3 \right) + \text{perm.} \,,
\ee
and in terms of power spectra
\bea
b^{ \int(\nabla\De_2\Psi_1)\nabla\Psi_1}_{\ell_1 \ell_2 \ell_3 } &=& -
A_{\ell_1 \ell_2 \ell_3 } \sqrt{\ell_2 \left( \ell_2 +1 \right) \ell_3 \left( \ell_3 + 1 \right)}\left[ \ell_2 \left( \ell_2 + 1 \right) + \ell_3 \left( \ell_3 + 1 \right) \right]
\nonumber \\
&&
\times \int_0^{r_1} dr\frac{r_1-r}{{  r}r_1}c_{\ell_2 }^{\Psi_1 \delta} \left( z \left(r \right) , z_2 \right)c_{\ell_3}^{\Psi_1 \delta} \left( z \left( r \right) , z_3 \right) \nonumber \\
&&-A_{\ell_2 \ell_1 \ell_3 }
\sqrt{\ell_1 \left( \ell_1 +1 \right) \ell_3 \left( \ell_3 + 1 \right)}
 \left[ \ell_1 \left( \ell_1 + 1 \right) + \ell_3 \left( \ell_3 + 1 \right) \right]
\nonumber \\
&&
\times \int_0^{r_2} dr\frac{r_2-r}{{  r }r_2}c_{\ell_2 }^{\Psi_1 \delta} \left( z \left(r \right) , z_1 \right)c_{\ell_3}^{\Psi_1 \delta} \left( z \left( r \right) , z_3 \right) \nonumber \\
&&-A_{\ell_3 \ell_1 \ell_2 }
\sqrt{\ell_1 \left( \ell_1 +1 \right) \ell_2 \left( \ell_2 + 1 \right)}
 \left[ \ell_1 \left( \ell_1 + 1 \right) + \ell_2 \left( \ell_2 + 1 \right) \right]
 \nonumber \\
 &&
 \times \int_0^{r_3} dr\frac{r_3-r}{{  r}r_3}  c_{\ell_2 }^{\Psi_1 \delta} \left( z \left(r \right) , z_1 \right)c_{\ell_3}^{\Psi_1 \delta} \left( z \left( r \right) , z_2 \right) \,.   \label{ea:term1}
\eea

Next we consider the second term in Eq.~(\ref{ea:66}),
\be
\left\langle \left( -\frac{1}{2}\int_0^r dr' \frac{r-r'}{{ r r'}}
  \spart^{*2} \Psi_1 \spart^2\Psi_1 \right) \left( \bn_1, z_1 \right) \delta\left( \bn_2, z_2 \right) \delta \left( \bn_3, z_3 \right) \right\rangle_c + \text{perm.} \, .
\ee
From~(\ref{spin-up}) and~(\ref{spin-down}) it follows
\bea
\spart\spart e^{i \bk \cdot \bn r} &=& 4 \pi \sum_{\ell m } i^\ell j_\ell \left( k r \right) \sqrt{\frac{\left( \ell +2 \right)!}{\left( \ell -2 \right)!}}  \ _{2}\!Y_{\ell m } \left( \bn \right) Y^*_{\ell m } \left( \hat \bk \right) \, , \\
\spart^*\spart^* e^{i \bk \cdot \bn r} &=& 4 \pi \sum_{\ell m } i^\ell  j_\ell \left( k r \right) \sqrt{\frac{\left( \ell +2 \right)!}{\left( \ell -2 \right)!}}  \ _{-2}\!Y_{\ell m } \left( \bn \right) Y^*_{\ell m } \left( \hat \bk \right) \, .
\eea
With this  we find
\bea
&& \hspace{-1cm}B^{ \int_0^rdr' \frac{r-r'}{r r'}
  \spart^{*2} \Psi_1 \spart^2\Psi_1  } \left( \bn_1, \bn_2 ,\bn_3, z_1,z_2 , z_3 \right)
\nonumber \\
&& =\frac{4 }{\pi^2} \sum_{\substack{\ell, \ell'\\ m, m'}} {  \frac{1}{2}}
\left[ \  _2 \!Y_{\ell m }\left( \bn_1 \right) \ _{-2}\! Y_{\ell' m'} \left( \bn_1 \right)
+ \  _{-2} \!Y_{\ell m }\left( \bn_1 \right) \ _{2}\! Y_{\ell' m'} \left( \bn_1 \right)
 \right]
Y^*_{\ell m } \left( \bn_2 \right) Y^*_{\ell' m'}\left( \bn_3 \right)
\nonumber
\\
& &
\quad
\times  \ Z^{\int_0^r dr' \frac{r-r'}{rr'}
  \spart^{*2} \Psi_1 \spart^2\Psi_1 }_{\ell \ell'} \left( z_1 , z_2 ,z_3 \right)
+ \text{perm.} \, ,
\eea
where
\bea
Z^{\int_0^rdr' \frac{r-r'}{r r'}
  \spart^{*2} \Psi_1 \spart^2\Psi_1 }_{\ell \ell'} \left( z_1 , z_2 ,z_3 \right) &&
  \nonumber \\
  &&\hspace{-5.5cm}   =
   -\int_0^{r_1} dr\frac{r_1-r}{r^3r_1}    \int dk_2 k_2^2 dk_3 k_3^2 \int_{0}^{r} dr'  \int_0^r dr''
     T_{\Psi+ \Phi} \left( k_2, \eta' \right) T_{\Psi+ \Phi} \left( k_3, \eta'' \right)  T_\delta \left( k_2 , \eta_2 \right) T_\delta \left( k_3 , \eta_3 \right)
 \nonumber \\
 &&\hspace{-5.0cm}\times \
P_R  \left( k_2 \right)P_R  \left( k_3 \right)
  j_{\ell} \left( k_2r' \right) j_{\ell'} \left( k_3 r'' \right) j_{\ell} \left( k_2 r_2 \right) j_{\ell'}\left( k_3 r_3 \right) \sqrt{\frac{\left( \ell + 2 \right)!}{\left( \ell - 2 \right)!}}\sqrt{\frac{\left( \ell' + 2 \right)!}{\left( \ell' - 2 \right)!}} \, .
  \label{ZA13p2}
\eea
We now define
\bea
\label{def:Cfactor}
C_{\ell_1 \ell_2 \ell_3}\equiv {  \frac{1}{2}}\frac{
\left( \begin{array}{ccc} \ell_1 & \ell_2 & \ell_3 \\ 0 & 2 & -2 \end{array} \right) +\left( \begin{array}{ccc} \ell_1 & \ell_2 & \ell_3 \\ 0 & -2 & 2 \end{array} \right)  }{\left( \begin{array}{ccc} \ell_1 & \ell_2 & \ell_3 \\ 0 & 0 & 0 \end{array} \right)} \, .
\eea
With this, the reduced bispectrum of the second term is
\be
b^{ \left(\int_0^r dr' \frac{r-r'}{{  rr'}}
  \spart^{*2} \Psi_1 \spart^2\Psi_1 \right) \delta \delta }_{\ell_1 \ell_2 \ell_3 } = \frac{4}{\pi^2} C_{\ell_1 \ell_2 \ell_3 } Z^{\int_0^rdr' \frac{r-r'}{{ r r'}}
  \spart^{*2}\Psi_1 \spart^2\Psi_1 }_{\ell_2 \ell_3} \left( z_1 , z_2, z_3 \right) + \text{perm.} \,,
\ee
and in terms of angular power spectra
\bea
b^{\int\frac{r-r'}{{  r r'}}
  \spart^{*2} \Psi_1 \spart^2\Psi_1 }_{\ell_1 \ell_2 \ell_3 } &=& -C_{\ell_1 \ell_2 \ell_3} \sqrt{\frac{\left( \ell_2 + 2 \right)!}{\left( \ell_2 - 2 \right)!}}\sqrt{\frac{\left( \ell_3 + 2 \right)!}{\left( \ell_3 - 2 \right)!}}
 \int_0^{r_1}dr \frac{r_1-r}{{  r}r_1}  c_{\ell_2 }^{\Psi_1 \delta} \left( z  , z_2 \right)c_{\ell_3}^{\Psi_1 \delta} \left( z , z_3 \right)
\nonumber \\
&&
 -C_{\ell_2 \ell_1 \ell_3} \sqrt{\frac{\left( \ell_1 + 2 \right)!}{\left( \ell_1 - 2 \right)!}}\sqrt{\frac{\left( \ell_3 + 2 \right)!}{\left( \ell_3 - 2 \right)!}}
\int_0^{r_2} dr\frac{r_2-r}{{  r}r_2}c_{\ell_1 }^{\Psi_1 \delta} \left( z, z_1 \right)c_{\ell_3}^{\Psi_1 \delta} \left( z , z_3 \right)
\nonumber \\
&&
 - C_{\ell_3 \ell_1 \ell_2} \sqrt{\frac{\left( \ell_1 + 2 \right)!}{\left( \ell_1 - 2 \right)!}}\sqrt{\frac{\left( \ell_2 + 2 \right)!}{\left( \ell_2 - 2 \right)!}}
\int_0^{r_3}dr \frac{r_3-r}{{  r}r_3}c_{\ell_1 }^{\Psi_1 \delta} \left( z , z_1 \right)c_{\ell_2}^{\Psi_1 \delta} \left( z, z_2 \right)\,. \nonumber \\
  \label{ea:term2}
 \eea

Finally, we compute also the third term of Eq.~(\ref{ea:66}),
\be
\left\langle \left( -\frac{1}{2} \int_0^{r} dr' \frac{r-r'}{r r'}
\left( \Delta_2 \Psi_1 \right)^2\right) \left( \bn_1, z_1 \right) \delta \left( \bn_2, z_2 \right) \delta \left( \bn_3, z_3 \right) \right\rangle_c + \text{perm.} \, .
\ee
We then obtain
\bea
b^{\int\frac{r-r'}{r r'}
\left( \Delta_2 \Psi_1 \right)^2 }_{\ell_1 \ell_2 \ell_3 } &=& \frac{4}{\pi^2} Z^{\int\frac{r-r'}{r r'}
\left( \Delta_2 \Psi_1 \right)^2 }_{\ell_2 \ell_3 }\,,
\eea
with
\bea
Z^{\int\frac{r-r'}{r r'}
\left( \Delta_2 \Psi_1 \right)^2 }_{\ell \ell' }  \left( z_1 , z_2 ,z_3 \right) &&
  \nonumber \\
  &&
\hspace{-4.5cm}= -\!\int_0^{r_1} \!dr\frac{r_1-r}{r^3r_1}   \! \int \!dk_2 k_2^2 dk_3 k_3^3 \!\int_{0}^{r}\! dr' \! \int_0^r \!dr''
     T_{\Psi+ \Phi} \left( k_2, \eta' \right) T_{\Psi+ \Phi} \left( k_3, \eta'' \right)  T_\delta \left( k_2 , \eta_2 \right) T_\delta \left( k_3 , \eta_3 \right)
 \nonumber \\
 &&\hspace{-4.0cm}\times \
P_R  \left( k_2 \right)P_R  \left( k_3 \right)
  j_{\ell} \left( k_2r' \right) j_{\ell'} \left( k_3 r'' \right) j_{\ell} \left( k_2 r_2 \right) j_{\ell'}\left( k_3 r_3 \right) \ell (\ell+1) \ell' (\ell'+1) \, .
   \label{ZA13p3}
   \eea
which leads directly to
\bea
b^{\int\frac{r-r'}{r r'}
\left( \Delta_2 \Psi_1 \right)^2 }_{\ell_1 \ell_2 \ell_3 } &=&
- \ell_2 \left( \ell_2+1 \right) \ell_3 \left( \ell_3 + 1\right)   \int_0^{r_1}dr \frac{(r_1-r)}{{  r}r_1}c^{\Psi_1 \delta}_{\ell_2}\left( z, z_2 \right) c^{\Psi_1 \delta}_{\ell_3}\left( z , z_3 \right)
\nonumber \\
&&
- \ell_1 \left( \ell_1+1 \right) \ell_3 \left( \ell_3 + 1\right)  \int_0^{r_2} dr\frac{(r_2-r)}{{  r}r_2}c^{\Psi_1 \delta}_{\ell_2}\left( z  , z_1 \right) c^{\Psi_1 \delta}_{\ell_3}\left( z, z_3 \right)
\nonumber \\
&&
- \ell_1 \left( \ell_1+ 1\right) \ell_2 \left( \ell_2 + 1\right) \int_0^{r_3}dr \frac{(r_3-r)}{{  r}r_3} c^{\Psi_1\delta}_{\ell_1}\left( z , z_1 \right) c^{\Psi_1 \delta}_{\ell_3}\left( z , z_3 \right) \,.    \qquad  \label{ea:term3}
\eea
The sum of the bispectra given in Eqs.~(\ref{ea:term1}),~(\ref{ea:term2}) and~(\ref{ea:term3}) yields the second integrated lensing term
$b^{\int\Delta_2(\nabla \Psi_1\nabla\Psi_1)}_{\ell_1\ell_2\ell_3} $,  given in Eq.~(\ref{e:last}).

Let us now use the Limber approximation on Eqs.~(\ref{ZA13p1}),~(\ref{ZA13p2}) and~(\ref{ZA13p3}) to simplify these results. After some manipulation we
obtain
\bea
& &\hspace{-1.5cm} \left[Z^{\int_0^{r_s} dr\frac{r_s-r}{r_s r}\left(\nabla^b  \Delta_2   \Psi_1\right) \nabla_b \Psi_1 }_{\ell \ell'} \left( z_1 , z_2 ,z_3 \right) \right]^{(\text{Limber})}
\nonumber \\
& =&
-\left[ \Theta(z_1-z_2) \Theta(z_2-z_3) \frac{(r_1-r_2)^2}{2 r_1^2 r_2^2}
+\Theta(z_1-z_3) \Theta(z_3-z_2) \frac{(r_1-r_3)^2}{2 r_1^2 r_3^2} \right]
\nonumber \\
& &\times
\left[ \sqrt{\ell' \left( \ell' +1 \right)} \left( \ell \left( \ell + 1 \right)\right)^{3/2}+ \sqrt{\ell \left( \ell +1 \right)} \left( \ell' \left( \ell' + 1 \right)\right)^{3/2}\right]
 \frac{\pi^2}{4}
\frac{1}{r_2^2 r_3^2} \nonumber \\
&&\times \
P_R\left(\frac{\ell+1/2}{r_2}\right)
P_R\left(\frac{\ell'+1/2}{r_3}\right)
T_{\Psi + \Phi} \left(\frac{\ell+1/2}{r_2}, \eta_2 \right)
T_{\Psi + \Phi} \left(\frac{\ell'+1/2}{r_3}, \eta_3 \right)
\nonumber \\
&&\times \
T_{\delta} \left(\frac{\ell+1/2}{r_2}, \eta_2 \right)
T_{\delta} \left(\frac{\ell'+1/2}{r_3}, \eta_3 \right)\,,
\eea
\bea
& & \hspace{-1.5cm} \left[  Z^{\int_0^rdr' \frac{r-r'}{r r'}
  \spart^{*2} \Psi_1 \spart^2\Psi_1 }_{\ell \ell'} \left( z_1 , z_2 ,z_3 \right) \right]^{(\text{Limber})}
\nonumber \\
& =&
-\left[ \Theta(z_1-z_2) \Theta(z_2-z_3) \frac{(r_1-r_2)^2}{2 r_1^2 r_2^2}
+\Theta(z_1-z_3) \Theta(z_3-z_2) \frac{(r_1-r_3)^2}{2 r_1^2 r_3^2} \right]
\nonumber \\
& &\times
\sqrt{\frac{(\ell+2)!}{(\ell-2)!}}\sqrt{\frac{(\ell'+2)!}{(\ell'-2)!}}
 \frac{\pi^2}{4}
\frac{1}{r_2^2 r_3^2} \nonumber \\
&&\times \
P_R\left(\frac{\ell+1/2}{r_2}\right)
P_R\left(\frac{\ell'+1/2}{r_3}\right)
T_{\Psi + \Phi} \left(\frac{\ell+1/2}{r_2}, \eta_2 \right)
T_{\Psi + \Phi} \left(\frac{\ell'+1/2}{r_3}, \eta_3 \right)
\nonumber \\
&&\times \
T_{\delta} \left(\frac{\ell+1/2}{r_2}, \eta_2 \right)
T_{\delta} \left(\frac{\ell'+1/2}{r_3}, \eta_3 \right)\,,
\eea
\bea
& & \hspace{-1.5cm} \left[ Z^{\int\frac{r-r'}{r r'}
\left( \Delta_2 \Psi_1 \right)^2 }_{\ell \ell' }  \left( z_1 , z_2 ,z_3 \right) \right]^{(\text{Limber})}
  \nonumber \\
  &=&
-\left[ \Theta(z_1-z_2) \Theta(z_2-z_3) \frac{(r_1-r_2)^2}{2 r_1^2 r_2^2}
+\Theta(z_1-z_3) \Theta(z_3-z_2) \frac{(r_1-r_3)^2}{2 r_1^2 r_3^2} \right]
\nonumber \\
& &\times \
\ell (\ell+1) \ell' (\ell'+1)
 \frac{\pi^2}{4}
\frac{1}{r_2^2 r_3^2} \nonumber \\
&&\times \
P_R\left(\frac{\ell+1/2}{r_2}\right)
P_R\left(\frac{\ell'+1/2}{r_3}\right)
T_{\Psi + \Phi} \left(\frac{\ell+1/2}{r_2}, \eta_2 \right)
T_{\Psi + \Phi} \left(\frac{\ell'+1/2}{r_3}, \eta_3 \right)
\nonumber \\
&&\times \
T_{\delta} \left(\frac{\ell+1/2}{r_2}, \eta_2 \right)
T_{\delta} \left(\frac{\ell'+1/2}{r_3}, \eta_3 \right)\,.
\eea

\section{Magnification Bias}
\label{app:B}

In  an experiment  the observed number of galaxies does not only depend on the true number of galaxies but also on their luminosity, since the experiment has a limiting sensitivity.  It can therefore happen that an intrinsically less luminous galaxy which is magnified by foreground matter concentrations makes it into a survey whereas another intrinsically more luminous one which is de-magnified is below the flux limit of the experiment. This phenomenon, called magnification bias, is relevant as long as an experiment cannot see 'all galaxies'.

Denoting the number of galaxies per redshift bin and solid angle above a flux limit $F$ by
$N \left( \vn, z, F \right) $ and the number of galaxies with intrinsic luminosity $L$ by $N \left( \vn, z, L \right)$, following \cite{DiDio:2014lka}, we have to second order
\bea
\hspace*{-1cm}N \left( \vn, z, \bar F \right)  &=& N \left( \vn, z,F \right)  + \frac{\dd}{\dd L}N \left( \vn, z, \bar L \right) (\de L^{(1)} +\de L^{(2)})  + \frac{1}{2} \frac{\dd^2}{\dd L^2}N \left( \vn, z, \bar L \right)(\de L^{(1)})^2 \\
 &=& N(\bar z, \bar L)\left[1 + \De^{(1)}+ \De^{(2)} + \frac{\partial_L \bar \rho}{\bar \rho} \left(\delta L^{(1)}+\delta L^{(2)}\right)  +\frac{1}{2} \frac{\partial_L^2\bar \rho}{\bar \rho}
\left(\delta L^{(1)}\right)^2  + \right.  \nonumber \\   &&
  \qquad \qquad  \left. + \frac{(\partial_L \rho  -\partial_L\bar\rho)^{(1)}}{\bar \rho}\delta L^{(1)}
  + \frac{\partial_\eta\left(\partial_L \bar \rho\right)}{\bar \rho} \delta L^{(1)} \frac{\de z^{(1)}}{\HH}
      \right]\,,
\label{GenDefrho2order}
\eea
where $\de z^{(1)}$ is the first order redshift perturbation given by (see \cite{DiDio:2014lka})
\be
\de z^{(1)} = -\dd_rv^{(1)}  -\Psi^{(1)}  -2\int_0^{r(z)}dr\dd_\eta\Phi^{(1)} _W \,.
\ee
Using that the flux is given by $F=L/(4 \pi d_L^2) = \bar L/(4 \pi \bar d_L^2)$, we have that at fixed flux the fluctuation of the intrinsic luminosity $L$ is given by the fluctuation of the luminosity distance squared.
The perturbation of the luminosity distance to first order  has been calculated in Ref.~\cite{Bonvin:2005ps} and the result to second order can be found in~\cite{BenDayan:2012wi,Fanizza:2013doa,Marozzi:2014kua}.
Introducing also
\be
\left(\frac{\partial \ln \bar{\rho}}{\partial \ln L}\right)({z,\bar{L}})=-\frac{5}{2} s\left(z, \bar{L}\right)\,,  \quad \quad
\frac{\partial^2}{\partial (\ln L)^2}\left( \ln \bar{\rho}\right)(z,\bar{L})=-\frac{5}{2} t\left(z, \bar{L}\right)\,,
\label{MB-p1}
\ee
\be
(1+\de^{(1)})\frac{\partial \ln{\rho}}{\partial \ln L}-\frac{\partial \ln \bar{\rho}}{\partial \ln L} =-\frac{5}{2}(\de s)^{(1)}(z, \bar{L})\,,
\label{MB-p2}
\ee
one obtains the following result to first order in perturbations (we neglect anisotropic stresses so that $\Phi^{(1)}=\Psi^{(1)}$, for the leading terms this make no difference)
\begin{eqnarray}
\Delta^{(1)}(\bn,z) &=&  \left( \frac{2-5 s}{\mathcal{H} r(z)}+5s+\frac{\mathcal{H}'}{\mathcal{H}^2} \right) \!\!\! \left( \partial_r v^{(1)}
+ \Psi^{(1)} + 2 \int_0^{r(z)} dr \partial_{\eta} \Psi^{(1)}  \right)
+(5 s - 1)\Psi^{(1)}   \nonumber \\
& & + (2- 5 s)\left(2  \Psi_1
- \ka^{(1)} \right)
+ \frac{1}{\mathcal{H}} \left( \partial_\eta\Psi^{(1)}+ \partial^2_r  v^{(1)} \right)
+\delta^{(1)}\,.
\label{Exp-MB-Order1}
\end{eqnarray}
This result has been originally derived in~\cite{Challinor:2011bk}, see also~\cite{DiDio:2013bqa} for details.

At second order one then obtains the following result for the
 leading  contributions
\bea
\Sigma^{(2)}(\bn,z) &=& \delta^{(2)} + \HH^{-1} \partial^2_r v^{(2)} -2\left(1-\frac{5}{2} s\right) \ka^{(2)}
+ \HH^{-2} \left[\left( \partial^2_r v\right)^2+ \dd_rv \partial_r^3v \right]
\nonumber \\
&&  + \HH^{-1} \left(\dd_r v \partial_r \delta+\partial^2_r v \, \delta\right)
-2 \delta \ka +5\,(\de s)^{(1)} \ka
+ \nabla_a \delta \nabla^a \psi
 \nonumber
 \\
 &&
  +\HH^{-1}\left[-2\left(1-\frac{5}{2} s\right) \partial^2_r v \,  \ka+\nabla_a \partial^2_r v \nabla^a \psi \right]
+2 \left(1-5 s +\frac{25}{4} s^2-\frac{5}{2} t \right)  \ka^2
\nonumber
\\
& &
-2\left(1-\frac{5}{2} s\right)\nabla_b \ka \nabla^b \psi
 - \left(1-\frac{5}{2} s\right)\frac{1}{2r(z)} \int_0^{r(z)} \!\!dr\frac{r(z)-r}{r} \Delta_2 \left( \nabla^b \Psi_1 \nabla_b \Psi_1 \right)
\nonumber
\\
& &
- 2\left(1-\frac{5}{2} s\right)\int_0^{r(z)} \frac{dr}{ r^2}  \nabla^a \Psi_1\nabla_a \ka
\,,
\label{Exp-MB-Order2}
\eea

If the number of galaxies depend on luminosity as a power law, $\rho \propto L^p$,
Eqs.~(\ref{MB-p1}) and~(\ref{MB-p2}) imply $s=-2p/5$, $t=0$ and $(\de s)^{(1)} = s \de^{(1)} =  -2p\de^{(1)}/5$.
It is interesting to note that the lensing terms vanish with $p=-1$, or equivalently $s=2/5$ at first order.
Whereas, at second order there are non-vanishing lensing terms which contribute to the galaxy number counts. Indeed the terms $\nabla_a \delta \nabla^a \psi +\HH^{-1} \nabla_a \partial^2_r v \nabla^a \psi$  are not affected by magnification bias.

\section{Intensity mapping}
\label{app:C}

In this section we show how to use the formalism developed in this work in the context of radio surveys measuring the intensity of the 21cm emission line of neutral hydrogen, which can be used as a biased tracer of the dark matter field. This complementary observable does not resolve individual galaxies. It is based on the measurement of photon flux intensity, or equivalently on surface brightness temperature mapping.

To extend the calculation presented in~\cite{DiDio:2014lka} and in the current work we use the fact that, as shown in~\cite{Hall:2012wd}, the surface brightness temperature $T_b$ is directly proportional to the density distribution of the neutral atomic hydrogen $\Delta_{n_\text{HI}}$ and inverse proportional to the square of the angular diameter distance, namely
\be
T_b \left( \bn , z \right) \propto \frac{\Delta_{n_\text{HI}} \left( \bn , z \right)}{d_A^2 \left( \bn , z \right)} =  \frac{\Delta_{n_\text{HI}} \left( \bn , z \right)}{\left( 1 + z \right)^4 d_L^2 \left( \bn , z \right)} \, .
\ee
Hence, we need consider the luminosity distance to second order~\cite{BenDayan:2012wi,Fanizza:2013doa,Marozzi:2014kua} to compute the temperature fluctuations at the same order in perturbation theory.
At first order one obtains~\cite{Hall:2012wd}
\be
\Delta^{(1)}_T=  \left( 2+\frac{\mathcal{H}'}{\mathcal{H}^2} \right) \!\!\! \left( \partial_r v^{(1)} + \Psi^{(1)} + 2 \int_0^{r(z)} \!\! dr \partial_{\eta} \Psi^{(1)}  \right)
+ \Psi^{(1)}
+ \frac{1}{\mathcal{H}} \left( \partial_\eta \Psi^{(1)} + \partial^2_r  v^{(1)} \right)
+\delta^{(1)} \,.
\label{Exp-MB-Order1-No-Lensing}
\ee
(we have again neglected anisotropic stress), and at second order we find for the leading terms
\bea
\Sigma^{(2)}_T(\bn,z) &=& \delta^{(2)} + \HH^{-1} \partial^2_r v^{(2)}
+ \HH^{-2} \left[\left( \partial^2_r v\right)^2+ \dd_rv \partial_r^3v \right]
\nonumber \\
&&  + \HH^{-1} \left(\dd_r v \partial_r \delta+\partial^2_r v \, \delta\right)
+ \nabla_a \delta \nabla^a \psi
  +\HH^{-1} \nabla_a \partial^2_r v \nabla^a \psi
\,.
\label{Leading-Exp-MB-Order2-almost-no-Lensing}
\eea
This expression agrees with~\cite{Umeh:2015gza}. Let us note that the result for
2nd order intensity mapping can be obtained from the one for galaxy number counts by simply setting $s=2/5$, $(\de s)^{(1)}=2/5\,\de^{(1)}$ and $t=0$.
From Eqs.~(\ref{Exp-MB-Order1-No-Lensing}, \ref{Leading-Exp-MB-Order2-almost-no-Lensing}) it is evident that there is no lensing term for intensity mapping at first order (as a consequence of photons conservation), whereas there are  non-vanishing lensing terms at second order.
These terms, given by the product of the deflection angle $ \nabla^a \psi$ with gradients of density and redshift space distortion, are analogous ones of those obtained when studying lensing of the cosmic microwave background (CMB),  one just has to replace the temperature fluctuations, $\de T/T$ by $\delta^{(1)} + \HH^{-1} \partial^2_r v^{(1)}$.  They simply stem from the fact that we have to evaluate the density and redshift space distortion in the unlensed direction $\bn +\bnabla\psi$, see e.g.~\cite{book}.
However, contrary to CMB analysis, when computing the second order power spectrum or the bispectrum the correlations between the deflection angles and density fluctuations can not be neglected in HI intensity mapping.

\bibliographystyle{JHEP}
\bibliography{biblio_bispectrum}

\end{document}